\documentclass[floatfix,twocolumn,showpacs,preprintnumbers,amsmath,amssymb,pre,superscriptaddress]{revtex4-1}
\usepackage{color}
\usepackage[usenames,dvipsnames,svgnames,table]{xcolor}
\usepackage[colorlinks=true,linkcolor=blue,urlcolor=blue,citecolor=blue]{hyperref}

\usepackage{graphicx}
\usepackage{dcolumn}
\usepackage{bm}
\usepackage{subfigure}
\usepackage{amssymb}
\usepackage{multirow}
\usepackage{amsmath}
\usepackage{braket}
\graphicspath{{plots/}}

\begin{document}

\title{\textit{Ab initio} Path Integral Monte Carlo Simulations of Quantum Dipole Systems in Traps: Superfluidity, Quantum Statistics, and Structural Properties}

\author{Tobias Dornheim}
\email{t.dornheim@hzdr.de}

\affiliation{Center for Advanced Systems Understanding (CASUS), G\"orlitz, Germany}








\begin{abstract}
We present extensive \textit{ab initio} path integral Monte Carlo (PIMC) simulations of two-dimensional quantum dipole systems in a harmonic confinement, taking into account both Bose- and Fermi-statistics. This allows us to study the nonclassical rotational inertia, which can lead to a negative superfluid fraction in the case of fermions [Phys.~Rev.~Lett.~\textbf{112}, 235301 (2014)]. Moreover, we study in detail the structural characteristics of such systems, and are able to clearly resolve the impact of quantum statistics on density profiles and the respective shell structure. Further, we present results for a more advanced center-two particle correlation function [Phys.~Rev.~E \textbf{91}, 043104 (2015)], which allows to detect differences between Fermi- and Bose-systems that do not manifest in other observables like the density.
Overall, we find that bosonic systems sensitively react to even small values of the dipole--dipole coupling strength, whereas such a weak interaction is effectively masked for fermions by the Pauli exclusion principle. In addition, the abnormal superfluid fraction for fermions is not reflected by the structural properties of the system, which are equal to the bosonic case even though the moments of inertia diverge from each other. Lastly, we have demonstrated that fermionic PIMC simulations of quantum dipole systems are feasible despite the notorious fermion sign problem, which opens up new avenues for future investigations in this field.

\end{abstract}

\maketitle

\section{Introduction}

Quantum dipole systems are of high current interest for many applications, with indirect excitons in quantum wells~\cite{snoke,cohen,exciton1,alex_PRL,exciton2,exciton3,exciton4}, Rydberg-dressed atoms~\cite{Ryd1,Ryd2}, and ultracold dipolar gases~\cite{stuhler,griesmaier,dynamic_alex1,dynamic_alex2} being arguably the three most important examples.
These systems are known to exhibit a plethora of remarkable physical effects, such as superfluidity~\cite{cep,exciton1,alex_PRL,jain} and possibly even supersolid behaviour~\cite{rotating,supersolid}, crystallization~\cite{boninsegni_dipole_crystal,crystal_review}, and collective excitations~\cite{dynamic_alex1,dynamic_alex2}.

From a theoretical perspective, the accurate description of quantum dipole systems constitutes a formidable challenge, as it must simultaneously take into account i) the long-range dipole--dipole interaction, ii) thermal excitations, and iii) quantum degeneracy effects. More specifically, point i) possibly rules out mean-field approaches~\cite{mean_field,mean_field2} when the coupling strength is increased, and point iii) rules out classical methods like molecular dynamics and is crucial for a correct description of, e.g., Bose-Einstein-condensation~\cite{jain,bec_review}.

In this regard, a reliable theory for quantum-dipole systems faces similar challenges as for warm dense matter (WDM)~\cite{new_wdm_paper,wdm_book}---an exotic state that is at the forefront of plasma physics and material science~\cite{fortov_review,falk_wdm}. A suitable candidate are \textit{ab initio} path integral Monte Carlo (PIMC) methods~\cite{cep,berne3}, which, in principle can deal with the effects i)-iii) without any approximations. Indeed, quasi-exact simulations of up to $N\sim10^4$ bosons are feasible~\cite{boninsegni1,boninsegni2}, and the PIMC approach has been vital for our current understanding of, e.g., superfluidity~\cite{sf2,sindzingre,cep} and collective excitations~\cite{dynamic_alex1,supersolid_spectrum,dornheim_dynamic,dynamic_folgepaper,dynamic_FSC}.

Yet, PIMC simulations of fermions are severely limited by the notorious fermion sign problem (FSP)~\cite{loh,troyer}, which leads to an exponential increase in computation time with a) decreasing temperature $T$ and b) increasing the system size $N$, see Ref.~\cite{dornheim_sign_problem} for an accessible topical review article. For this reason, almost no (if any) PIMC simulations of fermionic quantum dipole systems have been reported so far. This is very unfortunate, as they offer many potentially interesting effects.

On the other hand, the high level of activity in WDM research has triggered a remarkable spark of new developments regarding the quantum Monte Carlo simulation of electrons (which, too, are fermions and thus afflicted with the FSP) at finite temperature~\cite{cpimc_original, brown_ethan, blunt1, schoof_prl, malone1, blunt2, malone2, dornheim, dornheim2, vladimir_UEG, groth, dornheim3, dornheim_prl, groth_prl, dubois, claes, dornheim_cpp, brenda, universe, dornheim_neu}, see Ref.~\cite{dornheim_pop} for a recent overview of some of these methods. Moreover, Dornheim~\cite{dornheim_sign_problem} has reported that fermions with dipolar interaction exhibit a substantially less severe FSP compared to electrons with Coulomb interaction, which is a strong indication that the success of PIMC simulation of WDM might be carried over into this field.

In this context, we have performed extensive \textit{ab initio} PIMC simulations of both bosonic and fermionic quantum dipole systems in a $2D$ harmonic confinement. In addition to its worth as a proof-of-principle study, we mention that the investigation of trapped quantum systems is quite interesting in its own right~\cite{blume_review} and constitutes an active research field, see, e.g., Refs.~\cite{dornheim,JWA1,JWA2,dornheim_c2p,alex_wigner, reimann_wigner, reimann_review,egger,ghosal,ilkka,dyuti1,dyuti2}. More specifically, we study the interesting and intricate interplay of the dipolar repulsion with quantum statistics by computing different quantities. First and foremost, we study both the total~\cite{sindzingre,yushi} and local superfluid fraction~\cite{kwon_lsf,mezza,filinov_lsf,lsf,dornheim_superfluid} in dependence of system size, temperature, and coupling strength. Remarkably, we find that the superfuid fraction can be negative in the case of fermions, which is in good agreement to previous results~\cite{blume} for ultracold atoms with a different pair potential. In addition, we investigate the impact of quantum statistics on the structural properties of the system, like the radial density distribution $n(r)$, and a somewhat more advanced center-two particle (C2P) distribution function~\cite{dornheim_c2p,thomsen_c2p,ott,Hauke_PHD}.
This allows us to defy common wisdom and demonstrate that PIMC simulations are indeed capable to clearly resolve the effect of quantum statistics on physical observables upon increasing the degree of quantum degeneracy despite the FSP.

While the present study is restricted to finite systems in a harmonic confinement, similar investigations can be performed for bulk systems in periodic boundary conditions, and we hope to spark more research in this direction. Furthermore, our highly accurate PIMC data can be used as a benchmark for approximate theories and to guide the development of new simulation methods~\cite{hirshberg}.

The paper is organized as follows: In Sec.~\ref{sec:theory}, we introduce the required theoretical background, including the Hamiltonian (Sec.~\ref{sec:Hamiltonian}), the PIMC method and how it is afflicted with a sign problem in the case of fermions~(\ref{sec:PIMC_theory}), how we estimate the superfluid fraction in terms of the non-classical rotational inertial (\ref{sec:ncri_theory}), and the C2P function that allows us to study the structural properties of trapped quantum systems (\ref{sec:c2p_theory}). In addition, we give some formulas for noninteracting systems for both bosons and fermions in Sec.~\ref{sec:ideal_theory}, which are helpful to interpret our results and as a benchmark for our implementation.
Sec.~\ref{sec:results} is devoted to the presentation of our extensive new PIMC results, starting with the in principle well-known, yet still interesting case of ideal particles in Sec.~\ref{sec:ideal_results}. Here, we compare PIMC data to exact theoretical results and demonstrate the utility of the C2P function as a diagnostic for quantum degeneracy effects. In Sec.~\ref{sec:structural_results}, we present PIMC data for correlated quantum dipole systems and investigate the temperature- and coupling-strength dependence of different structural properties. Finally, the superfluid fraction is investigated in Sec.~\ref{sec:superfluid_results} for both bosons and fermions, and put into the context of other observables.
The paper is concluded by a brief summary and outlook in Sec.~\ref{sec:summary}.

\section{Theory\label{sec:theory}}

\subsection{Hamiltonian\label{sec:Hamiltonian}}

The Hamiltonian of a harmonically confined quantum dipole system can be written as
\begin{eqnarray}\label{eq:Hamiltonian_trap}
\hat H = - \frac{1}{2} \sum_{k=1}^N \nabla_k^2 + \frac{1}{2} \sum_{k=1}^N \mathbf{\hat r}_k^2 + \sum_{k>l}^N \frac{ \lambda }{ |\mathbf{\hat r}_l - \mathbf{\hat r}_k|^3 } \quad ,
\end{eqnarray}
where we assume oscillator units, i.e., the characteristic length $l_0=\sqrt{\hbar/m\Omega}$ (with $\Omega$ being the trap frequency) and energy scale $E_0=\hbar\Omega$. As usual, the first term corresponds to the kinetic contribution and the last two terms to the external potential and the dipole--dipole interaction, respectively. Moreover, we note that the coupling constant $\lambda$ can, in principle, be tuned in experiments via different techniques~\cite{tune1,tune2}. All simulation results in this work have been obtained for strictly two-dimensional systems.

\subsection{Path integral Monte Carlo\label{sec:PIMC_theory}}

In statistical physics, all thermodynamic expectation values can be computed from the partition function, which, in the canonical ensemble (i.e., fixed particle number $N$, inverse temperature $\beta=1/k_\textnormal{B}T$, and trap frequency $\Omega$), can be expressed in coordinate space as
\begin{eqnarray}\label{eq:Z}
 Z^\textnormal{B/F}= \frac{1}{N!} \sum_{\sigma\in S_{N}} \textnormal{sgn}^\textnormal{B/F}(\sigma) 
 \int \textnormal{d}\mathbf{R}\ \bra{ \mathbf{R} } e^{-\beta\hat{H}} \ket{ \hat{\pi}_{\sigma}\mathbf{R}} \quad ,
\end{eqnarray}
where the sum is carried out over all possible permutations $\sigma$ of the permutation group $S_N$, and $\hat\pi_\sigma$ being the corresponding permutation operator acting on the $N$-particle state $\ket{\mathbf{R}}$. Moreover, the sign for bosons (B) and fermions (F) is given by
\begin{eqnarray}\label{eq:sign}
\textnormal{sgn}^\textnormal{B}(\sigma) &=& 1 \\ \nonumber
\textnormal{sgn}^\textnormal{F}(\sigma) &=& (-1)^{l_\sigma} \quad ,
\end{eqnarray}
with $l_\sigma$ being the number of pair-exchanges for a given $\sigma$. Note that we restrict ourselves to a single particle species of $N$ spin-polarized bosons or fermions throughout this work.

The main obstacle regarding Eq.~(\ref{eq:Z}) is that the matrix elements of the density operator cannot be directly evaluated as the kinetic ($\hat K$) and potential ($\hat V$) contributions to the Hamiltonian do not commute, 
\begin{eqnarray}\label{eq:primitive}
 e^{-\beta\hat H} =  e^{-\beta\hat K} e^{-\beta\hat V} + \mathcal O\left(\beta^2 \right) \quad .
\end{eqnarray}
To solve this problem, we recall the following semi-group property of the exponential function,
\begin{eqnarray}\label{eq:group}
 e^{-\beta\hat H} = \prod_{\alpha=0}^{P-1}e^{-\epsilon\hat H} \quad ,
\end{eqnarray}
which implies that the density matrix can be expressed as the integral over the product of $P$ density matrices, but evaluated at a $P$-times higher temperature, 
\begin{eqnarray}\label{eq:Z_PIMC}
 Z^\textnormal{B/F} &=& \frac{1}{N!} \sum_{\sigma\in S_{N}} \textnormal{sgn}^\textnormal{B/F}(\sigma) \\ \nonumber
 & &
 \int \textnormal{d}\mathbf{R}_0 \dots \textnormal{d}\mathbf{R}_{P-1} \prod_{\alpha=0}^{P-1} \bra{ \mathbf{R}_\alpha } e^{-\epsilon\hat{H}} \ket{ \hat\pi_{\sigma,P}\mathbf{R}_{\alpha+1}}  \quad ,
\end{eqnarray}
where the notation $\hat\pi_{\sigma,P}$ indicates that the permutation operator is only acting on $\ket{\mathbf{R}_P}$. Therefore, the factorization error in Eq.~(\ref{eq:primitive}) can be made arbitrarily small by increasing $P$ (this follows from the celebrated Trotter formula~\cite{trotter}), which is a convergence parameter within the PIMC formalism.

\begin{figure}
\includegraphics[width=0.4147\textwidth]{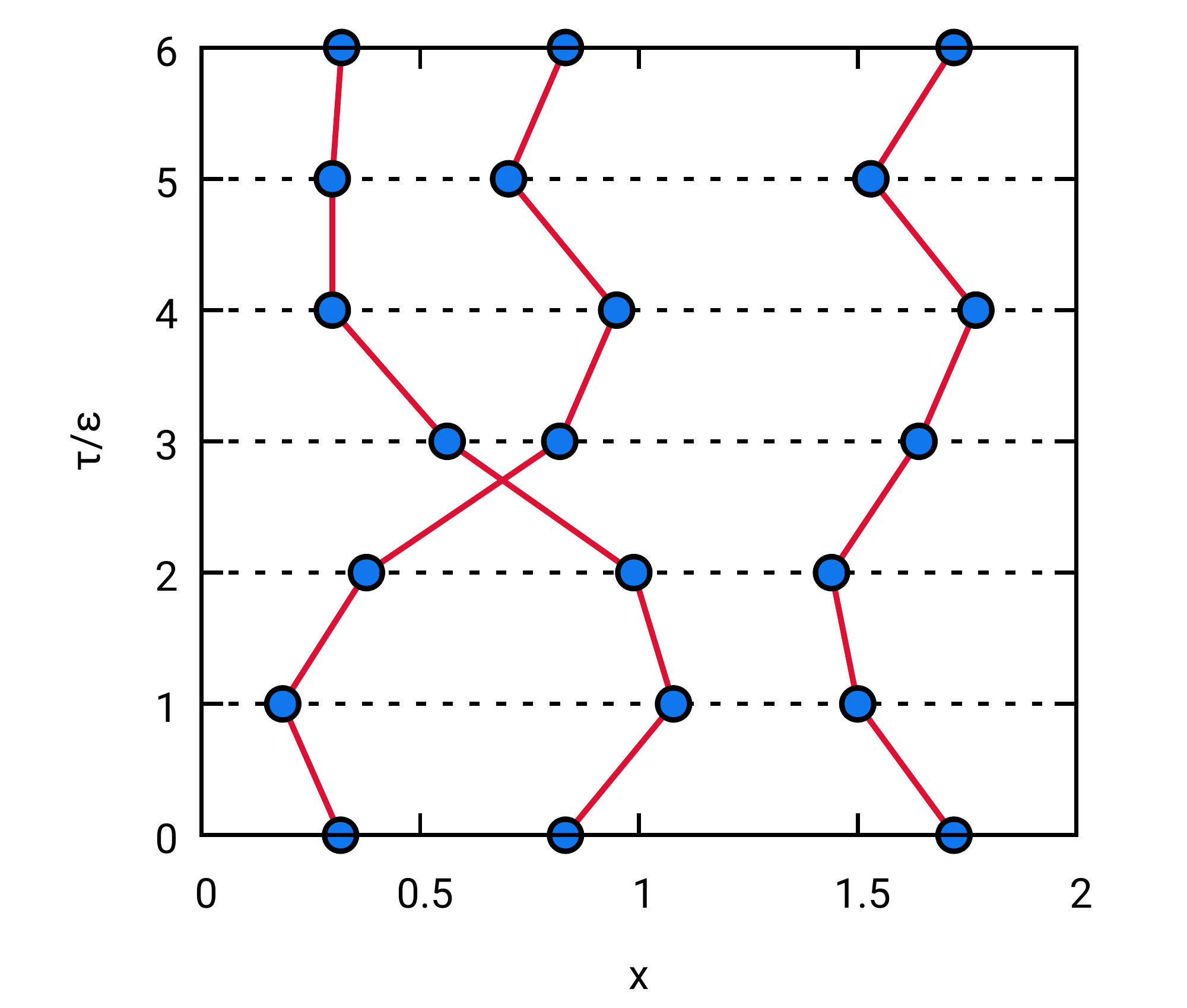}
\caption{\label{fig:PIMC}
Schematic illustration of Path Integral Monte Carlo---Shown is a configuration of $N=3$ electrons with $P=6$ imaginary--time propagators in the $x$-$\tau$ plane. Due to the single pair-exchange, the corresponding configuration weight $W(\mathbf{X})$ [cf.~Eq.~(\ref{eq:def:W})] is negative. Reprinted from T.~Dornheim \textit{et al.}, \textit{J.~Chem.~Phys.}~\textbf{151}, 014108 (2019) \cite{dornheim_permutation_cycles} with the permission of AIP Publishing.
}
\end{figure}  

In the end, the partition function can be written in a compact form as
\begin{eqnarray}\label{eq:def:W}
 Z^\textnormal{B/F} = \int \textnormal{d}\mathbf{X}\ W^\textnormal{B/F}(\mathbf{X}) \quad ,
\end{eqnarray}
where we are integrating over the meta-variable $\mathbf{X}=(\mathbf{R}_0,\dots,\mathbf{R}_{P-1})^T$, which is often interpreted as a \textit{configuration}. This is illustrated in Fig.~\ref{fig:PIMC}, where we show an example configuration of $N=3$ particles in the $\tau$-$x$-plane. First and foremost, we note that each particle is now represented by an entire \textit{path} of $P$ particle coordinates in the imaginary time $\tau\in[0,\beta]$ (with $\epsilon=\beta/P$ being the imaginary-time step), which is a direct consequence of Eq.~(\ref{eq:group}). While all paths are closed, there appear trajectories containing multiple particles, such as on the lhs.~of Fig.~\ref{fig:PIMC}, which are due to the permutation operator $\hat\pi_{\sigma,P}$.
The basic idea of the PIMC method is to use the Metropolis algorithm~\cite{metropolis} to generate a Markov chain of configurations $\{\mathbf{X}_i\}$ which are distributed proportionally to the configuration weight $W^\textnormal{B/F}(\mathbf{X})$, which is a function that can be readily evaluated.

For bosons (and for distinguishable particles, i.e., \textit{boltzmannons}), $W(\mathbf{X})$ is strictly positive and simulations of $N\sim10^{3}-10^{4}$ particles are feasible. In the case of fermions, on the other hand, the density matrix is anti-symmetric with respect to pair-exchanges, and the sign of the configuration weight changes, cf.~Eq.~(\ref{eq:sign}). This means that $P^\textnormal{F}(\mathbf{X})=W^\textnormal{F}(\mathbf{X})/Z^\textnormal{F}$ cannot be interpreted as a probability, and a straightforward sampling of the configurations $\mathbf{X}$ is not possible. To work around this issue, we switch to the modified configuration space defined by
\begin{eqnarray}
Z' &=& \int \textnormal{d}\mathbf{X}\ |W^\textnormal{F}(\mathbf{X})| \\ \nonumber
   &=& \int \textnormal{d}\mathbf{X}\ W^\textnormal{B}(\mathbf{X}) = Z^\textnormal{B} \quad ,
\end{eqnarray}
which, in the case of the standard PIMC methods as introduced above, is equal to the configuration space of the corresponding Bose system, see Ref.~\cite{dornheim_neu} for an extensive and accessible discussion.

It is easy to see that the exact fermionic expectation value of an arbitrary observable $\hat A$ is then given by
\begin{eqnarray}\label{eq:ratio}
\braket{\hat A}^\textnormal{F} = \frac{ \braket{\hat A\hat S}^\textnormal{B} }{\braket{\hat S}^\textnormal{B}} \quad .
\end{eqnarray}
In a nutshell, Eq.~(\ref{eq:ratio}) implies that PIMC results for a Fermi system are obtained from a simulation of a corresponding bosonic simulation at the same parameters by keeping track of Eq.~(\ref{eq:sign}) and, thus, taking into account all cancellations due to the antisymmetry of the density matrix under particle exchange.

The denominator in Eq.~(\ref{eq:ratio}) is commonly known simply as the average sign $S$, and constitutes a convenient measure for the degree of cancellation of positive and negative terms. In particular, the relative statistical uncertainty of the Monte Carlo expectation value is inversely proportional to $S$~\cite{ceperley_fermions}
\begin{eqnarray}\label{eq:fsp_error}
\frac{\Delta A_\textnormal{F}}{\braket{\hat A}^\textnormal{F}} \sim \frac{1}{S \sqrt{N_\textnormal{MC}}} \sim \frac{e^{\beta N (f_\textnormal{F}-f_\textnormal{B})}}{\sqrt{N_\textnormal{MC}}} \quad ,
\end{eqnarray}
which is the origin of the notorious fermion sign problem~\cite{dornheim_sign_problem,loh,troyer}. 
More specifically, it is straightforward to see that it holds $S\sim e^{-\beta N (f_\textnormal{F}-f_\textnormal{B})}$ (with $f_\textnormal{B,F}$ being the free energy density of bosons and fermions), which leads to an exponential increase in $\Delta A_\textnormal{F}$ both towards low temperature (increasing $\beta$) and with increasing system size $N$. The increasing error bar can only be reduced by increasing the number of Monte Carlo samples $N_\textnormal{MC}$ as $\Delta A_\textnormal{F}\sim 1/\sqrt{N_\textnormal{MC}}$, which at some point becomes computationally too expensive. In practice, one eventually runs into an exponential wall regarding $\beta$ or $N$ and the simulations become unfeasible, see Ref.~\cite{dornheim_sign_problem} for a recent review article. 
In fact, the sign problem constitutes the main bottleneck in this work, and limits our results to $N<10$.

For completeness, we mention that all PIMC simulations in this work have been carried out using an implementation of the worm algorithm introduced in Refs.~\cite{boninsegni1,boninsegni2}.

\subsection{Superfluidity and non-classical rotational inertia\label{sec:ncri_theory}}

For a finite system, the superfluid fraction is typically defined by the response of the system to an infinitesimal rotation. More specifically, one assumes a two-fluid model, where the total particle density is decomposed into a normal part that reacts to the rotation, and a superfluid component that does not, $n=n_\textnormal{n}+n_\textnormal{sf}$.
The superfluid fraction is then readily defined as the ratio of $n_\textnormal{sf}$ and $n$,
\begin{eqnarray}\label{eq:sf_definition}
\gamma_\textnormal{sf} = \frac{n_\textnormal{sf}}{n} = 1 - \frac{I}{I_\textnormal{cl}} \quad ,
\end{eqnarray}
(with $I$ and $I_\textnormal{cl}$ denoting the moment of inertia in the quantum and classical case, respectively) which, in the path integral picture, can be expressed as~\cite{sindzingre}
\begin{eqnarray}\label{eq:area_estimator}
\gamma_\textnormal{sf} = \frac{4m^2 \braket{A_z^2} }{\beta\hbar^2 I_\textnormal{cl}} \quad .
\end{eqnarray}
Eq.~(\ref{eq:area_estimator}) is often referred to as the \textit{area estimator}, as it depends on the expectation value of the area enclosed by the paths in the PIMC simulations,
\begin{eqnarray}\label{eq:area}
\mathbf{A} = \frac{1}{2} \sum_{k=1}^N\sum_{i=1}^P \left( 
\mathbf{r}_{k,i} \times \mathbf{r}_{k,i+1}
\right) \quad .
\end{eqnarray}
Note that our system is located in the $x$-$y$ plane, and, hence, the $z$-component of Eq.~(\ref{eq:area}) denotes the area therein.

In an inhomogeneous system, the superfluid density is typically not distributed uniformly throughout the system. In that situation, it is highly desirable to obtain a local measure of $n_\textnormal{sf}$, which can be defined as~\cite{kwon_lsf}
\begin{eqnarray}\label{eq:nsf}
n_\textnormal{sf}(\mathbf{r}) = \frac{4m^2}{\beta\hbar^2 I_\textnormal{cl}(\mathbf{r})} \braket{A_z A_{z,loc}(\mathbf{r})} \quad ,
\end{eqnarray}
with $I_\textnormal{cl}(\mathbf{r})=m r^2$ and $A_\textnormal{loc}(\mathbf{r})$ being the contribution to Eq.~(\ref{eq:area}) around the position $\mathbf{r}$.
For completeness, we mention that the estimator from Eq.~(\ref{eq:nsf}) is consistent in the sense that it integrates to the correct quantum mechanical moment of inertia,
\begin{eqnarray}
\int \textnormal{d}\mathbf{r}\ n_\textnormal{sf}(\mathbf{r}) \mathbf{r}^2 = \gamma_\textnormal{sf}I_\textnormal{cl} \quad .
\end{eqnarray}
This is in contrast to an alternative estimator presented in Ref.~\cite{draeger}, which is normalized differently.

\subsection{Structural properties\label{sec:c2p_theory}}

\begin{figure}
\includegraphics[width=0.35\textwidth]{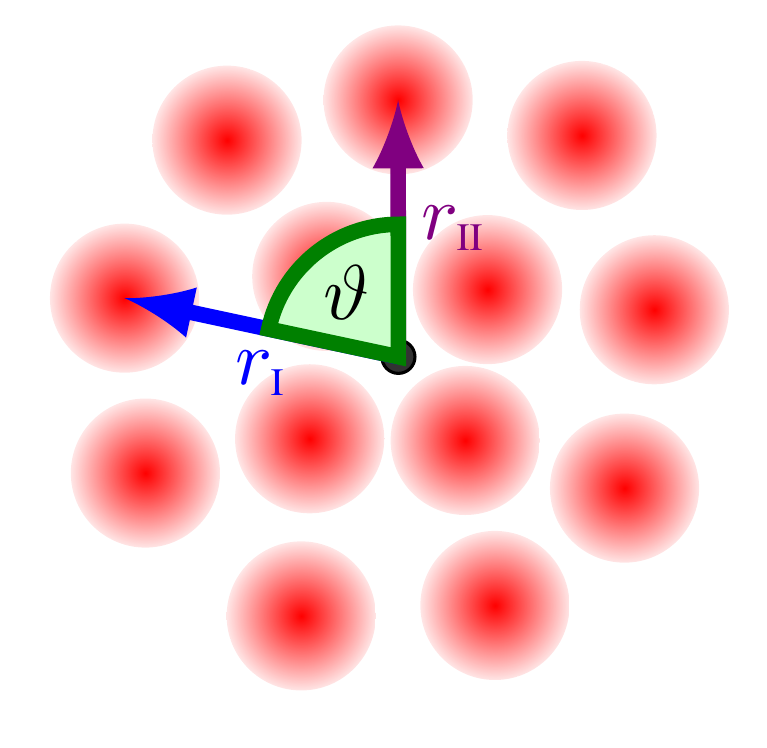}
\caption{\label{fig:c2p_illustration}
Illustration of the center-two particle correlation function $\rho_2(r_1,r_2,\vartheta)$. Due to the rotational symmetry of the Hamiltonian Eq.~\ref{eq:Hamiltonian_trap}), the two-particle correlations only depend on the relative angle $\vartheta$ (denoted as $\alpha$ throughout this work) and the respective distances to the center of the trap.
Taken from Ref.~\cite{dornheim_c2p}. Reprinted with permission of WILEY-VCH Verlag GmbH $\&$ Co.~KGaA, Weinheim.}
\end{figure}

The spatial correlations within a system are fully characterized by the two-particle distribution function $\rho_2(\mathbf{r}_1,\mathbf{r}_2)$, which gives the probability to find two particles at the coordinates $\mathbf{r}_1$ and $\mathbf{r}_2$. In a uniform system, $\rho_2$ only depends on the modulus of the distance, and it is sufficient to consider the one-dimensional function $\rho_2(|\mathbf{r}_1-\mathbf{r}_2|)$. In a harmonic confinement, such a simplification is not possible or, to be more precise, information about correlation is averaged out and, therefore, lost.

Still, $\rho_2(\mathbf{r}_1,\mathbf{r}_2)$ can be simplified as the system exhibits rotational symmetry. This is illustrated in Fig.~\ref{fig:c2p_illustration}, where a schematic configuration of $N=13$ quantum particles is shown with the smeared out red circles illustrating the quantum delocalization of the particles. In particular, one can define a center-two particle distribution function $\rho_2(r_1,r_2,\alpha)$ (observe that the relative angle $\alpha$ is denoted as $\vartheta$ in Fig.~\ref{fig:c2p_illustration}), which gives the probability to find two atoms at the distances to the center of the trap of $r_1$ and $r_2$ with a relative angle of $\alpha$ towards each other. 

\par 

In order to analyze angular correlations and to filter out effects that are purely caused by the inhomogeneous density profile $n(r)$, it is quite useful to define a center-two particle correlation function (hereafter referred to as C2P)~\cite{thomsen_c2p}
\begin{eqnarray}
g_\textnormal{c2p}(r_1,r_2,\alpha) = \frac{\rho_2(r_1,r_2,\alpha)}{\rho^0_2(r_1,r_2,\alpha)} \quad .
\end{eqnarray}
Here the denominator corresponds to the two-particle density of a hypothetical uncorrelated (ideal) system, but with the exact, fully correlated inhomogeneous density profile $n(r)$,
\begin{eqnarray}
\rho^0_2(r_1,r_2,\alpha) = \frac{N-1}{N} 4\pi r_1 r_2 n(r_1)n(r_2) \quad ,
\end{eqnarray}
which, by design, does not depend on the relative angle $\alpha$ between two atoms.
As the handling, visualization, and interpretation of a three-dimensional function is quite cumbersome, we define the integrated C2P function
\begin{eqnarray}
g^\textnormal{int}_\textnormal{c2p}(r_1,\alpha;r_{2,\textnormal{min}},r_{2,\textnormal{max}}) = 
\frac{ \int_{r_{2,\textnormal{min}}}^{r_{2,\textnormal{max}}} \textnormal{d}r_2\ \rho_2(r_1,r_2,\alpha) }{ \int_{r_{2,\textnormal{min}}}^{r_{2,\textnormal{max}}} \textnormal{d}r_2\ \rho^0_2(r_1,r_2,\alpha) }\ , \label{eq:c2p_int}
\end{eqnarray}
which allows for a straightforward interpretation: given that one atom is located at a distance $r_{2,\textnormal{min}}\leq r_2 \leq r_{2,\textnormal{max}}$ to the center of the trap (typically $r_{2,\textnormal{min}}$ and $r_{2,\textnormal{max}}$ are chosen as the boundaries of a shell), Eq.~(\ref{eq:c2p_int}) constitutes a measure for the relative probability to find a second particle at $r_1$ with a relative angular difference of $\alpha$ between the pair.

\subsection{Ideal bosons and fermions\label{sec:ideal_theory}}

The partition function of $N$ spin-polarized noninteracting bosons or fermions at the inverse temperature $\beta$ is readily expressed in terms of permutation cycle frequencies~\cite{dornheim_permutation_cycles,krauth_book},
\begin{eqnarray}\nonumber
Z^\textnormal{B,F}(N,\beta) &=& \frac{1}{N!} \sum_{\{C_q\}_\textnormal{r}} \sigma^\textnormal{B,F}\left(\{C_q\} \right) M\left(\{C_q\} \right) \\ & & \times \prod_{q=1}^N Z(1,q\beta)^{C_q}\ , \label{eq:ideal_partition}
\end{eqnarray}
with $\{C_q\}_\textnormal{r}$ denoting the set of all permutation cycle occupations that are possible for $N$ particles,
\begin{eqnarray}\label{eq:permutation_norm}
\sum_{q=1}^N q C_q = N \quad .
\end{eqnarray}
Here $M\left(\{C_q\} \right)$ denotes a combinatorial factor of the form
\begin{eqnarray}
M\left(\{C_q\} \right) = \frac{N!}{\prod_{q=1}^N C_q! q^{C_q}} \quad ,
\end{eqnarray}
and the sign is given by 
\begin{eqnarray}
\sigma^\textnormal{B,F}\left(\{C_q\} \right) = (\pm 1)^{\sum_{q=1}^N \left( q-1\right) C_q } \quad ,
\end{eqnarray}
with the plus and minus signs corresponding to bosons and fermions, respectively. Moreover, the single-particle partition function appearing in Eq.~(\ref{eq:ideal_partition}) is equal for Bose- and Fermi-statistics and is known from the literature,
\begin{eqnarray}
Z(1,\beta) = \left( 
\frac{e^{-\beta/2}}{1-e^{-\beta}}
\right)^2 \quad .
\end{eqnarray}

To compute the classical and quantum mechanical expectation values of the moment of inertia, which are needed to estimate the superfluid fraction [see Eq.~(\ref{eq:sf_definition})], we introduce the ancilla function
\begin{eqnarray}
\Gamma^\textnormal{B,F}_{N,\beta}(q) = \sum_{\{C_q\}_\textnormal{r}} \sigma^\textnormal{B,F}\left(\{C_q\} \right) \prod_{r=1}^N \frac{Z(1,r\beta)^{C_r}}{C_r! r^{C_r}} C_q \quad ,
\end{eqnarray}
which leads to 
\begin{eqnarray}
I = \frac{2\hbar^2\beta}{Z^\textnormal{B,F}(N,\beta)} \sum_{q=1}^N \left(
\Gamma^\textnormal{B,F}_{N,\beta}(q) \frac{ q^2 e^{-q\beta\hbar\omega}}{\left( 1-e^{-q\beta\hbar\omega} \right)^2}
\right)\ ,
\end{eqnarray}
and
\begin{eqnarray}
I_\textnormal{cl} = \frac{1}{Z^\textnormal{B,F}(N,\beta)} \frac{\hbar}{\omega} \sum_{q=1}^N \left(
\Gamma^\textnormal{B,F}_{N,\beta}(q) q \frac{1 + e^{-q\beta\hbar\omega}}{1 - e^{-q\beta\hbar\omega}}
\right) \ .
\end{eqnarray}
The final result for $\gamma_\textnormal{sf}$ is then obtained by evaluating Eq.~(\ref{eq:sf_definition}).

Furthermore, we introduce the permutation cycle frequency (i.e., the probability to find a trajectory with $l$ particles in it) as
\begin{eqnarray}\label{eq:ideal_Pl}
P(l) = \frac{Z^\textnormal{B}(1,l\beta) Z^\textnormal{B}(N-l,\beta)}{l Z^\textnormal{B}(N,\beta)}\ ,
\end{eqnarray}
with the bosonic partition functions obeying the recursion relation~\cite{krauth_book}
\begin{eqnarray}
Z^\textnormal{B}(N,\beta) = \frac{1}{N} \sum_{q=1}^N Z^\textnormal{B}(1,q\beta)Z^\textnormal{B}(N-q,\beta)\ .
\end{eqnarray}

\section{Results\label{sec:results}}

\subsection{Ideal bosons and fermion\label{sec:ideal_results}}

\begin{figure}\centering
\includegraphics[width=0.4147\textwidth]{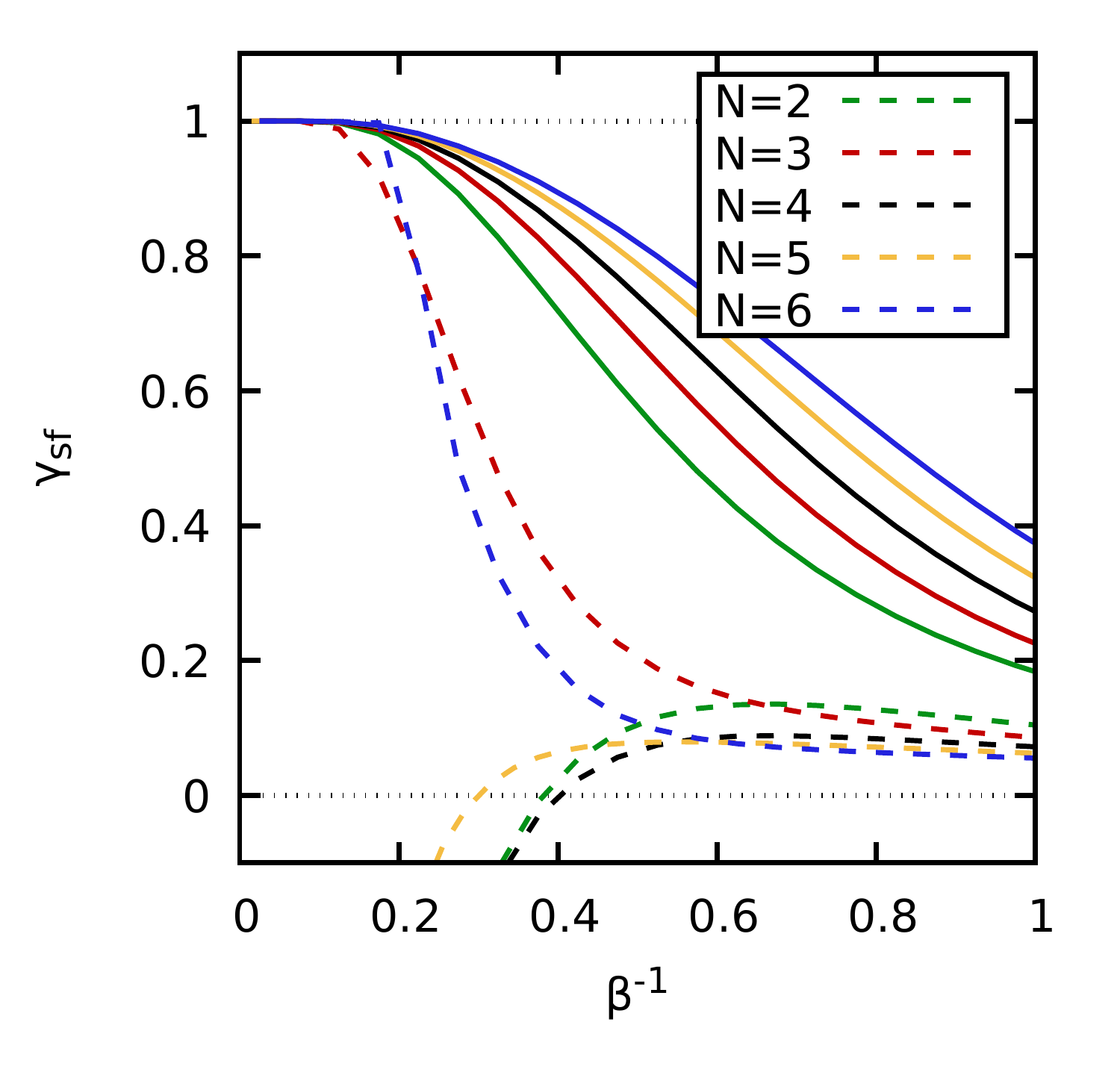}
\caption{\label{fig:ideal_sf}
Temperature dependence of the superfluid fraction $\gamma_\textnormal{sf}$ of $N=2,3,4,5,6$ ideal (noninteracting) bosons (solid) lines and fermions (dashed lines) in a $2D$ harmonic confinement. The corresponding formulas are given in Sec.~\ref{sec:ideal_theory}.
}
\end{figure}  

Let us start our investigation by revisiting the behavior of ideal, i.e., noninteracting bosons and fermions in a harmonic trap, which corresponds to setting $\lambda=0$ in Eq.~(\ref{eq:Hamiltonian_trap}). In this case, the partition function and all derivative thermodynamic properties can be expressed in terms of permutation cycle distributions, which are known from theory, see Sec.~\ref{sec:ideal_theory}. In Fig.~\ref{fig:ideal_sf}, we show the temperature dependence of $\gamma_\textnormal{sf}$ for bosons (solid lines) and fermions (dashed lines) for five different particle numbers $N$. For completeness, we note that a similar plot has been presented in Ref.~\cite{blume} for the same system, but in $3D$. In the case of bosons, we observe the expected crossover from the classical regime at large temperature, where $\gamma_\textnormal{sf}$ vanishes, to the ground-state limit where the system is fully superfluid. Moreover, the curves are ordered with ascending $N$, as the systems are more degenerate at larger density. In addition, we mention that the crossover will eventually approach a real phase transition for substantially larger system sizes, cf.~Ref.~\cite{dornheim_superfluid}.

In contrast, the corresponding fermionic results exhibit a significantly more complicated behaviour. For $N=3$ (red) and $N=6$ (blue), the system also becomes completely superfluid, although at significantly lower temperature than for bosons. Further, $\gamma_\textnormal{sf}$ becomes negative for $N=2$ (green), $N=4$ (black), and $N=5$ (yellow), and, in fact, even diverges towards $-\infty$ in those cases. This odd behaviour is an artifact of the definition of $\gamma_\textnormal{sf}$, Eq.~(\ref{eq:sf_definition}), and indicates a diverging moment of inertia. This feature was vividly explained in Ref.~\cite{blume} by the topology of the density matrix: without an energetically low eigenstate with finite rotation, the system cannot respond to an infinitesimal rotation.

At this point, we feel that a note of caution regarding the terminology is pertinent. While we do refer to the quantity defined in Eq.~(\ref{eq:sf_definition}) as the \textit{superfluid fraction}, the behaviour shown in Fig.~\ref{fig:ideal_sf} does not indicate the onset of a frictionless flow such as in He$^4$ and other bulk materials~\cite{cep}. In particular, it is known that the latter emerges as a consequence of an off-diagonal long-range order of the density matrix~\cite{yushi}, which, by definition, cannot occur in a few-particle system. Therefore, it is more accurate to speak of \textit{non-classical rotational inertial} (NCRI) in the present case.

\begin{figure}
\includegraphics[width=0.4147\textwidth]{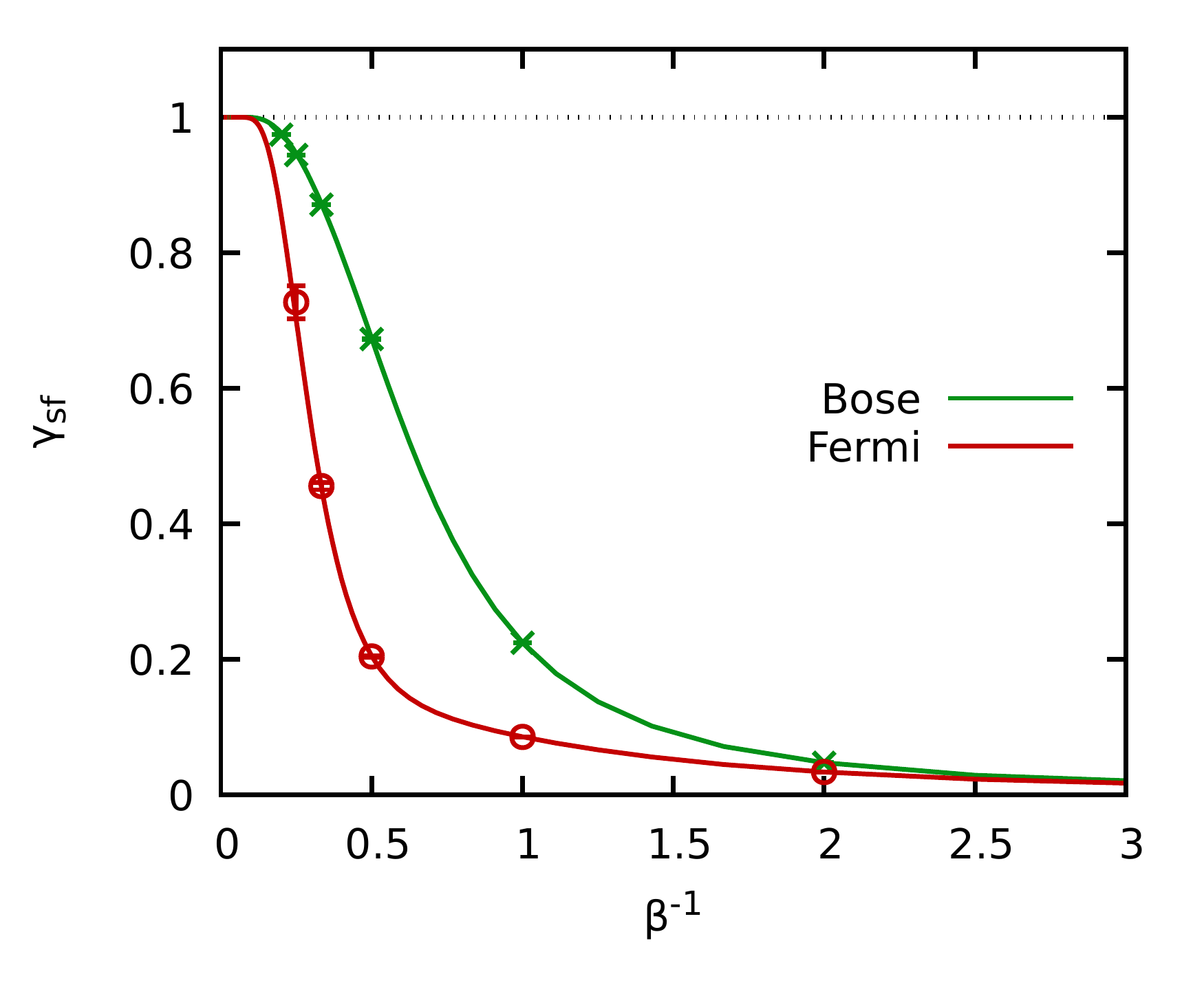}
\caption{\label{fig:ideal_2D_N3}
Temperature dependence of the superfluid fraction of $N=3$ ideal bosons (green) and fermions (red) in a $2D$ harmonic confinement. The lines and points correspond to the exact result known from theory, cf.~Sec.~\ref{sec:ideal_theory}, and our PIMC data calculated via the area estimator from Eq.~(\ref{eq:area_estimator}).
}
\end{figure}

Let us next use the exact data for the superfluid fraction to demonstrate the correctness and consistency of our PIMC simulations. To this end, we show the temperature dependence of $\gamma_\textnormal{sf}$ in Fig.~\ref{fig:ideal_2D_N3} for $N=3$ ideal bosons (green) and fermions (red), again in $2D$. More specifically, the solid lines depict the exact curves, and the symbols the PIMC data that was obtained using the area estimator defined in Eq.~(\ref{eq:area}), see Sec.~\ref{sec:ncri_theory}. First and foremost, we note that the PIMC data is in perfect agreement to the theoretical prediction for all temperatures. We stress that this is a striking validation of our code since $\gamma_\textnormal{sf}$ is highly sensitive to the distribution of permutation cycles within the PIMC simulation. This is particularly true in the case of fermions, where the expectation value for both the area estimator and the classical moment of inertia are strongly dependent on the cancellation of positive and negative terms [cf.~Eq.~(\ref{eq:ratio})] and, consequently, on the respective permutation lengths, see also Ref.~\cite{dornheim_permutation_cycles} for a topical discussion.

Secondly, we observe a significantly increased statistical  uncertainty (error bars) in the case of fermions. This is a direct consequence of the fermion sign problem (cf.~Sec.~\ref{sec:PIMC_theory}), as the average sign $S$ monotonically decreases towards low temperature. More specifically, we find $S=3.5(2)\cdot10^{-4}$ for $\beta=4$, which is the lowest depicted temperature in the case of fermions, and the sign nearly vanishes for $\beta=5$. In contrast, we find $S\approx0.028$ at $\beta=2$, which means that the simulations are more involved for fermions as compared to bosons, but that simulations are still feasible and the corresponding error bar is relatively small. However, a more extensive discussion of the sign problem has been presented elsewhere~\cite{dornheim_sign_problem}, and need not be repeated here.

\begin{figure*}
\includegraphics[width=0.40447\textwidth]{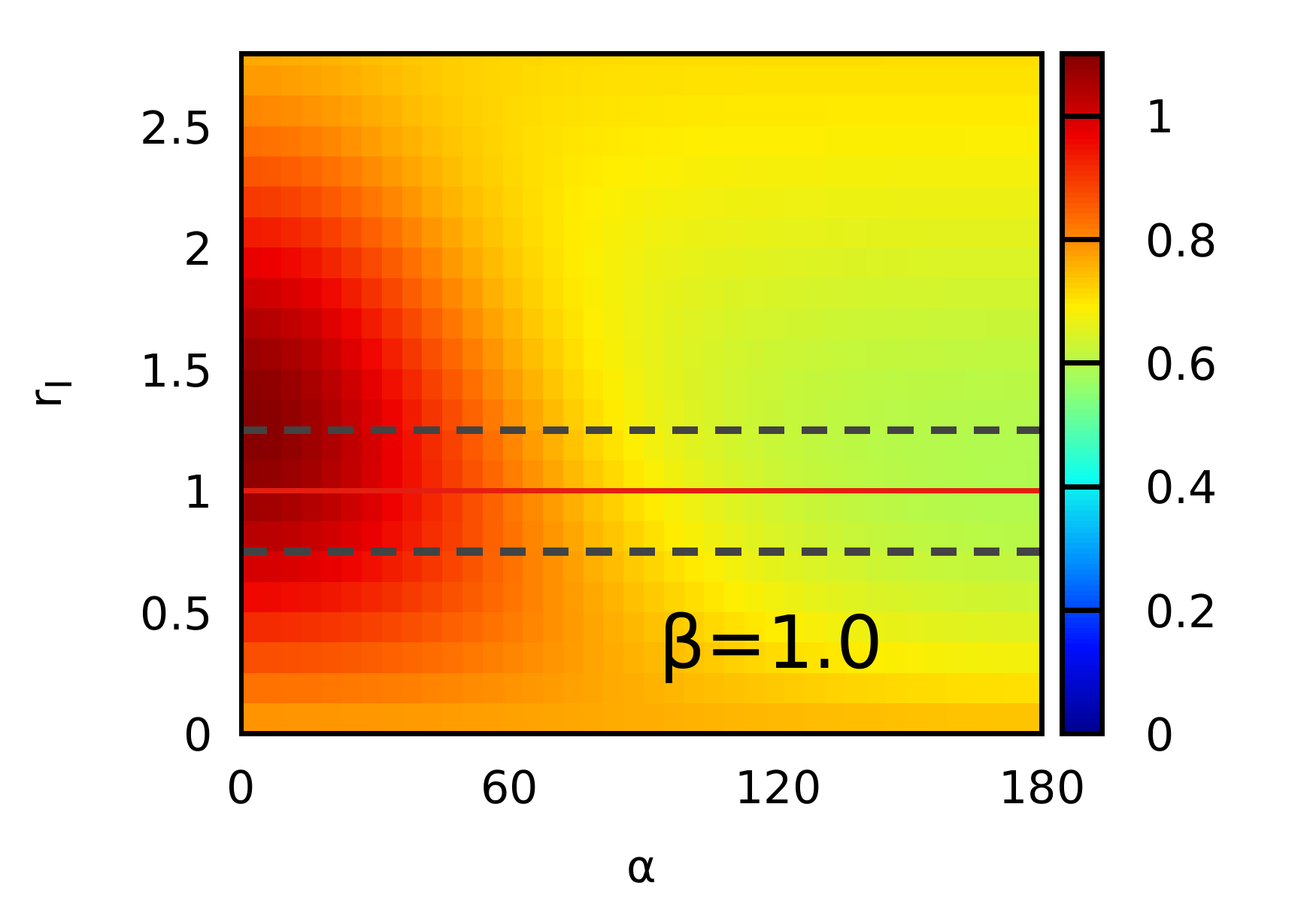}
\includegraphics[width=0.40447\textwidth]{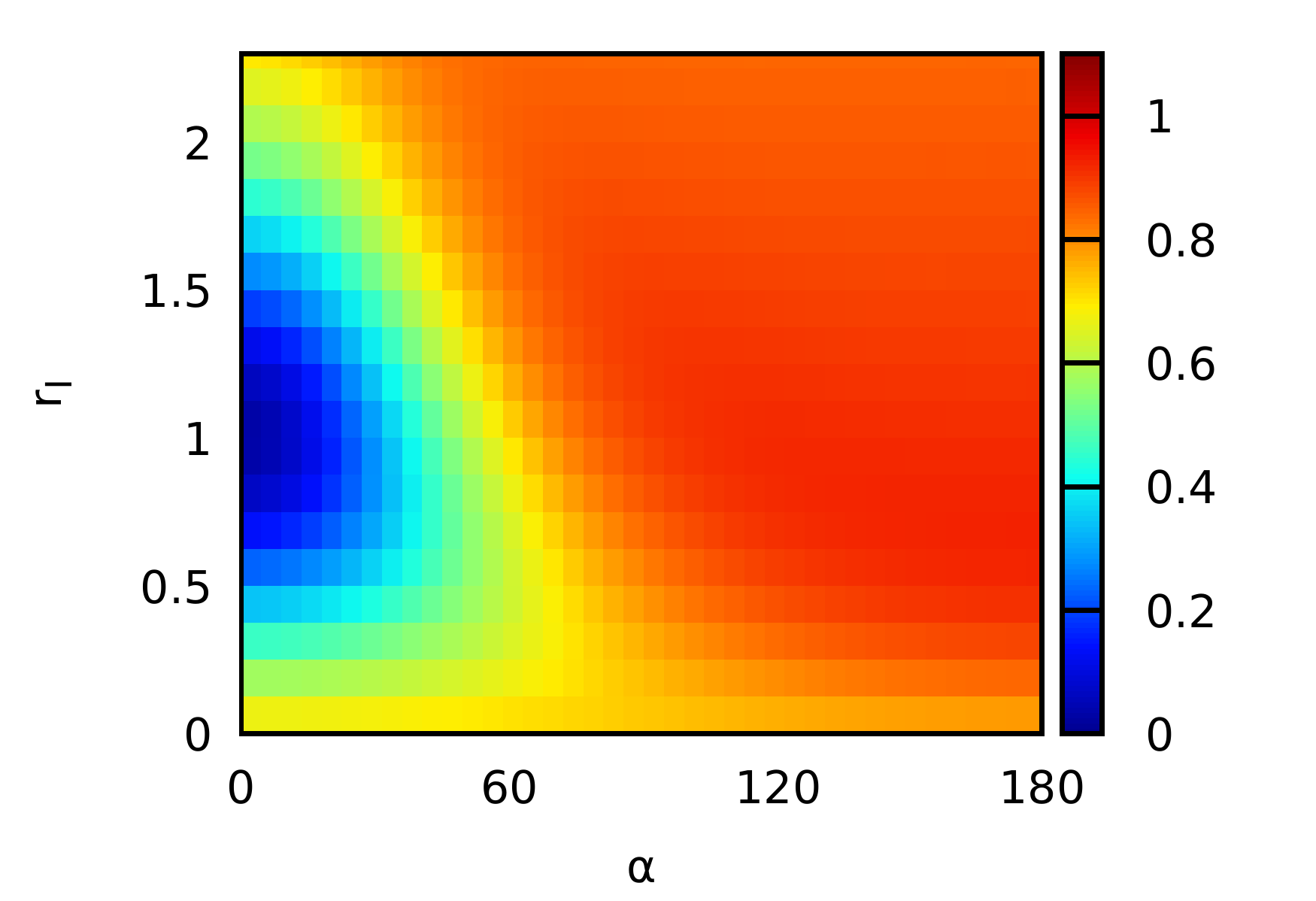}\\
\includegraphics[width=0.40447\textwidth]{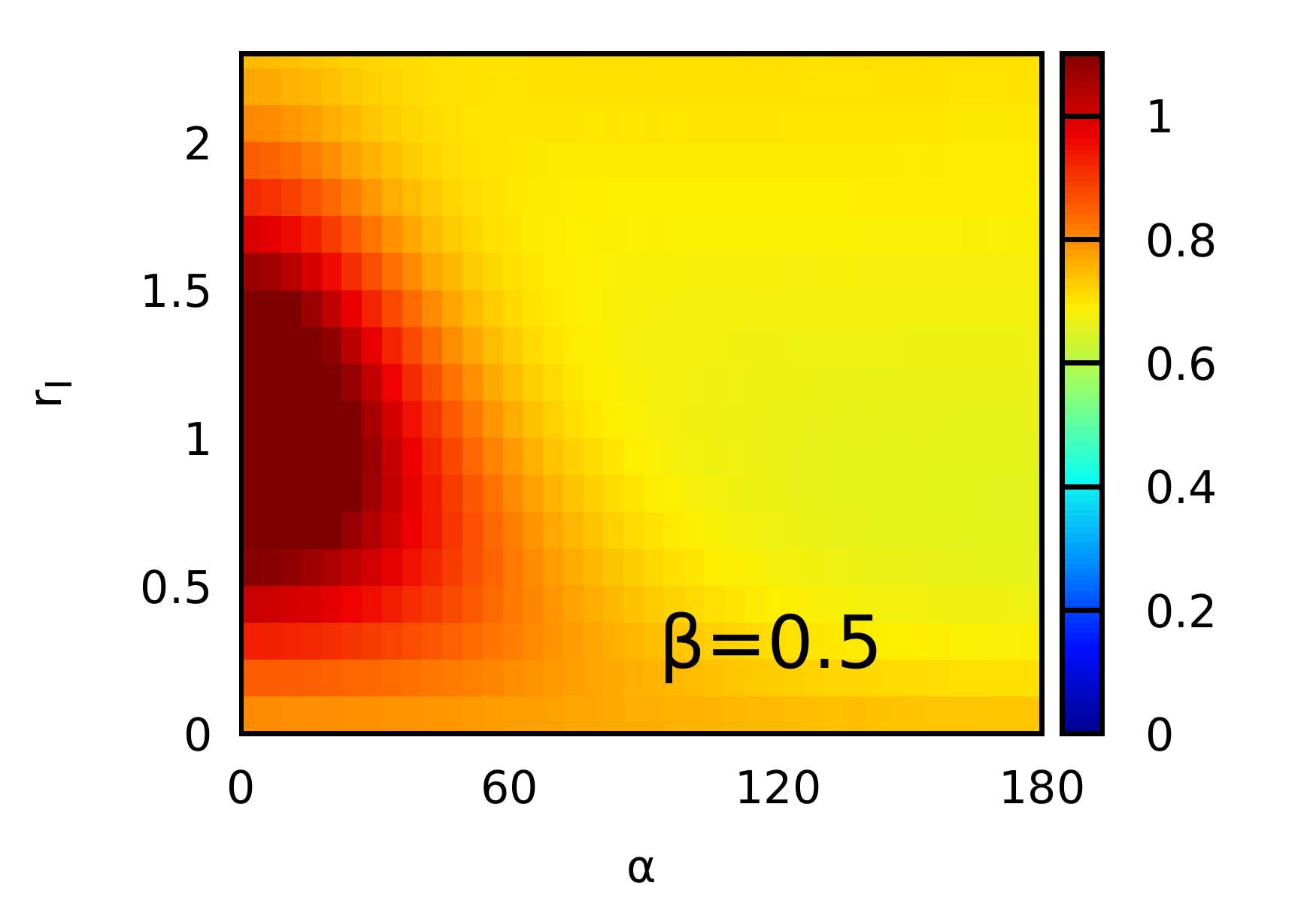}
\includegraphics[width=0.40447\textwidth]{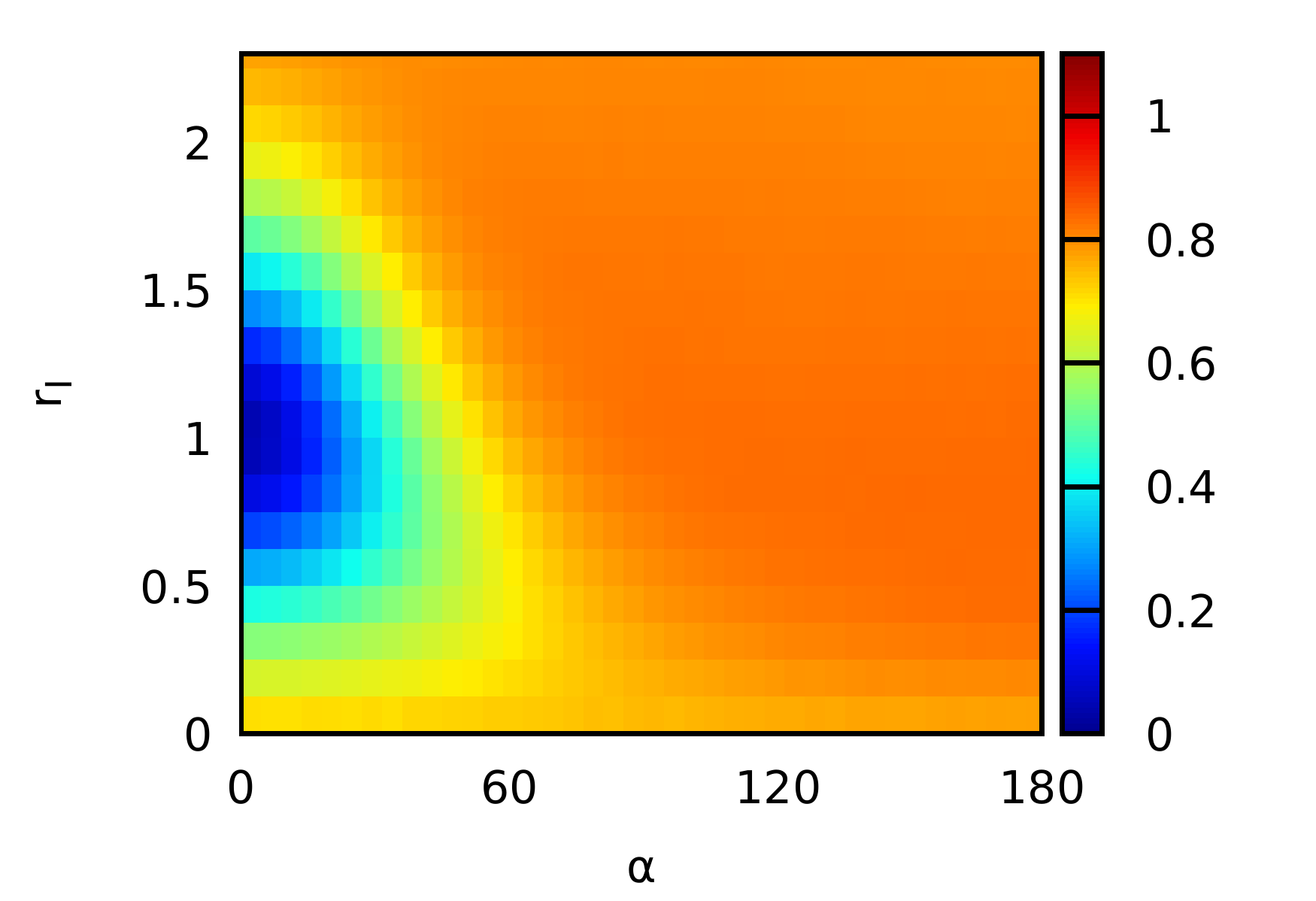}\\
\includegraphics[width=0.40447\textwidth]{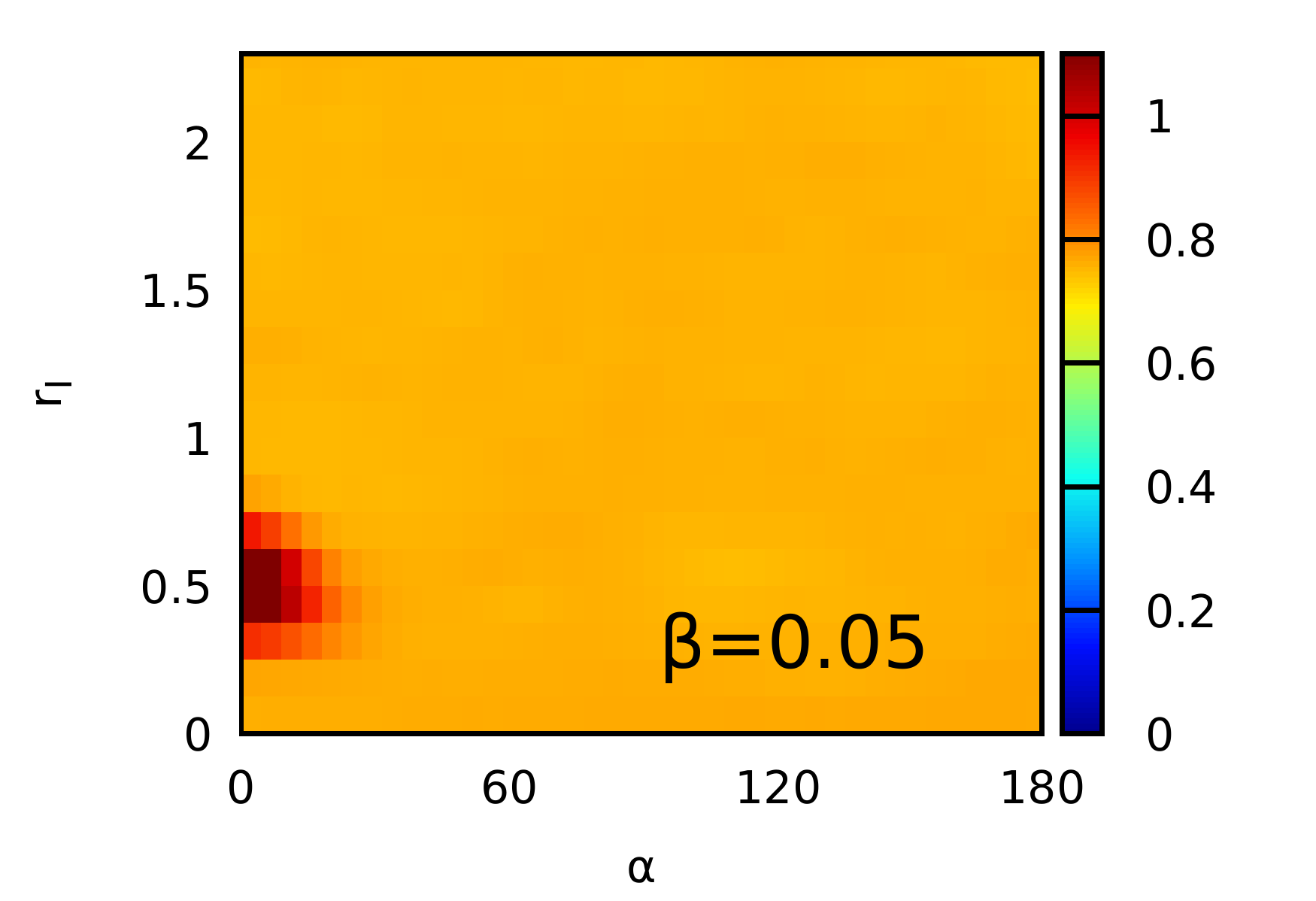}
\includegraphics[width=0.40447\textwidth]{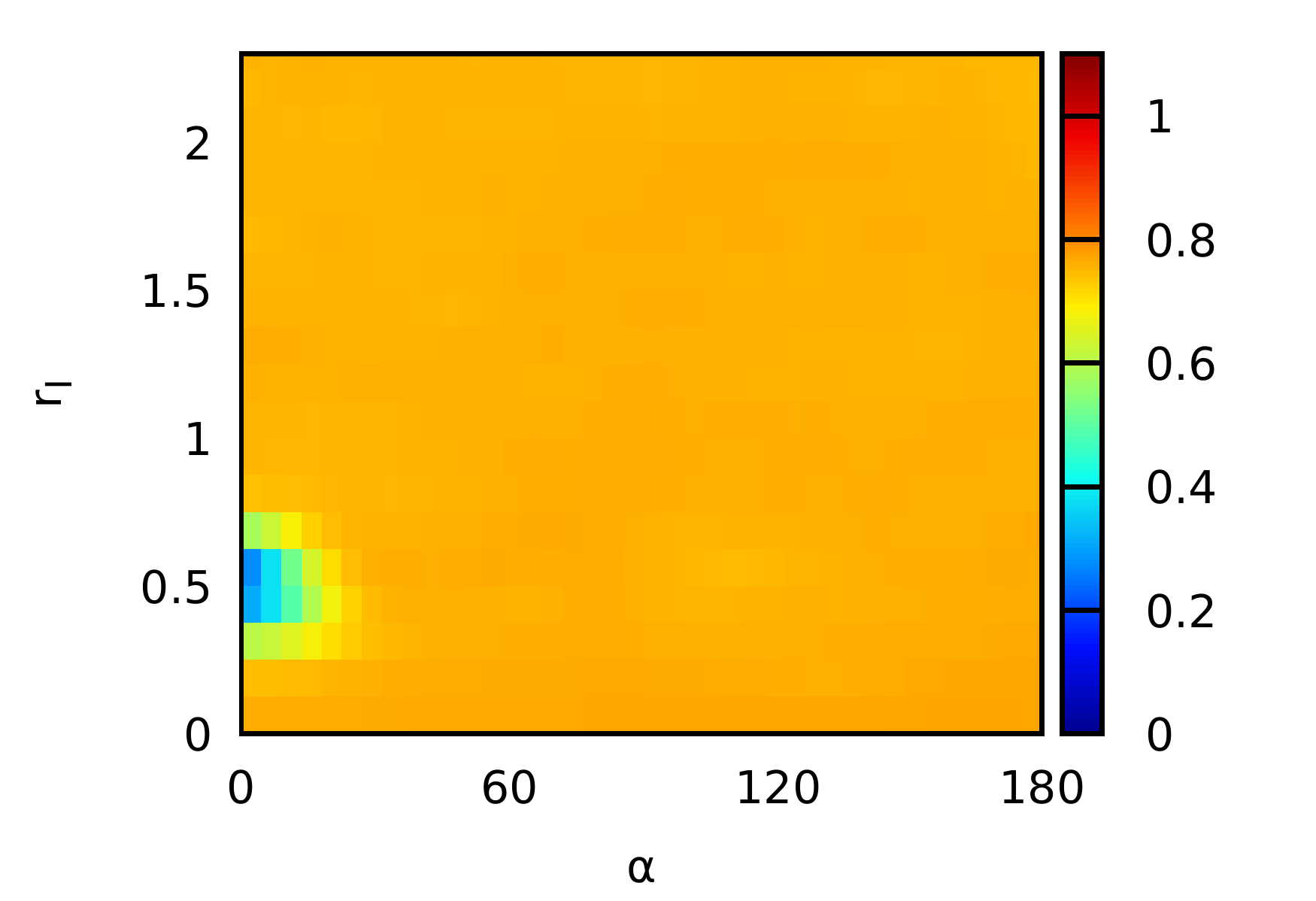}
\caption{\label{fig:c2p_idea_N4}
Integrated center-two particle correlation function $g^\textnormal{int}_\textnormal{c2p}(r_1,\alpha,0.75,1.25)$ [cf.~Eq.~(\ref{eq:c2p_int})] for $N=4$ and $\lambda=0$ (ideal). The left and right columns correspond to Bose- and Fermi-statistics. Top: $\beta=1$, center: $\beta=0.5$, bottom: $\beta=0.05$. The dashed black lines in the top left panel indicate the integration boundaries for the reference particle, and the solid red line the scan line depicted in Fig.~\ref{fig:Scanline_N4_ideal}.
}
\end{figure*}

\begin{figure}
\includegraphics[width=0.4447\textwidth]{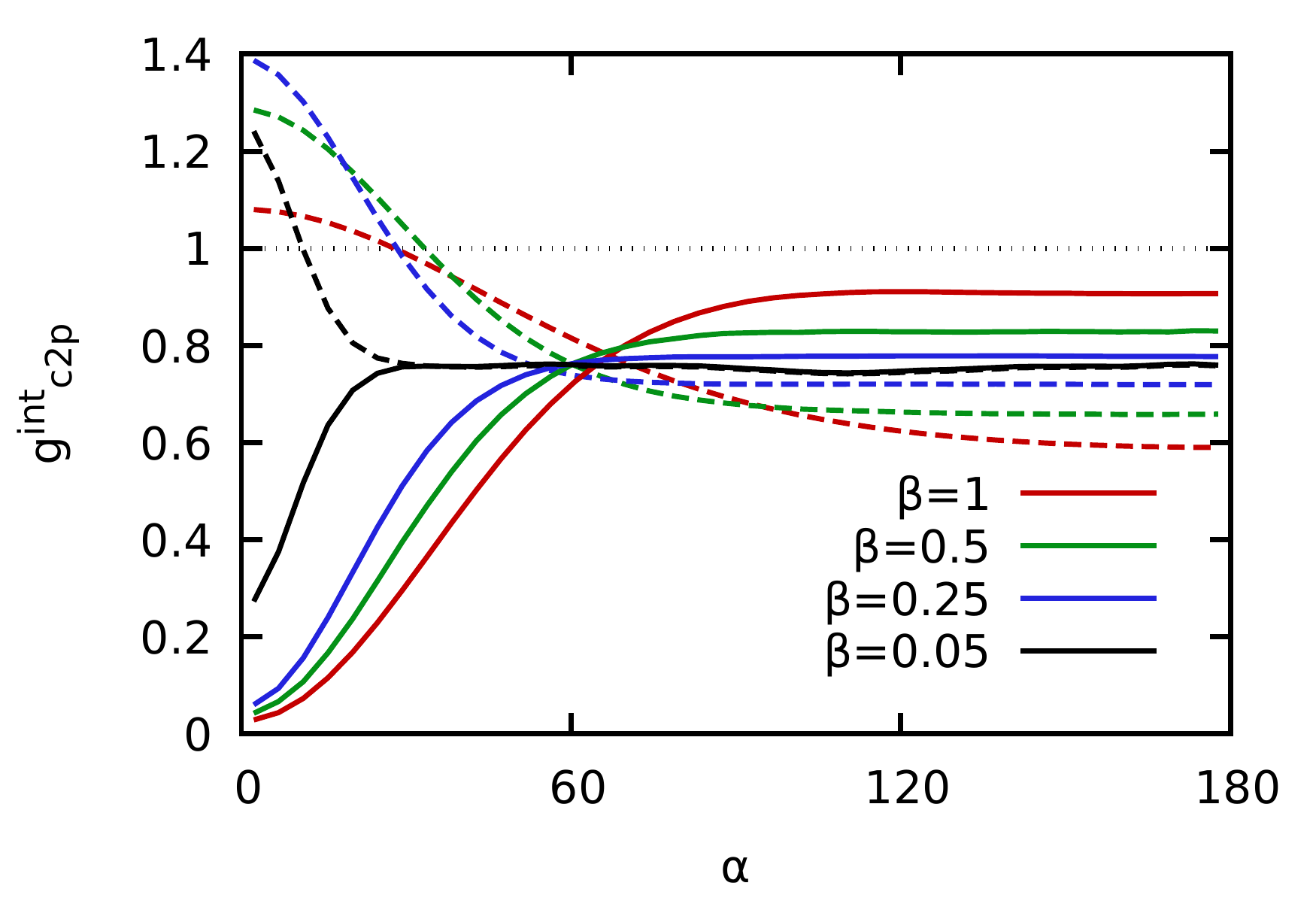}
\caption{\label{fig:Scanline_N4_ideal}
Scanline of the integrated center-two particle correlation function shown in Fig.~\ref{fig:c2p_idea_N4}, evaluated at $r_1=1$. The dashed and solid lines correspond to bosons and fermions.
}
\end{figure}

While the physics of ideal Bose- and Fermi-systems might seem almost trivial, it is still worthwhile to use them as a first test case to demonstrate the capability of the integrated C2P (see Sec.~\ref{sec:c2p_theory}).
In Fig.~\ref{fig:c2p_idea_N4}, we show results for $g^\textnormal{int}_\textnormal{c2p}(r_1,\alpha,0.75,1.25)$, i.e., for the probability to find one particle at $0.75\leq r_2 \leq 1.25$ (see the dashed dark grey lines in the top left panel), and a second particle at an angular distance $\alpha$ ($x$-axis) and a distance $r_1$ to the center of the trap ($y$-axis). The left and right columns correspond to bosons and fermions, and the different rows to different temperatures. At the lowest temperature ($\beta=1$, top row), the results for bosons and fermions vividly demonstrate the key difference between these two particle species. Ideal bosons tend to cluster around each other, and the probability to find two particles close to each other is significantly increased. In contrast, fermions are repelled by the Pauli blocking, and we find a distinct exchange--correlation hole around $\alpha=0$. This can be seen particularly well in Fig.~\ref{fig:Scanline_N4_ideal}, where we show scan lines over $g^\textnormal{int}_\textnormal{c2p}(r_1,\alpha,0.75,1.25)$ at $r_1=1$ (see the red solid line in the top left panel of Fig.~\ref{fig:c2p_idea_N4}). Let us first restrict ourselves to the red curves corresponding to $\beta=1$, i.e., to the same conditions as in the top row of Fig.~\ref{fig:c2p_idea_N4}. The solid curve corresponds to fermions, which experience a degeneracy pressure, whereas the dashed curve depicts bosonic results, which feel an effective attraction.

The center row in Fig.~\ref{fig:c2p_idea_N4} corresponds to $\beta=0.5$ and exhibits a qualitatively similar behavior as for $\beta=1$, which is also true for the scan lines depicted as the green curves in Fig.~\ref{fig:Scanline_N4_ideal}.
Lastly, the bottom row shows the integrated C2P for a high-temperature case, $\beta=0.05$. First and foremost, we note that almost all structure disappears from the system, and $g^\textnormal{int}_\textnormal{c2p}(r_1,\alpha,0.75,1.25)$ becomes nearly flat. Still, there remain distinct vestiges both of the fermionic exchange--correlation hole and the bosonic maximum for $0\leq \alpha \lesssim 30$, cf.~the black curves in Fig.~\ref{fig:Scanline_N4_ideal}. This is particularly remarkable as the bosonic and fermionic partition functions are almost equal, and we find an average sign of $S\approx0.99$.

\begin{figure}
\includegraphics[width=0.4447\textwidth]{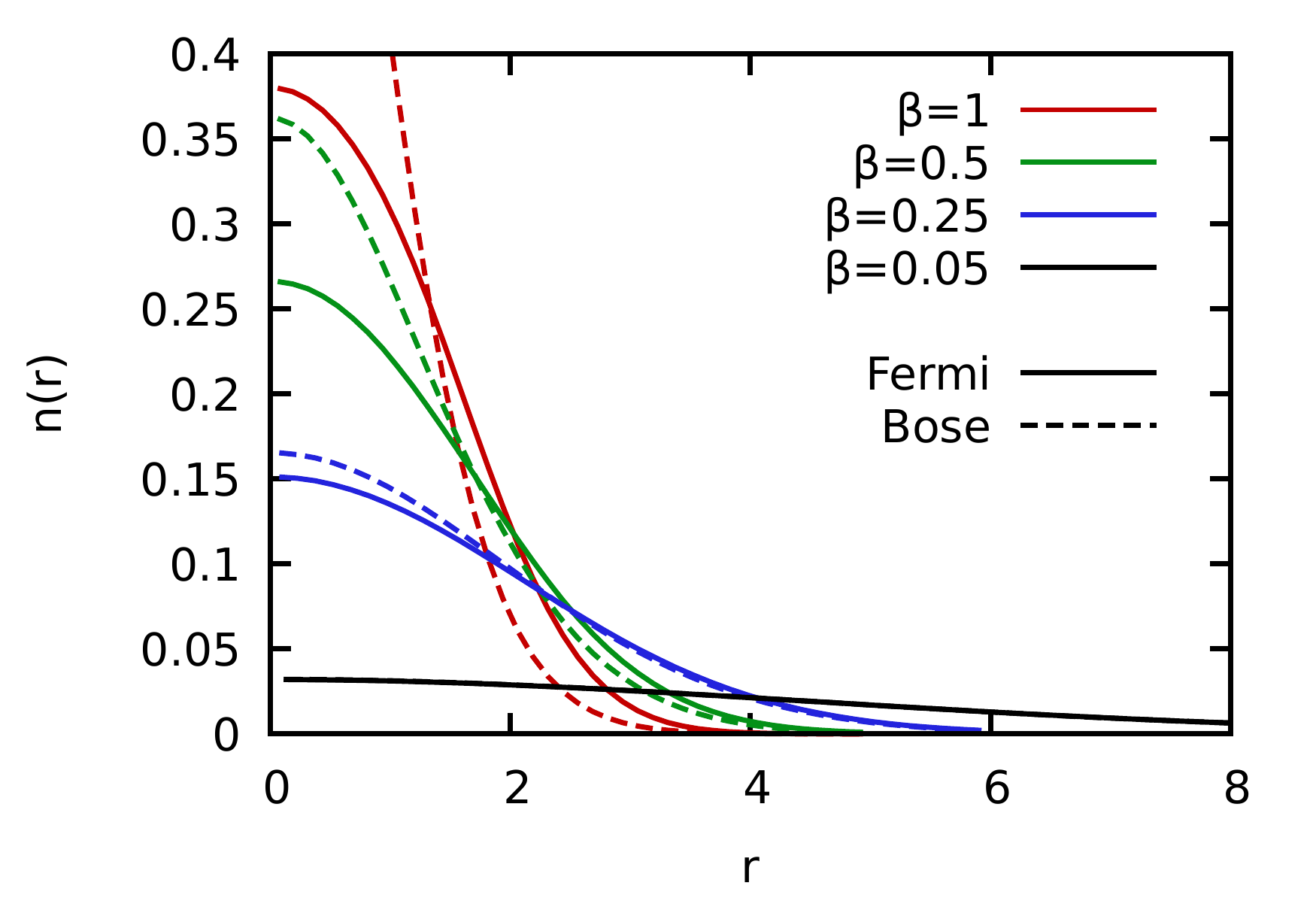}
\caption{\label{fig:Density_N4_ideal}
Radial density distribution $n(r)$ of $N=4$ ideal bosons (dashed) and fermions (solid) for $\beta=1$ (red), $\beta=0.5$ (green), $\beta=0.25$ (blue), and $\beta=0.05$ (black).
}
\end{figure}

To further illustrate these findings, we show the corresponding radial density distributions $n(r)$ in Fig.~\ref{fig:Density_N4_ideal}. Upon changing the temperature, there appear two main trends: i) with increasing temperature, the density is more smeared out and particles are more frequently found at larger distances to the center of the trap $r$, and ii) the effect of quantum statistics decays, and eventually vanishes. In particular, the difference between the fermionic and bosonic curves at $\beta=0.05$ cannot be resolved within the given statistical uncertainty.

\begin{figure}\centering
\includegraphics[width=0.4447\textwidth]{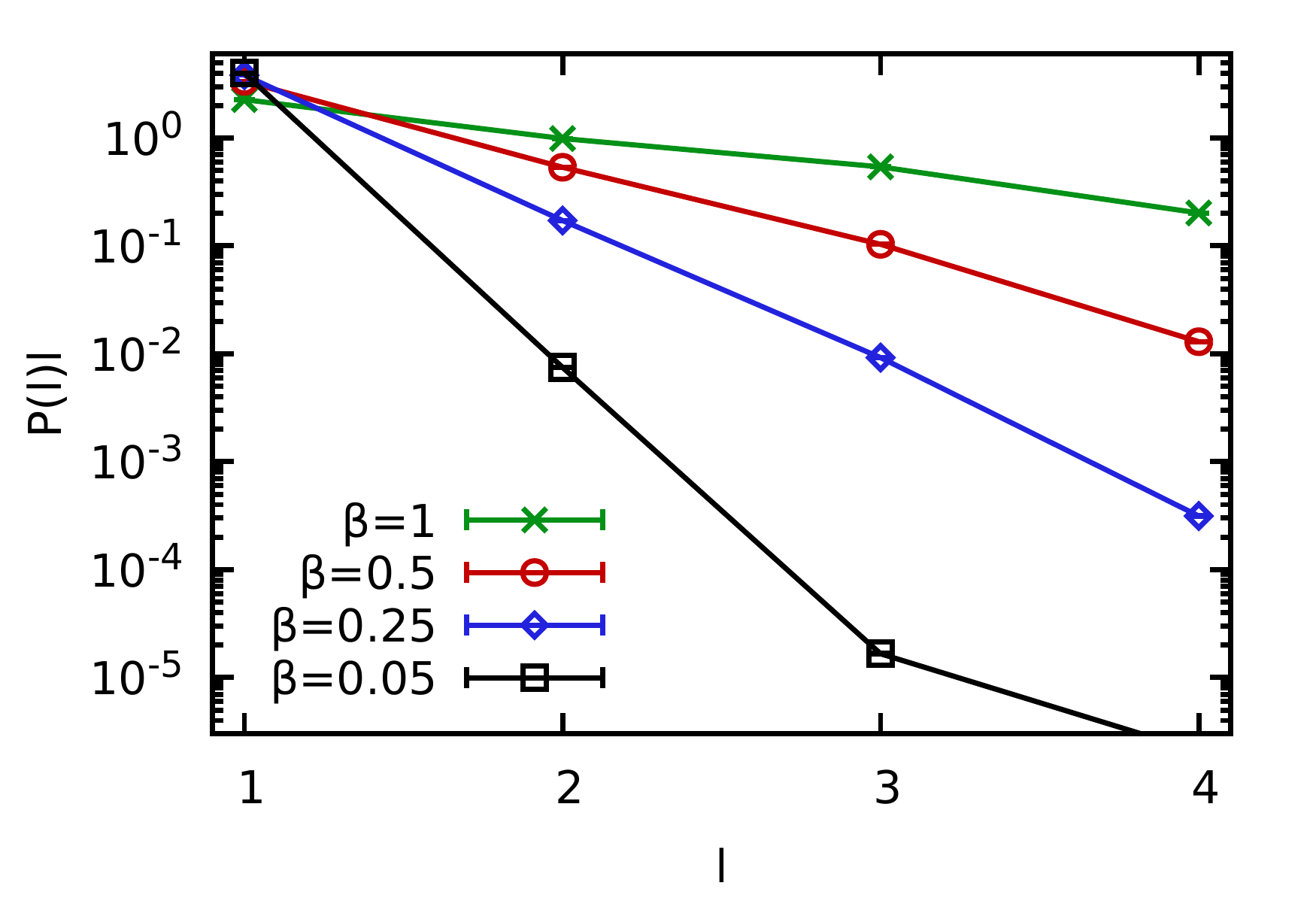}
\caption{\label{fig:ideal_permutations}
Probability to find a particle in a permutation cycle of length $l$ for $N=4$ ideal bosons in a $2D$ harmonic trap. The solid lines depict the exact result from Eq.~(\ref{eq:ideal_Pl}), and the symbols the corresponding data from our PIMC simulations.
}
\end{figure}

To explain the remarkable differences in $g^\textnormal{int}_\textnormal{c2p}(r_1,\alpha,0.75,1.25)$ even at high temperature, we directly examine the manifestation of quantum statistics in our PIMC simulations in Fig.~\ref{fig:ideal_permutations}. More specifically we show the permutation-cycle frequencies $P(l)l$, i.e., the probability to find a particle in a permutation cycle of length $l$ obeying Eq.~(\ref{eq:permutation_norm}). The solid lines correspond to the exact result known from theory [Eq.~(\ref{eq:ideal_Pl})], and the symbols to the results from our PIMC simulations at the same conditions. Again, we note the perfect agreement between theory and simulations, which further validates our implementation. At the lowest temperature, permutation cycles of all possible lengths occur with high probability in our simulation, which explains the relatively small average sign and the significant differences between bosons and fermions for both $g^\textnormal{int}_\textnormal{c2p}(r_1,\alpha,0.75,1.25)$ and $n(r)$. In fact, $P(l)l$ will eventually become completely flat in the low-temperature limit~\cite{krauth_book,dornheim_permutation_cycles}, which means that $S$ vanishes and fermionic PIMC simulations become impossible.

\begin{figure*}
\includegraphics[width=0.40447\textwidth]{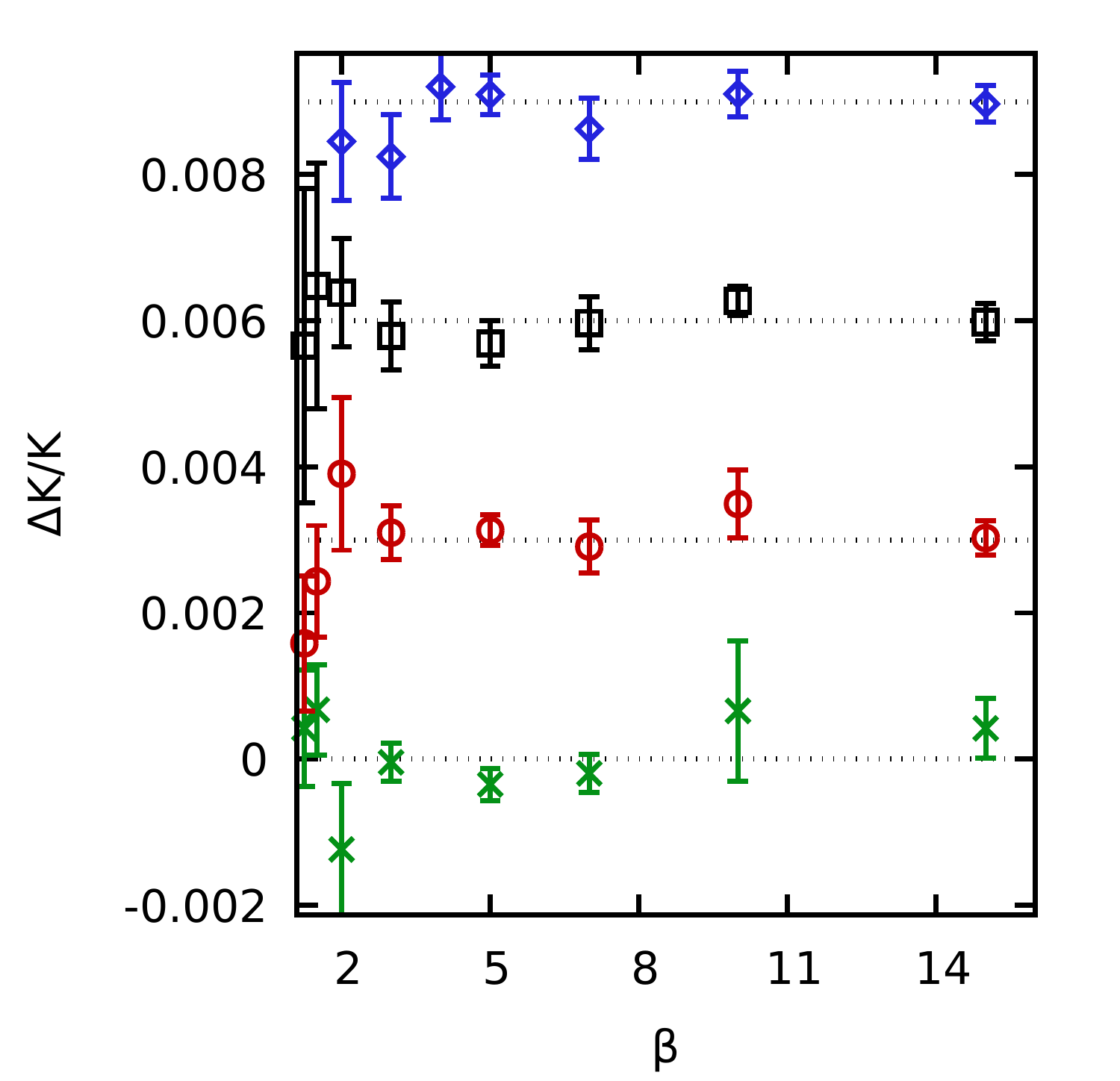}
\includegraphics[width=0.40447\textwidth]{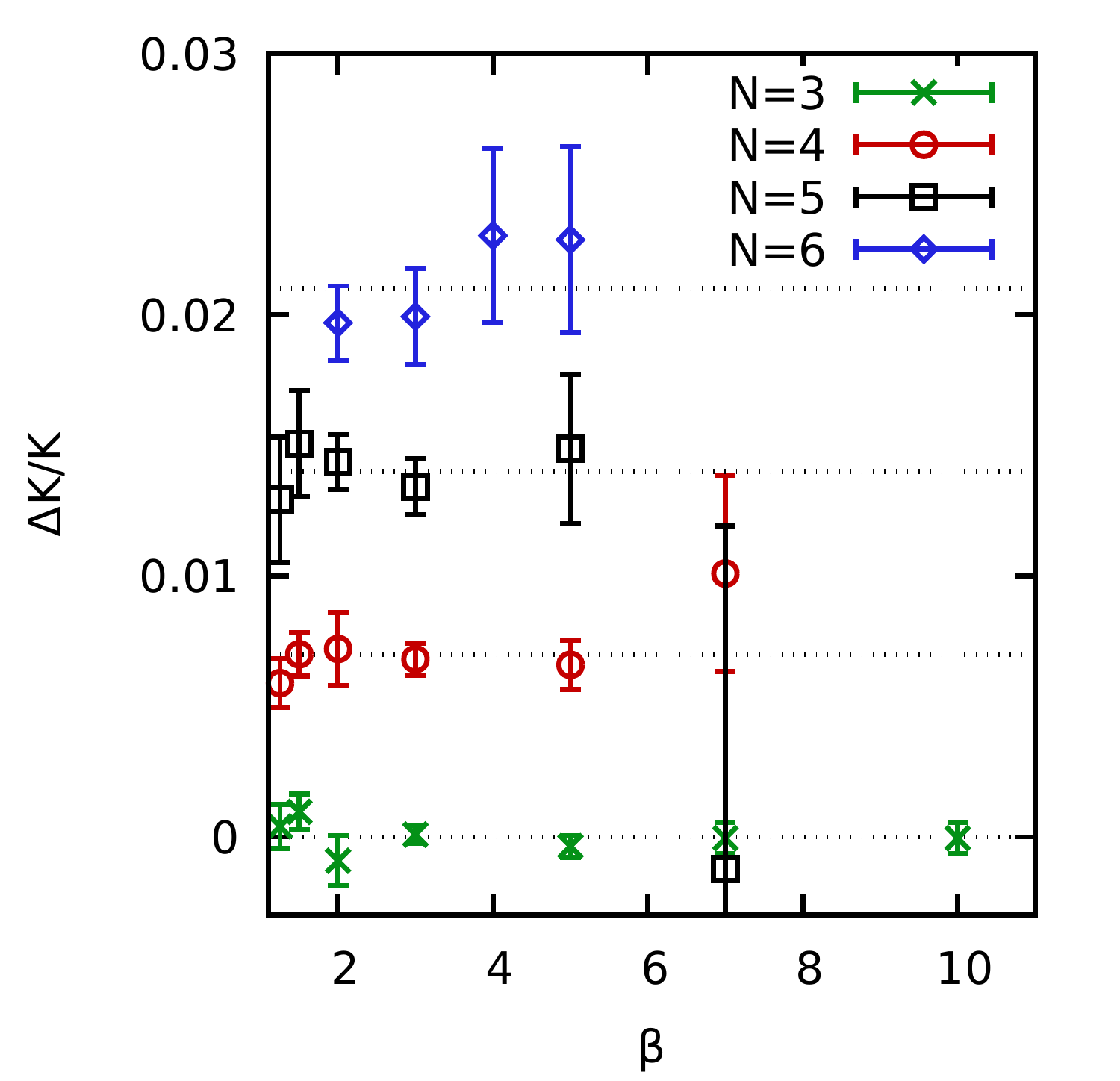}
\caption{\label{fig:virial_lambda3}
Verification of the virial theorem at $\lambda=3$ for Bose- (left) and Fermi-statistics (right). Shown is the relative deviation in our PIMC data for the kinetic energy between the standard thermodynamic estimator (e.g., Ref.~\cite{janke}) and the virial theorem, Eq.~(\ref{eq:K_vir}). The green crosses, red circles, black squares, and blue diamonds correspond to $N=3,4,5,$ and $6$ which have been shifted for better visibility, see the dotted grey lines.
}
\end{figure*}

Upon increasing the temperature, on the other hand, the probability to find particles not involved in any exchange-cycles, $P(1)$, increases whereas $P(l)$ decreases for all other $l\geq2$. At $\beta=0.05$, around $99\%$ of the particles within the PIMC simulation are involved in single-particle cycles, which explains the nearly vanishing impact of quantum statistics on $n(r)$ in that case. We stress that this is a consequence of i) the particles being spread out at larger $r$, which makes exchange less likely, and ii) the thermal wavelength $\lambda_\beta=\sqrt{2\pi\beta}$ being small. However, those configurations in which two particles do come close to each other have similarly large probabilities to have or not have a permutation cycle in it, which would result in a negative or positive configuration weight, respectively. Consequently, $g^\textnormal{int}_\textnormal{c2p}(r_1,\alpha,0.75,1.25)$ at small angular distances constitutes the perfect tool to resolve the resulting impact of quantum statistics, which nearly completely vanishes from averaged observables like the radial density.

We have thus demonstrated that the integrated C2P does indeed constitute a suitable tool for the investigation of the structural properties of trapped quantum systems. In the following section, it will be used to illuminate the interplay of quantum statistics and scattering, and thermal excitations for the more interesting case of ultracold atoms with dipole interaction.

\subsection{Structural properties of quantum dipole systems\label{sec:structural_results}}

Let us start our discussion of the \textit{ab initio} PIMC simulation of ultracold dipolar atoms with a further check of our implementation. While the partition function and, hence, all derivative thermodynamic properties are a-priori unknown for $\lambda\neq0$, the different contributions to the total energy are related by the virial theorem~\cite{greiner_book}. For example, it is possible to express the expectation value of the kinetic energy $K$ in terms of the potential energy due to the external potential $V_\textnormal{ext}$ and the interaction energy $W$ as,
\begin{eqnarray}\label{eq:K_vir}
K = V_\textnormal{ext} - 3 \frac{W }{2} \quad .
\end{eqnarray}
Note that Eq.~(\ref{eq:K_vir}) holds for all system parameters ($N$, $\beta$, and $\lambda$) and both for bosons and fermions. To use the virial theorem as a verification, we independently estimate the three different contributions to the energy in our PIMC simulations and compare the lhs.~of Eq.~(\ref{eq:K_vir}) to the expression on the rhs.

The results are shown in Fig.~\ref{fig:virial_lambda3}, where we plot the relative difference between the two expressions for four different particle numbers at moderate coupling ($\lambda=3$) versus the inverse temperature $\beta$. Note that the results for $N=4,5,6$ have been shifted upwards for better visibility, see the dashed grey lines. The left column corresponds to Bose statistics, and the results are of high quality. More specifically, the difference between the two different estimators for $K$ vanishes within the given statistical uncertainty for all data points with an accuracy of $\Delta K/K\sim10^{-4}$. For completeness, we mention that the comparably large fluctuations at high temperature for $N=4$ (red circles) and $N=5$ (black squares) are a consequence of the inherent large variance of the thermodynanmic estimator for the kinetic energy, see Ref.~\cite{janke} for an extensive discussion of this issue.

Let us now proceed to the right column corresponding to Fermi statistics. Recall that the fermionic expectation values are extracted from a standard bosonic PIMC simulation by keeping track of cancellations and the sign, and subsequently evaluating Eq.~(\ref{eq:ratio}), see Sec.~\ref{sec:PIMC_theory}. Again, $\Delta K$ vanishes within the Monte Carlo error bars, although the uncertainty is somewhat larger due to the fermion sign problem (see Ref.~\cite{dornheim_sign_problem} for an accessible topical discussion). In particular, the sign problem is the reason for the increasing error bars towards low temperature, which becomes even more pronounced for larger system sizes.

Still, we conclude that the virial theorem is perfectly fulfilled by our PIMC expectation values, and thus fully validates our implementation.

\begin{figure*}
\includegraphics[width=0.40447\textwidth]{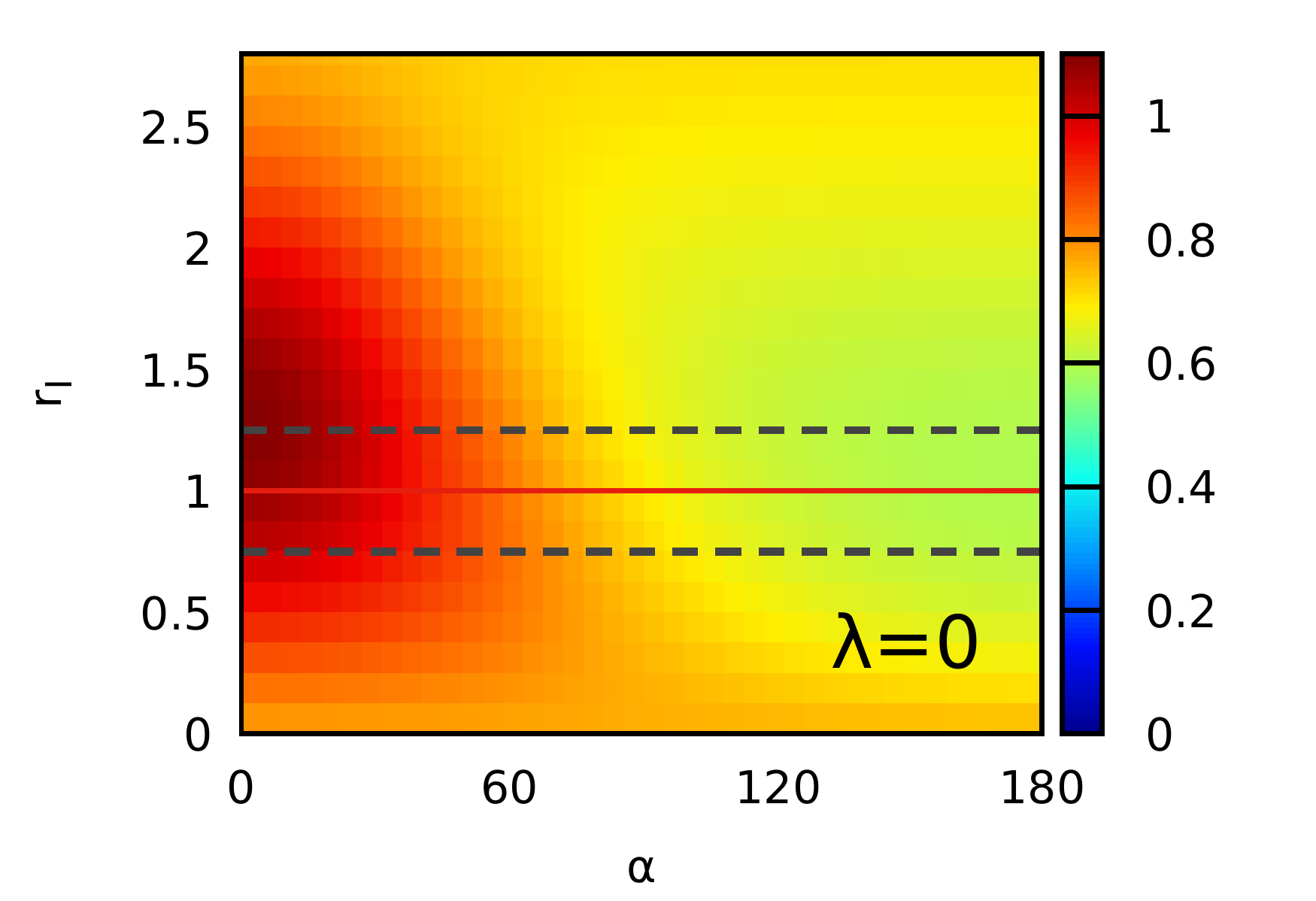}
\includegraphics[width=0.40447\textwidth]{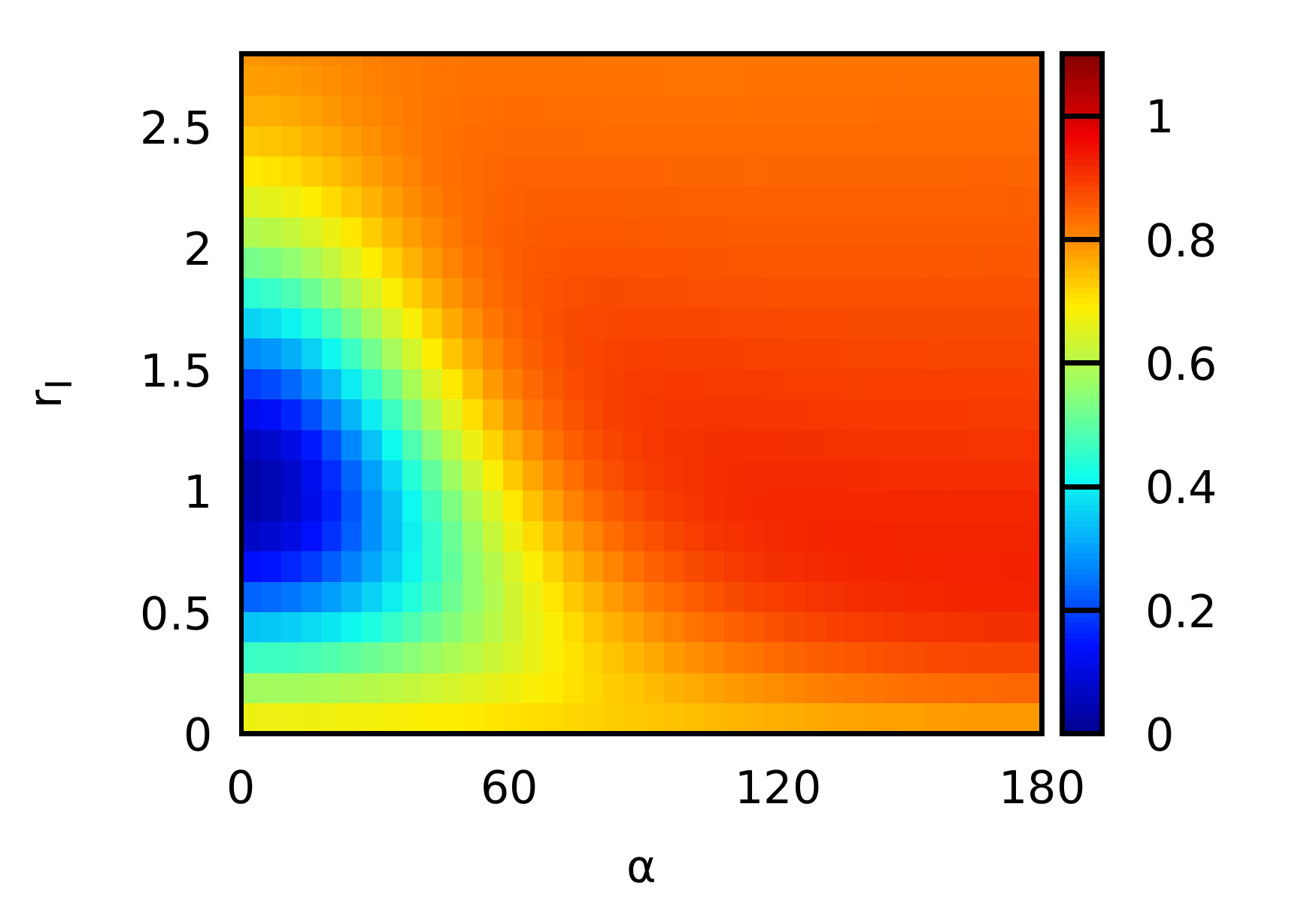}\\ \vspace*{-0.59cm}
\includegraphics[width=0.40447\textwidth]{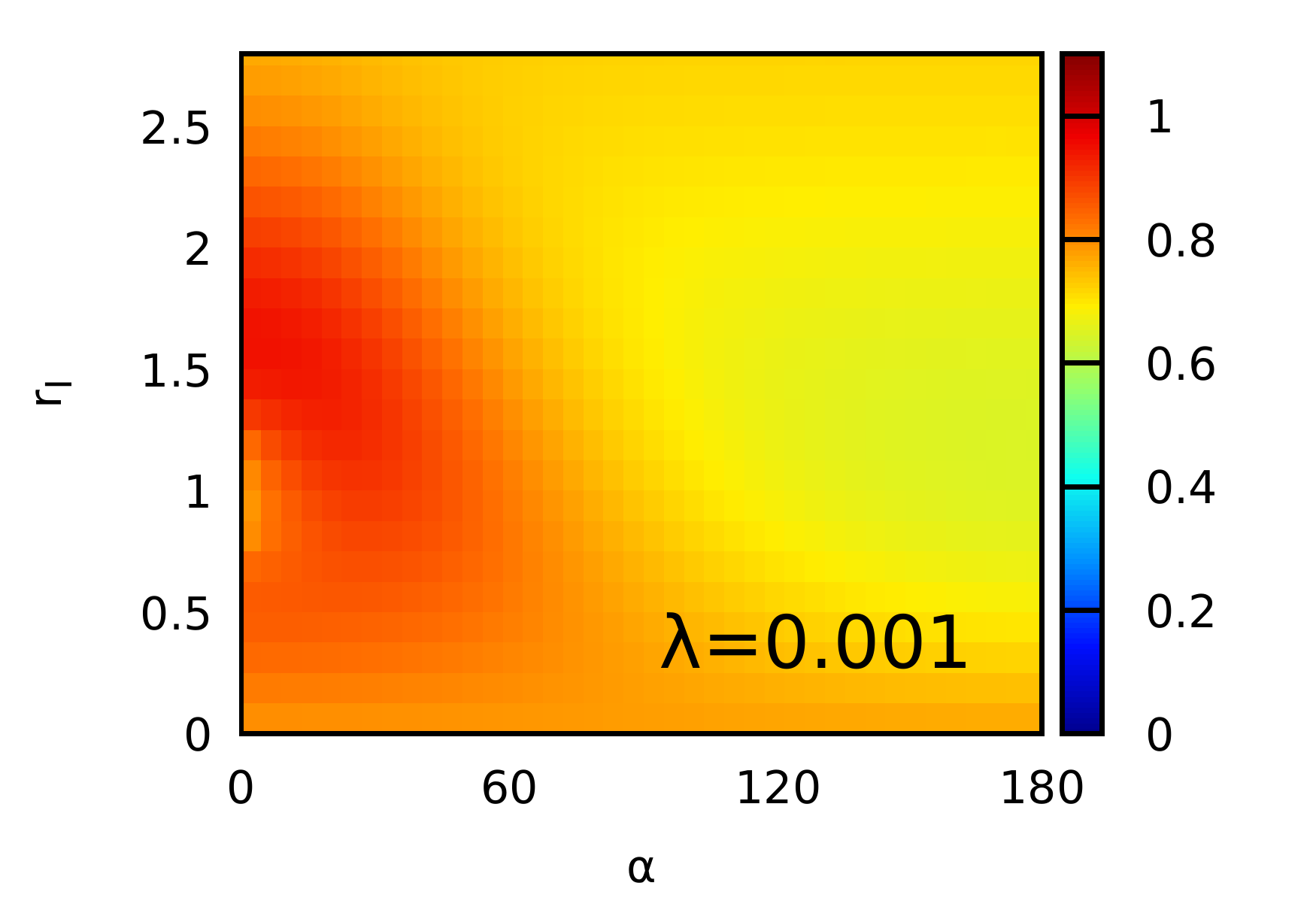}
\includegraphics[width=0.40447\textwidth]{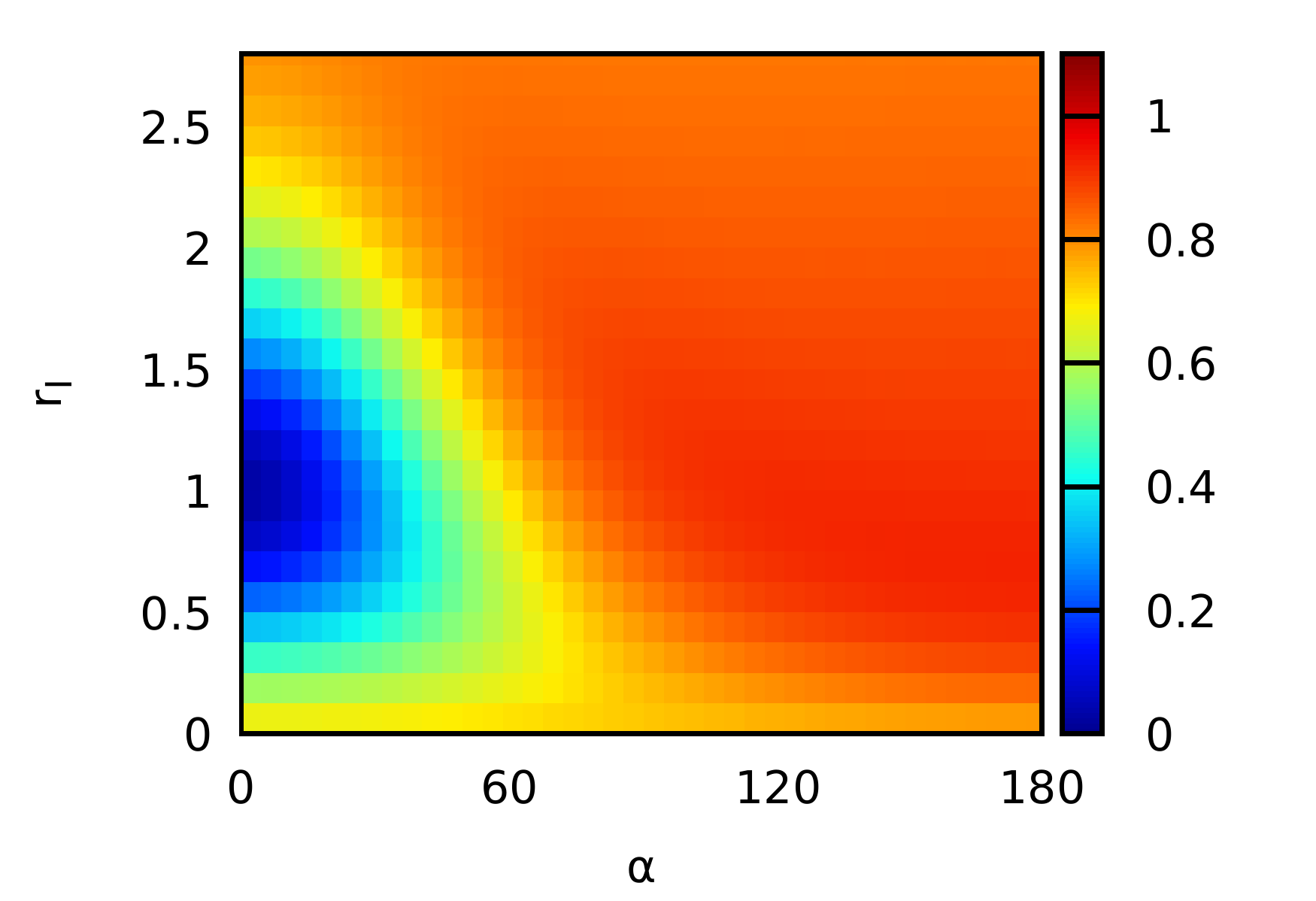}\\ \vspace*{-0.59cm}
\includegraphics[width=0.40447\textwidth]{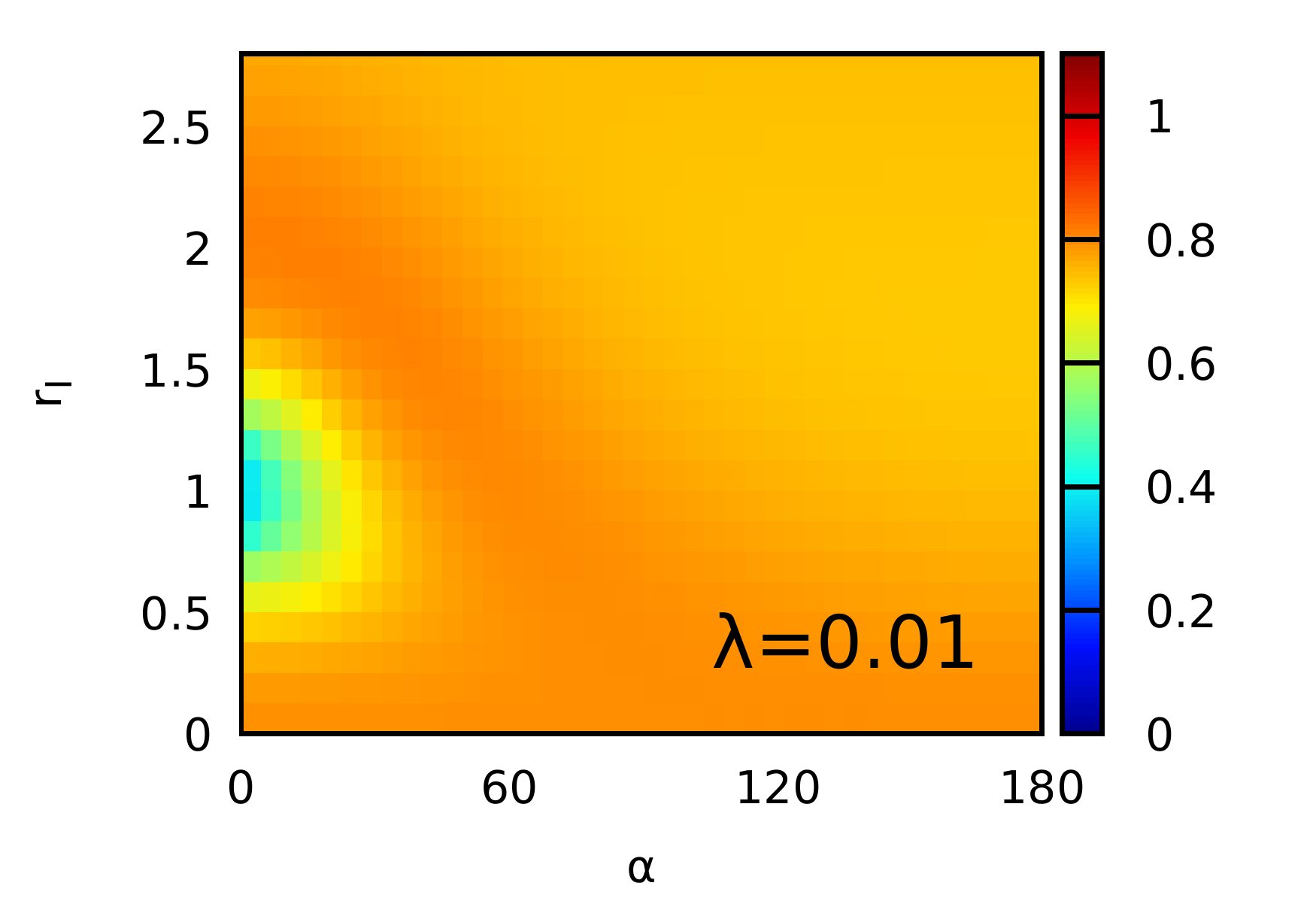}
\includegraphics[width=0.40447\textwidth]{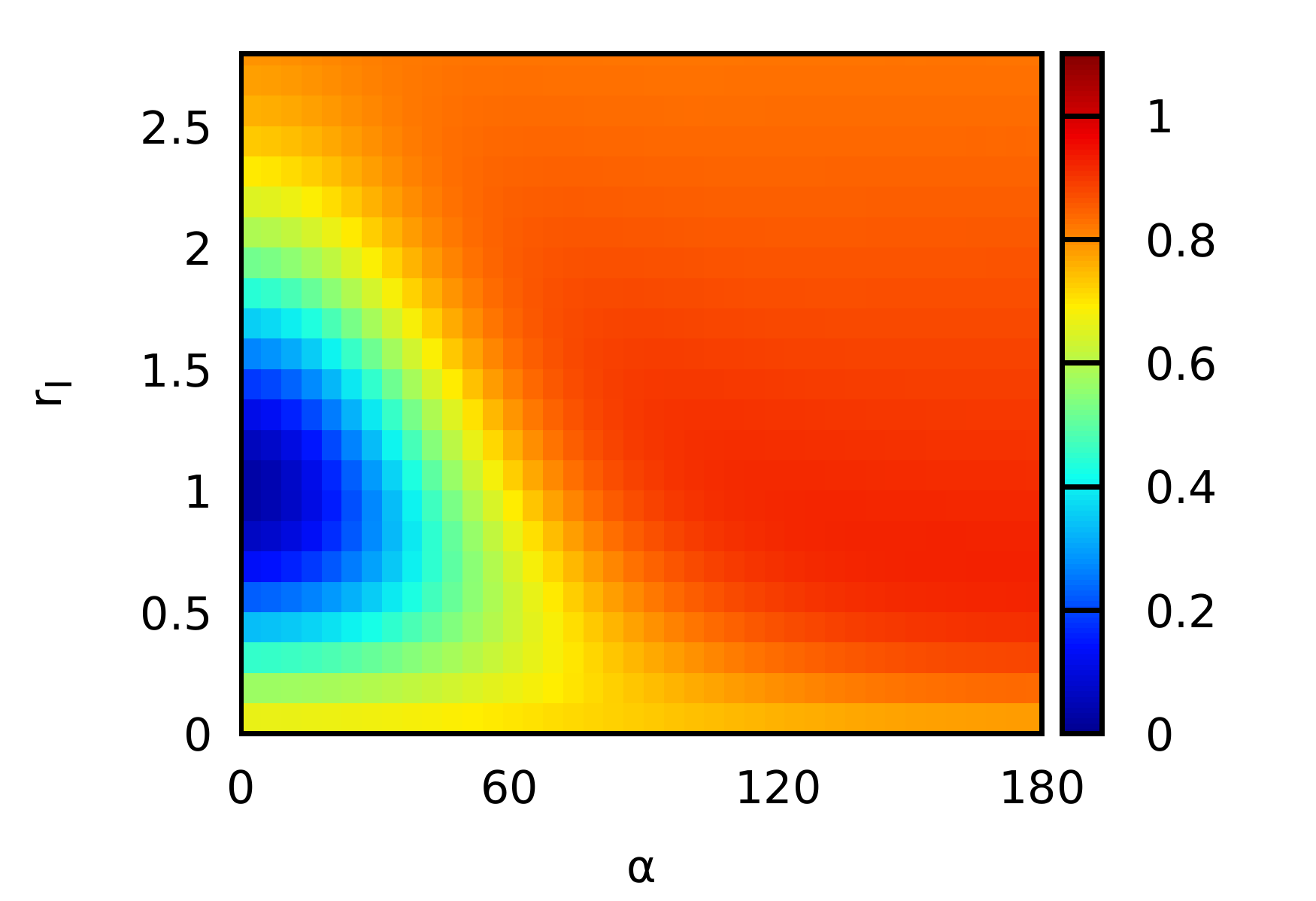}\\ \vspace*{-0.59cm}
\includegraphics[width=0.40447\textwidth]{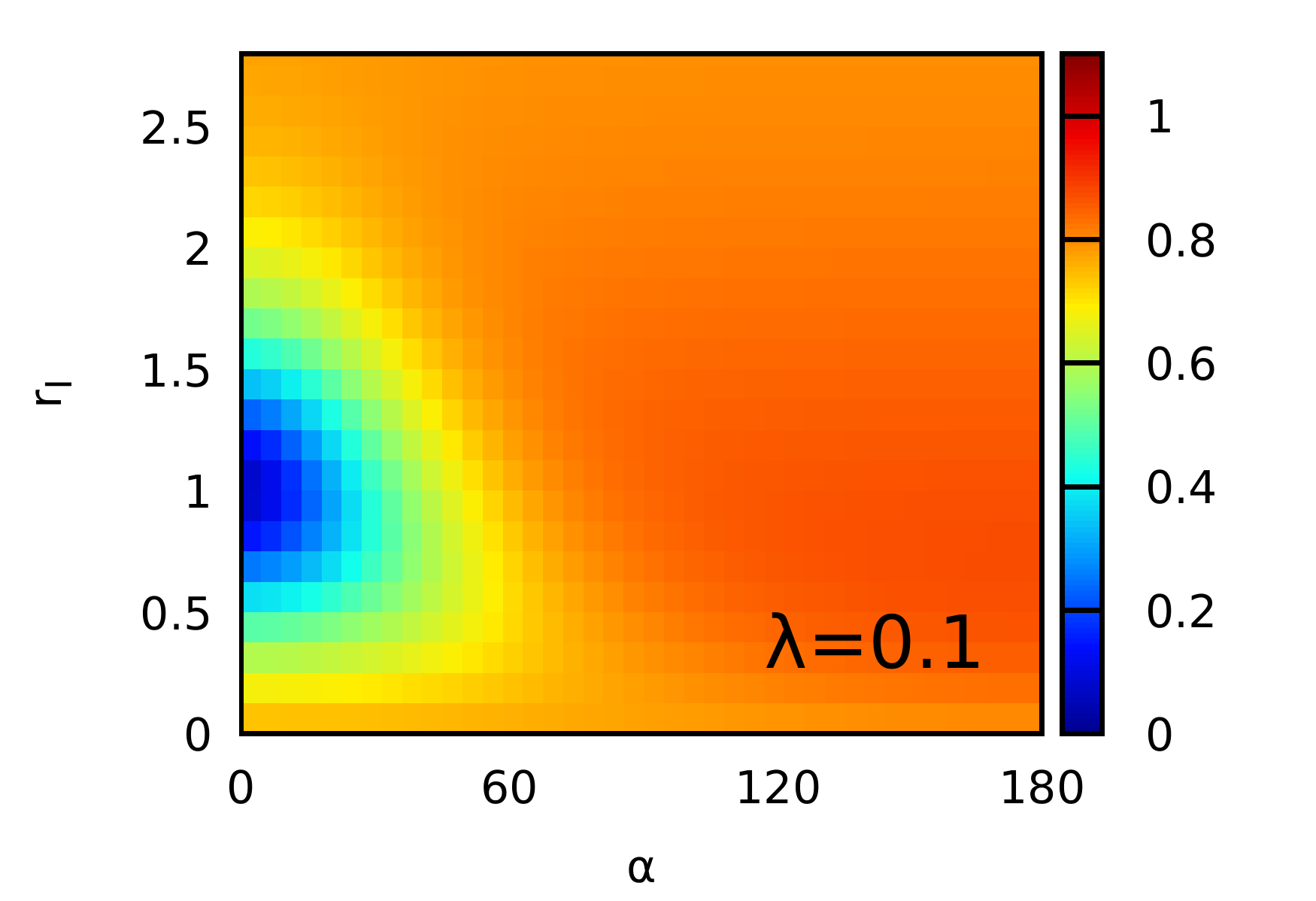}
\includegraphics[width=0.40447\textwidth]{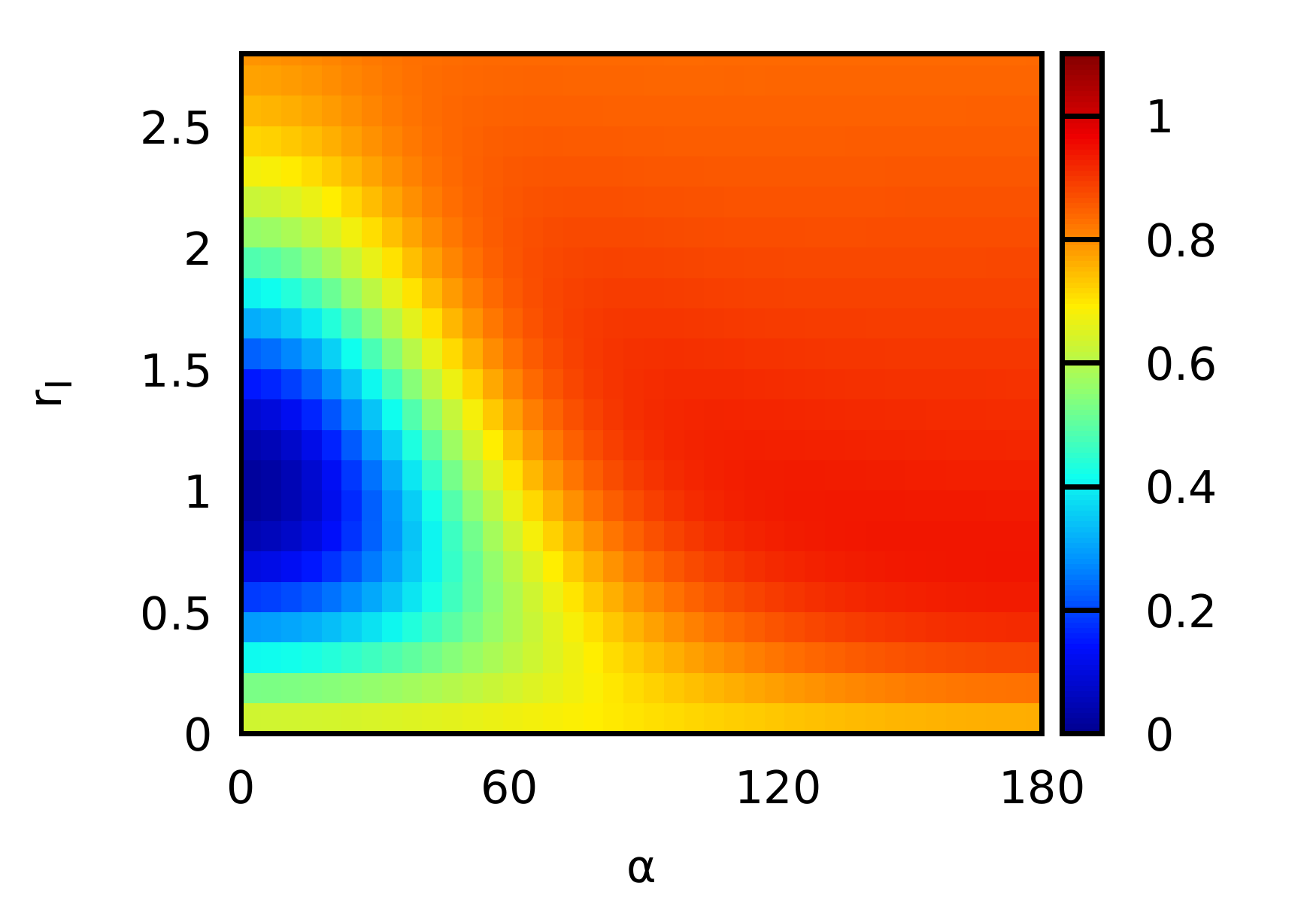}
\caption{\label{fig:c2p_N4_lambda}
Integrated center-two particle correlation function $g^\textnormal{int}_\textnormal{c2p}(r_1,\alpha,0.75,1.25)$ [cf.~Eq.~(\ref{eq:c2p_int})] for $N=4$ and $\beta=1$. The left and right columns correspond to Bose- and Fermi-statistics, and the rows to different values of the coupling parameter $\lambda$. The dashed dark grey lines in the top left panel indicate the integration boundaries for the reference particle, and the solid red line the scan line depicted in the bottom panel of Fig.~\ref{fig:N4_LAMDBA}.
}
\end{figure*}

Being equipped with the PIMC approach and the previously discussed integrated C2P, we are now in a position to address the first major point of this work: the transition from the ideal system ($\lambda=0$) to the interacting quantum dipole case upon increasing the coupling parameter $\lambda$, and how it is affected by quantum statistics. To this end, we show the integrated C2P $g^\textnormal{int}_\textnormal{c2p}(r_1,\alpha,0.75,1.25)$ in Fig.~\ref{fig:c2p_N4_lambda} for $N=4$ particles at a moderate temperature, $\beta=1$. As usual, the left and right columns correspond to Bose- and Fermi-statistics, and the four rows belong to $\lambda=0,0.001,0.01,0.1$ (in descending order).

\begin{figure}\centering
\includegraphics[width=0.4647\textwidth]{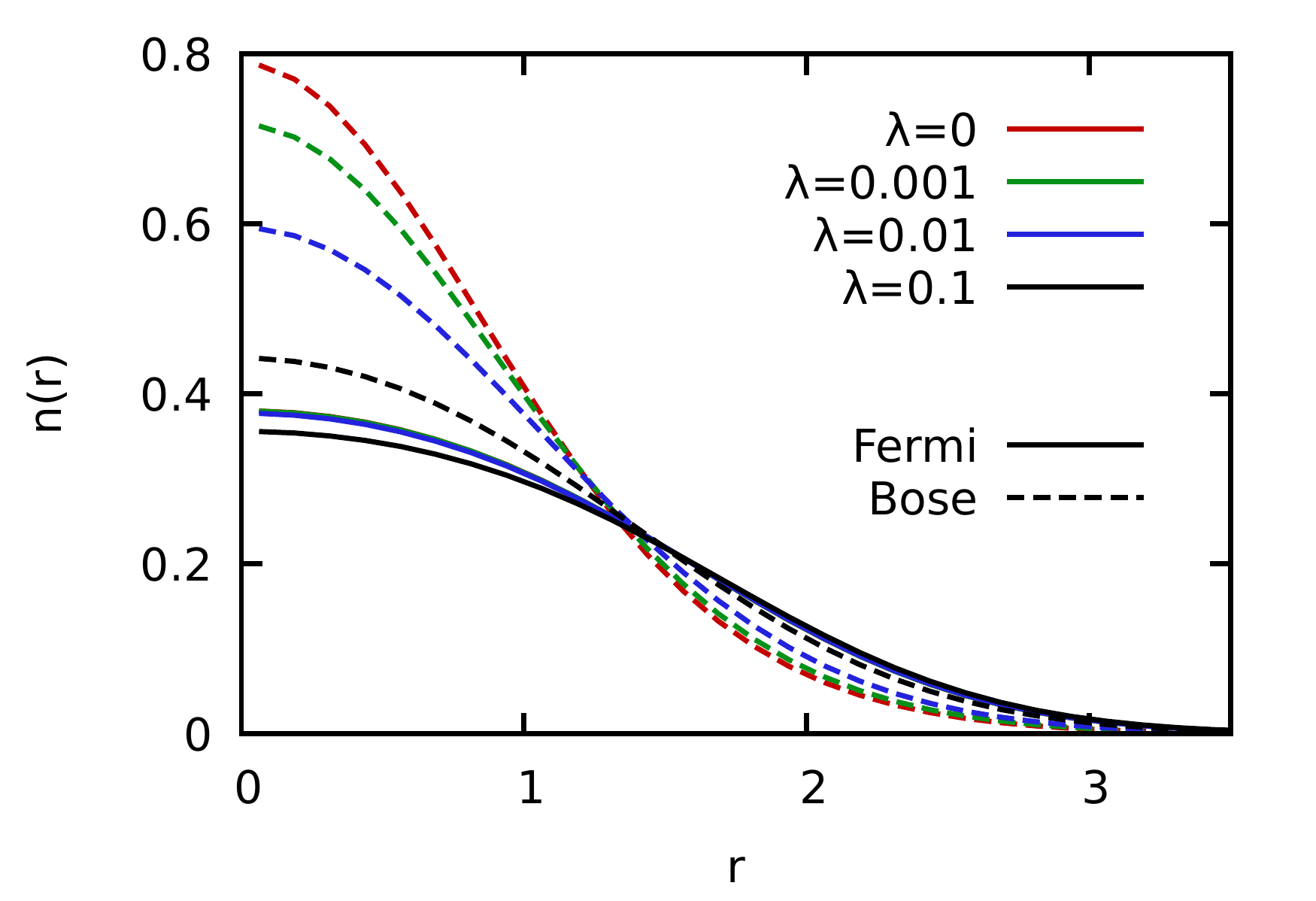}\\
\includegraphics[width=0.4647\textwidth]{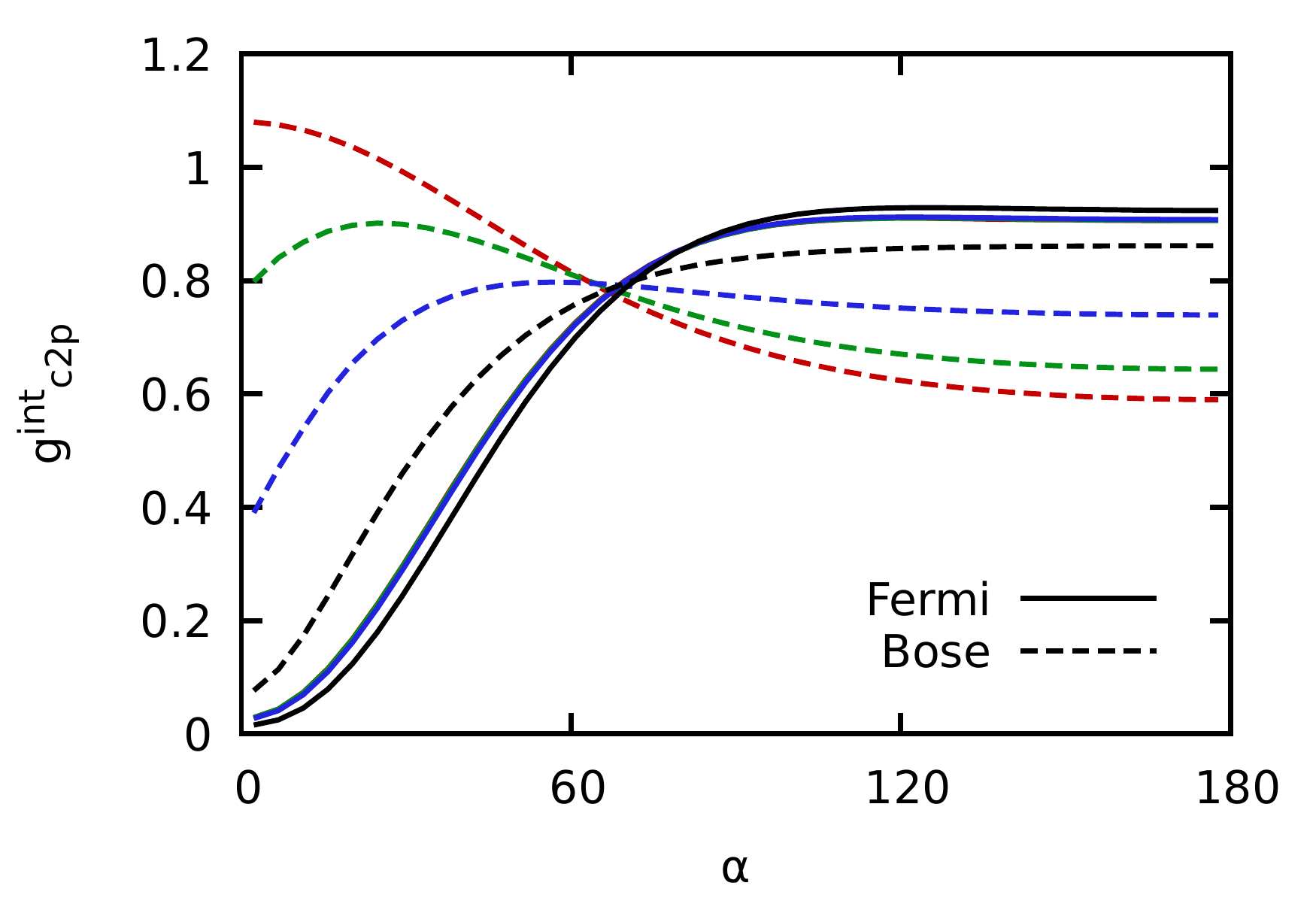}
\caption{\label{fig:N4_LAMDBA}
Density profile (top) and scanline over the integrated C2P (bottom) for the same conditions as in Fig.~\ref{fig:c2p_N4_lambda}.
}
\end{figure}

Let us start our discussion by revisiting the noninteracting case depicted in the top row, with an effective attraction of bosons and the exchange--correlation hole in the case of fermions. Again, this can be seen particularly well in the scan lines depicted in the bottom panel of Fig.~\ref{fig:N4_LAMDBA} as the dashed (Bose) and solid (Fermi) red curves.

The second row from the top in Fig.~\ref{fig:c2p_N4_lambda} corresponds to very weak coupling, $\lambda=0.001$, and the situation substantially changes. In particular, the dipole potential $W_\lambda(r) = \lambda / r^3$ diverges towards $r=0$ for every finite values of $\lambda$ and therefore counters the tendency of bosons to cluster around each other. Consequently, there appears a dip in $g^\textnormal{int}_\textnormal{c2p}(r_1,\alpha,0.75,1.25)$ for small $\alpha$, see also the dashed green curve in the bottom panel of Fig.~\ref{fig:N4_LAMDBA}. For completeness, we mention that the dip around $\alpha=0$ would be even more pronounced, if the integration interval of $r_2$ was decreased. Presently, it is potentially possible to have two particles at the same angle, but, say, $r_1=0.75$ and $r_2=1.25$, which are hardly affected by the divergent dipole potential, but would still contribute to $g^\textnormal{int}_\textnormal{c2p}(r_1,\alpha=0,0.75,1.25)$, thereby reducing the dip.

For fermions, on the other hand, the finite coupling strength has no discernible effect on the integrated C2P, as it is hidden by the exchange--correlation hole. Consequently, the scan line in the bottom panel of Fig.~\ref{fig:N4_LAMDBA} cannot be distinguished from the ideal curve with the bare eye.
In addition, the top panel of the same figure depicts the corresponding radial density distributions $n(r)$. Here, too, we find a significant difference between $\lambda=0$ and $\lambda=0.001$ for bosons, but none for fermions.

Upon further increasing the coupling strength to $\lambda=0.01$, the competition between bosonic clustering and the dipolar repulsion becomes even more pronounced and we observe the emergence of a correlation hole, albeit substantially less pronounced than in the case of fermions. The latter are still hardly affected by the interaction both regarding the integrated C2P and the radial density, whereas the bosons are further pushed away from the center of the trap.

Finally, the bottom row of Fig.~\ref{fig:c2p_N4_lambda} corresponds to $\lambda=0.1$. In this case, the correlation hole constitutes the most prominent feature for bosons, too, and $g^\textnormal{int}_\textnormal{c2p}(r_1,\alpha,0.75,1.25)$ resembles the case of fermions. The same is true for the radial density distribution, as the particles are more strongly separated, and the bosonic clustering is almost completely masked by the dipolar repulsion. Remarkably, the fermionic results are still hardly affected by the finite value of $\lambda$, and we find only small deviations from the ideal data both in the integrated C2P and the radial density.

We note that this might indicate that mean-field theories and weak-coupling expansions might be much better in the case of fermions.

\begin{figure}\centering
\includegraphics[width=0.4147\textwidth]{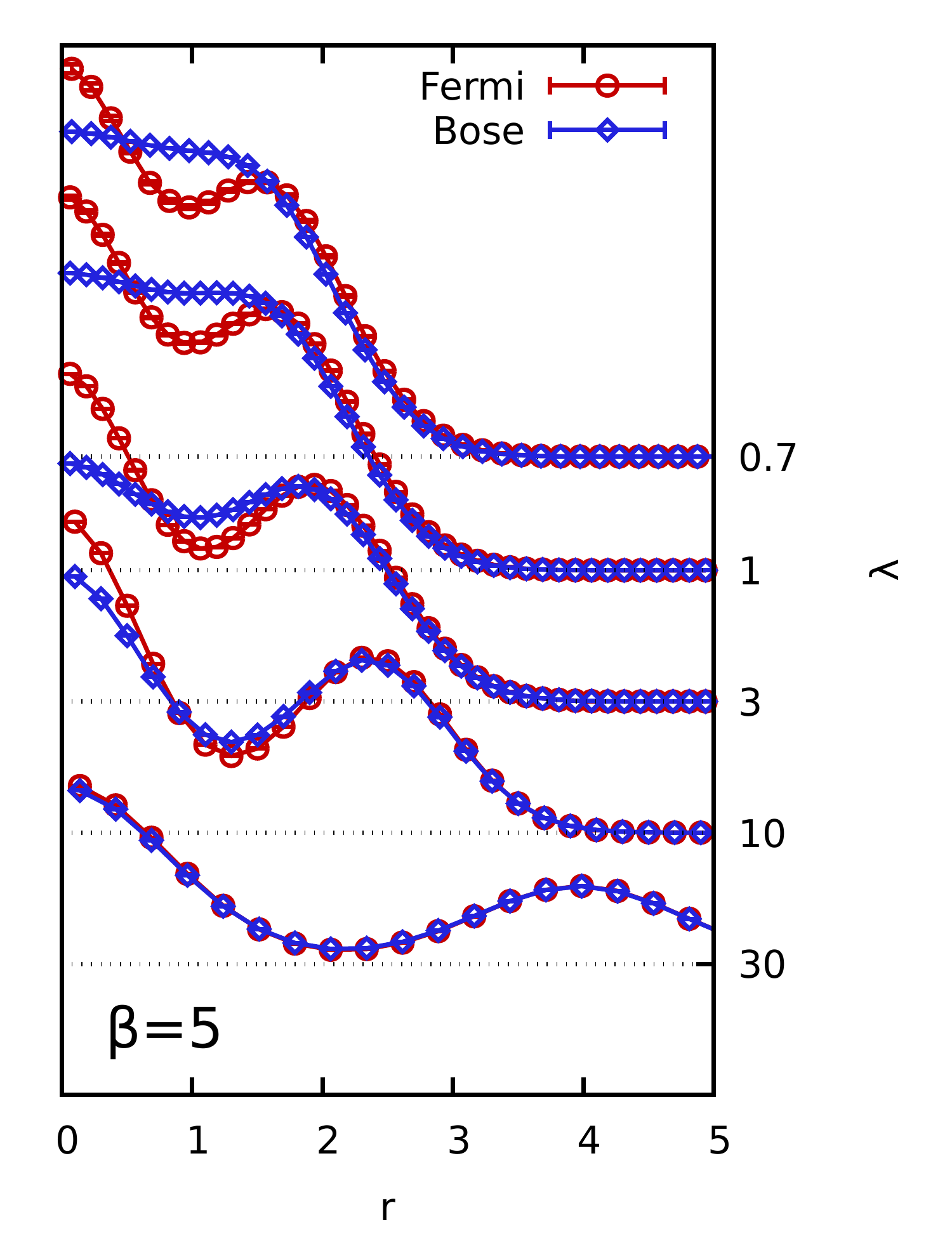}
\caption{\label{fig:densty_beta5_N6_lambda}
Effect of quantum statistics on the radial density in dependence of the coupling strength. Shown are PIMC results for $n(r)$ for $N=6$ ultracold atoms at $\beta=5$ for different values of the coupling parameter $\lambda$. The red circles and blue diamonds distinguish Fermi- and Bose-statistics.
}
\end{figure}

\begin{figure*}\centering
\includegraphics[width=0.4147\textwidth]{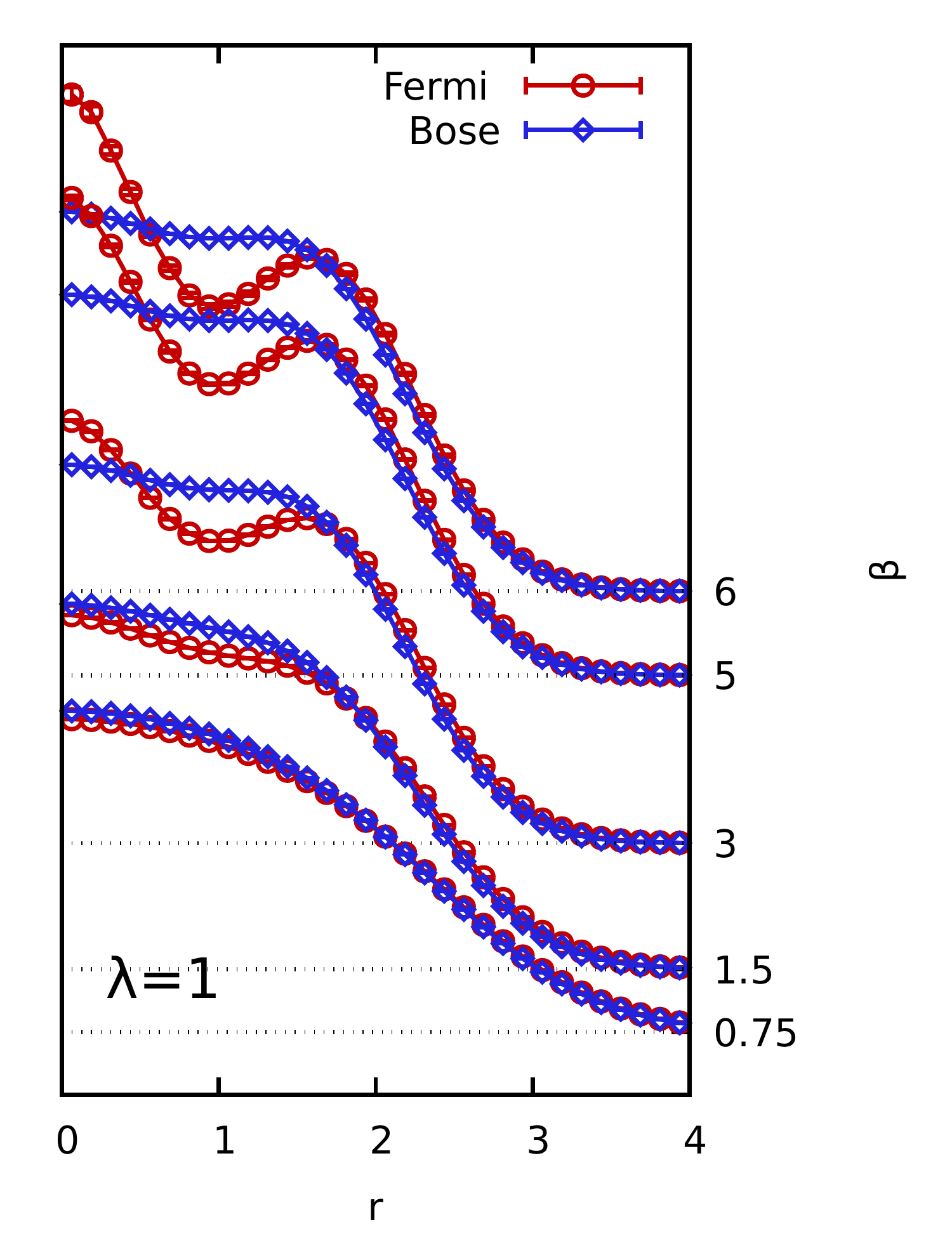}
\includegraphics[width=0.4147\textwidth]{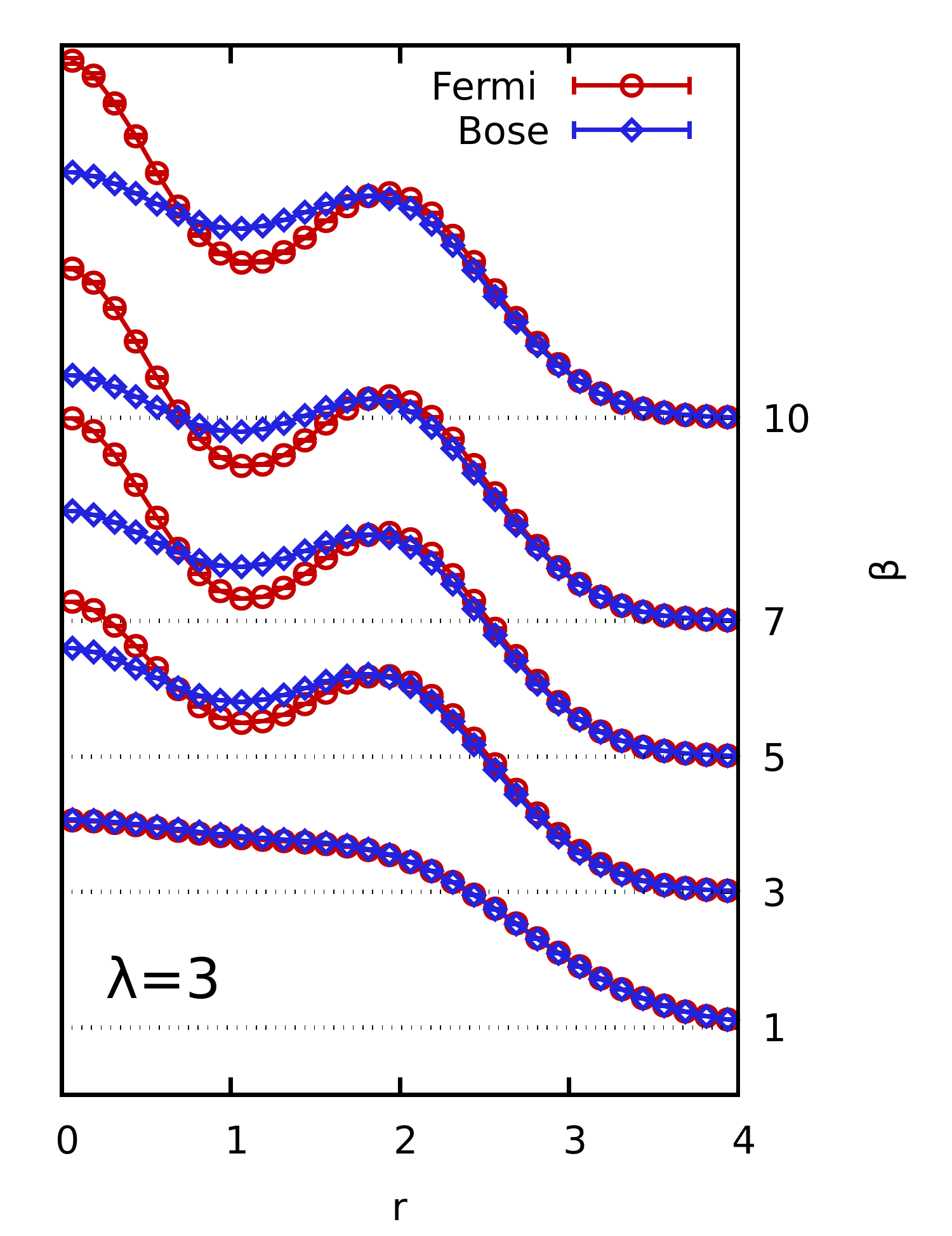}
\caption{\label{fig:densty_beta_N6_lambda3}
Effect of quantum statistics on the radial density. Shown are PIMC results for $n(r)$ for $N=6$ ultracold atoms at $\lambda=1$ (left panel) and $\lambda=3$ (right panel) for different temperatures ($T$ is increasing from top to bottom). The red circles and blue diamonds distinguish Fermi and Bose statistics. 
}
\end{figure*}

Let us next investigate the transition from the classical regime, where the particles are separated by the strong coupling, to the quantum regime. To this end, we simulate $N=6$ ultracold atoms at a relatively low temperature, $\beta=5$, where the expectation values are close to the respective ground state. For completeness, we mention that going to even lower temperature is not possible in the case of fermions due to the fermion sign problem, see Ref.~\cite{dornheim_sign_problem} for a review article, and the discussion below.

In Fig.~\ref{fig:densty_beta5_N6_lambda}, we show PIMC data for the radial density for different values of $\lambda$, with the red circles and blue diamonds corresponding to fermions and bosons. 
The bottom plot has been obtained for strong coupling, $\lambda=30$, and both curves cannot be distinguished with the naked eye. More specifically, we find a pronounced structure with two distinct shells, and the density almost completely vanishes in between.

For $\lambda=10$ (second from the bottom), we still find a shell structure for both kinds of particles, although there do appear significant differences around the center of the trap and the subsequent minimum. Moreover, there is substantially more overlap between the shells than for $\lambda=30$, and the particles are pushed less far away from the center of the trap. 
Thus, $\lambda=10$ might still be viewed as strong coupling, with quantum statistics acting as a perturbation.

Further decreasing the coupling parameter to $\lambda=3$ (center) brings us to the interesting transition regime, where both quantum statistics and the dipole interaction are important at the same time. As a consequence, the shell structure nearly fully disappears for bosons, but remains remarkably pronounced in the case of Fermi statistics. Hence, quantum exchange can no longer be interpreted as a small perturbation at such a moderate coupling strength, and we find an average sign of $S\approx0.063$ in our PIMC simulation.

Finally, we approach the more weakly coupled regime for $\lambda=1$ (second from the top) and $\lambda=0.7$ (top), and the shell structure has completely vanished for bosons in the latter case. Evidently, quantum statistics significantly shape the physical behaviour of the system, and we find average signs of $S\approx 2.3\cdot10^{-3}$ and $S\approx6.9\cdot10^{-4}$ for the two values of $\lambda$. Hence, the PIMC simulations are rendered computationally expensive by the fermion sign problem, and it takes $\mathcal{O}\left(10^4\right)$ CPU hours to accurately resolve the fermionic density. Remarkably, the shell structure is still clearly pronounced and, thus, constitutes a quantum exchange effect. For completeness, we mention that going to even lower values of the coupling parameter is not feasible at these conditions, again due to the fermion sign problem.

\begin{figure*}
\includegraphics[width=0.40447\textwidth]{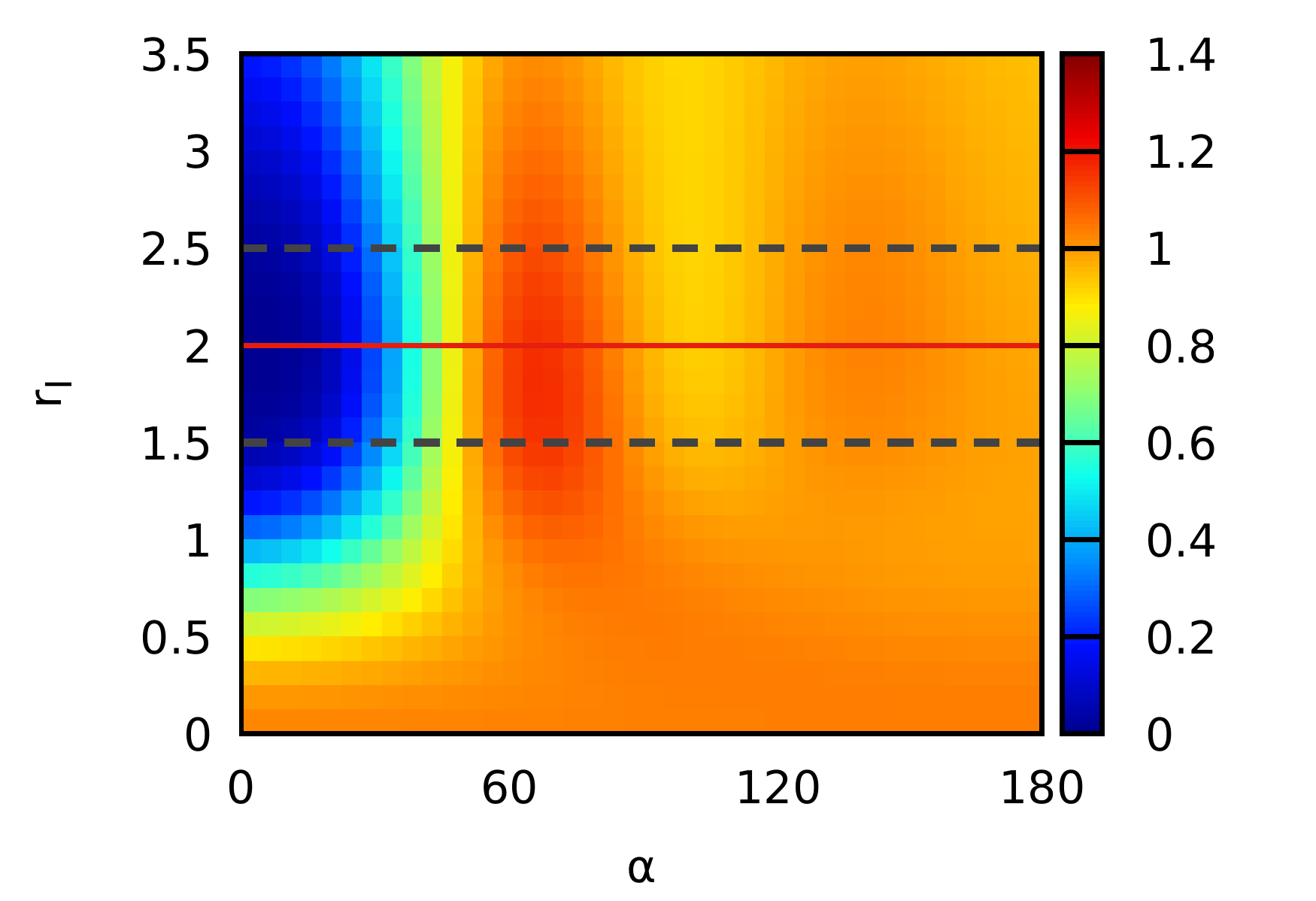}
\includegraphics[width=0.40447\textwidth]{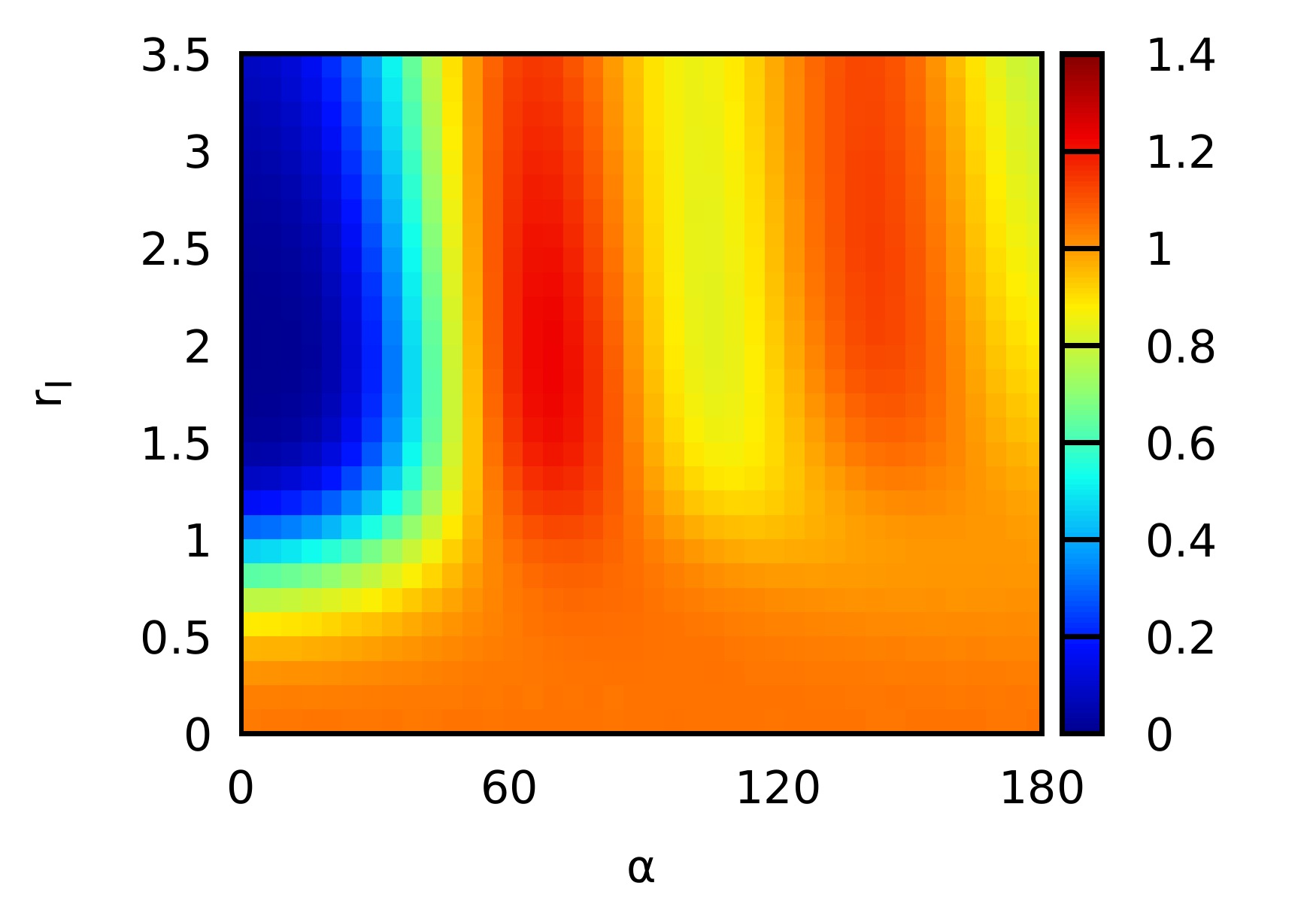}\\ \vspace*{-0.59cm}
\includegraphics[width=0.40447\textwidth]{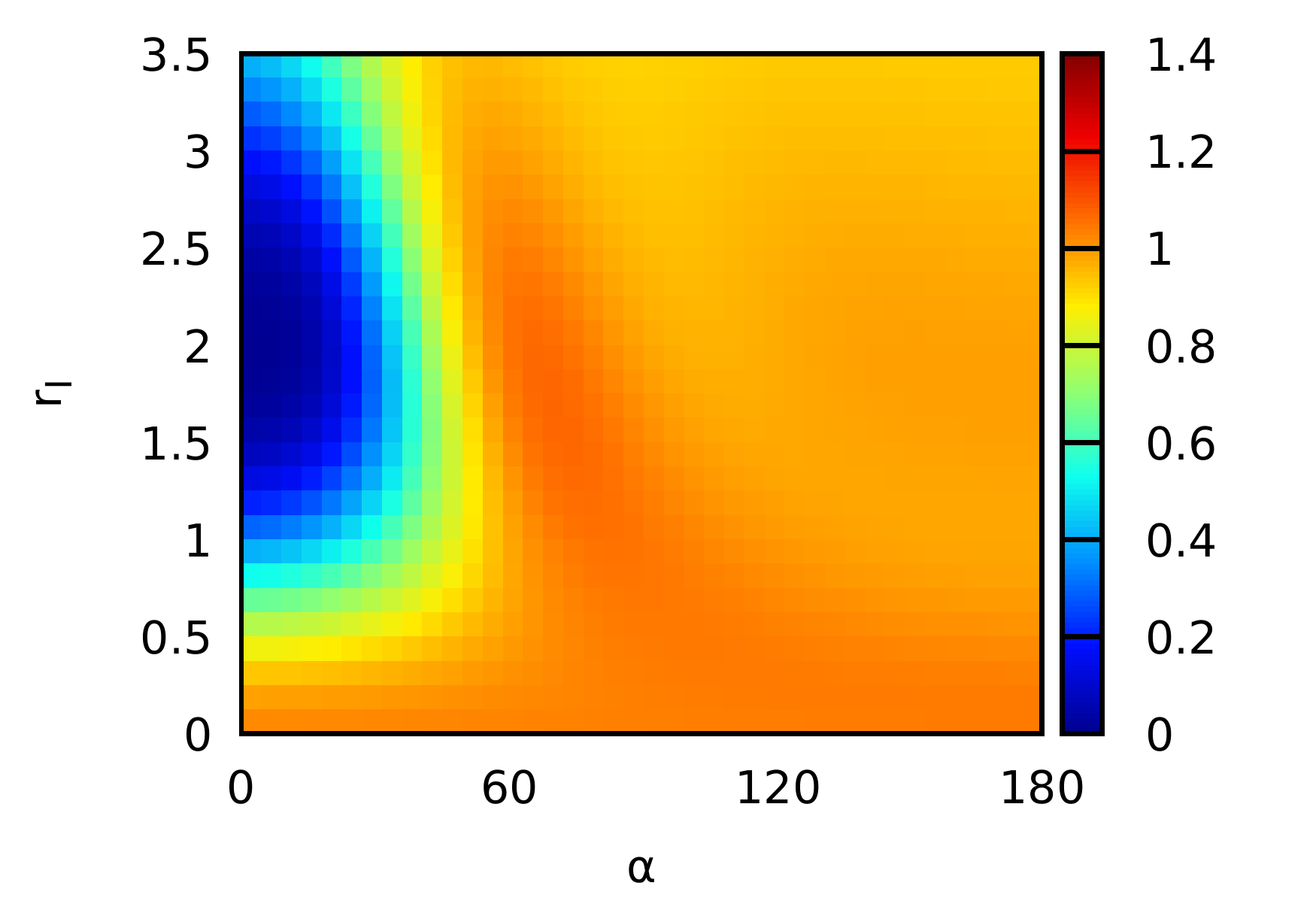}
\includegraphics[width=0.40447\textwidth]{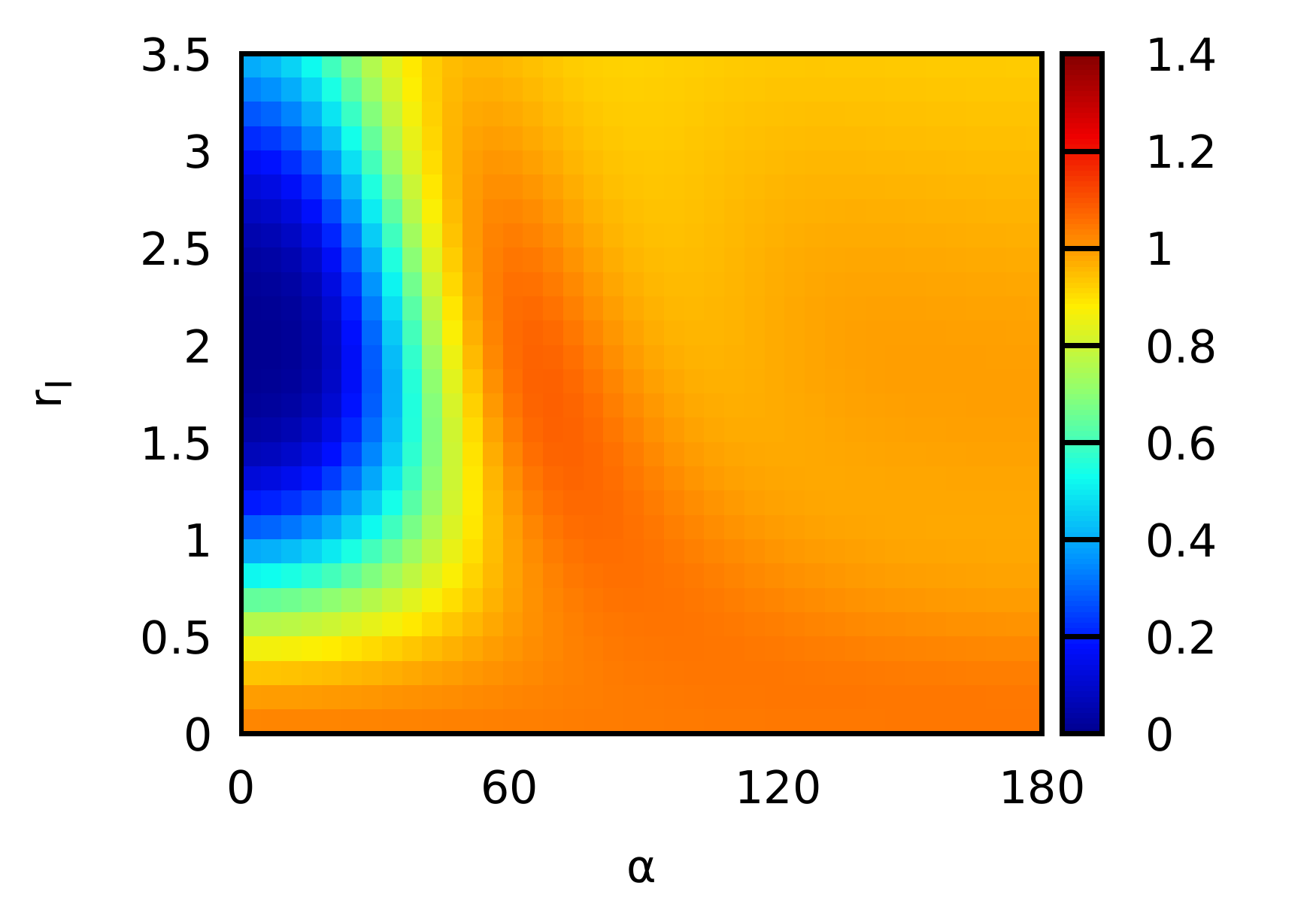}
\caption{\label{fig:c2p_lambda3_N6}
Integrated center-two particle correlation function $g^\textnormal{int}_\textnormal{c2p}(r_1,\alpha,1.5,2.5)$ [cf.~Eq.~(\ref{eq:c2p_int})] for $N=6$ and $\lambda=3$. The left and right columns correspond to Bose- and Fermi-statistics, and the top and bottom rows to $\beta=7$ and $\beta=1$. The dashed dark grey lines in the top left panel indicate the integration boundaries for the reference particle, and the solid red line the scan line depicted in Fig.~\ref{fig:Scanline_N6_lambda3}.
}
\end{figure*}

The second type of classical-to-quantum transition in harmonically confined ultracold atoms takes place upon increasing the inverse temperature $\beta$. This is investigated in Fig.~\ref{fig:densty_beta_N6_lambda3} for $N=6$ particles, with the left panel corresponding to $\lambda=1$. At the highest considered temperature, $\beta=0.75$, the system is nearly classical and the bosonic and fermionic curves almost coincide. Doubling the inverse temperature to $\beta=1.5$ makes the effect of quantum statistics more pronounced: the bosonic curve progresses very smoothly, whereas there appears a saddle point for fermions at $r\approx1$.
At $\beta=3$, the system already almost resembles the ground state, and the results are similar to the curves discussed in Fig.~\ref{fig:densty_beta5_N6_lambda}. Finally, for $\beta=5$ and $\beta=6$ (the two top curves), the bosonic curves remain almost structure-less, whereas the fermionic shell structure has become even more pronounced, and we find an average sign of $S\approx6.8\cdot10^{-4}$ in the latter case. Thus, going to even lower temperature is presently computationally too expensive.

The right panel of Fig.~\ref{fig:densty_beta_N6_lambda3} shows similar information, but at thrice the coupling strength, $\lambda=3$. As the general trend is similar as for $\lambda=1$, we restrict ourselves to a brief summary of the key differences: i) the stronger repulsive forces push the particles further away from the center of the trap for both types of quantum statistics; ii) quantum-statistical effects start to manifest at lower temperatures; iii) although the shell structure is significantly more pronounced for fermions, it does appear for bosons as well due to the moderate coupling strength.

\begin{figure}
\includegraphics[width=0.4447\textwidth]{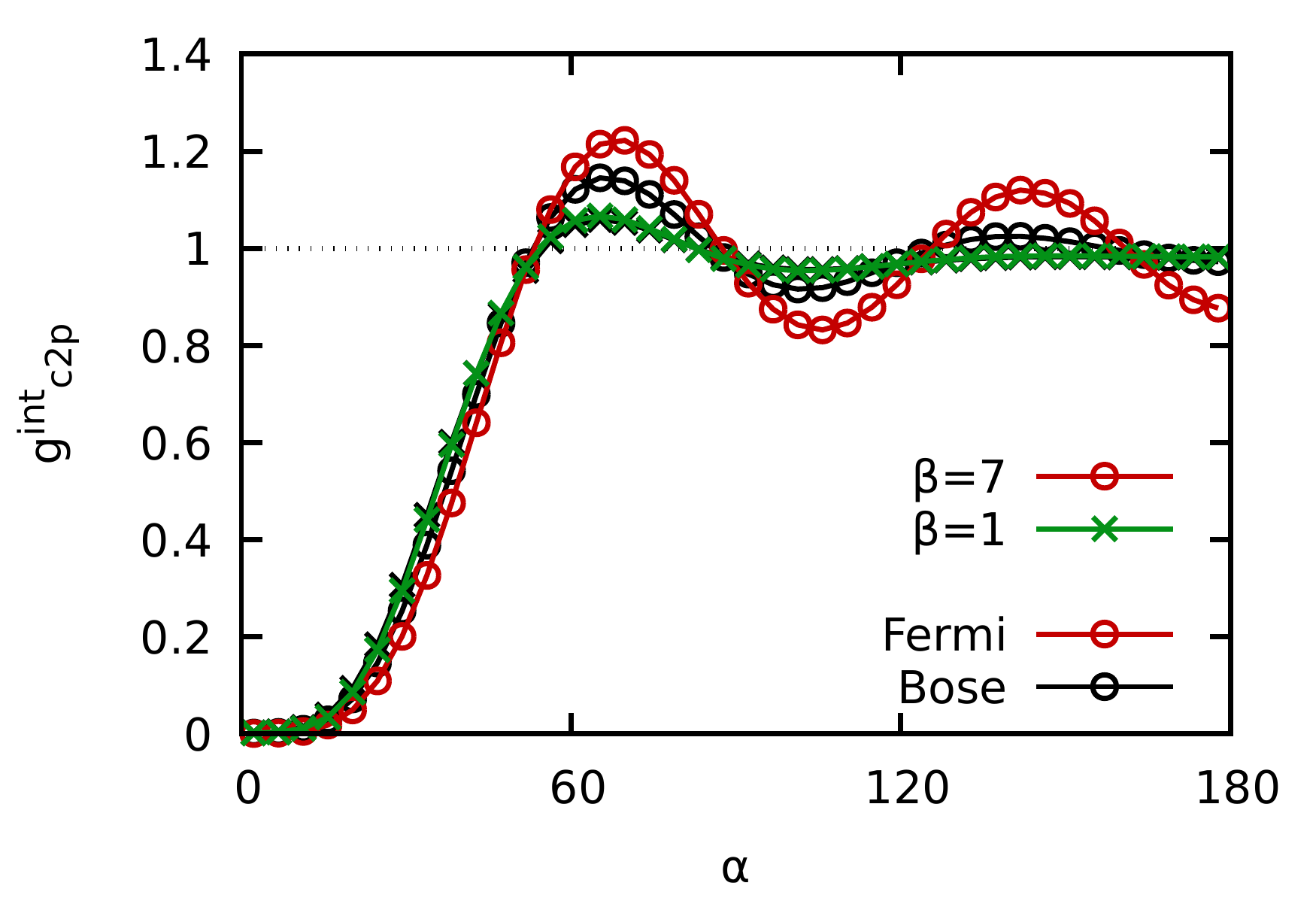}
\caption{\label{fig:Scanline_N6_lambda3}
Scanline of the integrated center-two particle correlation function shown in Fig.~\ref{fig:c2p_lambda3_N6}, evaluated at $r_1=2$.
}
\end{figure}

Let us conclude the investigation of the static properties of quantum dipole systems with results for the integrated C2P for the most interesting transition regime. To this end, we show $g^\textnormal{int}_\textnormal{c2p}(r_1,\alpha,1.5,2.5)$ in Fig.~\ref{fig:c2p_lambda3_N6}, which measures the correlation between a particle in the outer shell ($1.5\leq r_2 \leq 2.5$, see the dashed dark grey lines in the top left panel, and also the density profiles in the right panel of Fig.~\ref{fig:densty_beta_N6_lambda3}) and the rest of the system for $N=6$ and $\lambda=3$. As usual, the left and right columns correspond to bosons and fermions, and the top row has been computed for low temperature, $\beta=7$, which is close to the ground state. First and foremost, we note that the C2P exhibits a qualitatively similar behavior for both types of quantum statistics, namely a pronounced exchange--correlation hole around $\alpha=0$ followed by a maximum around $\alpha\approx75$. Yet, this structure is significantly more pronounced in the case of fermions, which can be seen particularly well in Fig.~\ref{fig:Scanline_N6_lambda3} where we show scanlines over $r_1=2$, see the solid red line in the top left panel.

\begin{figure*}
\includegraphics[width=0.40447\textwidth]{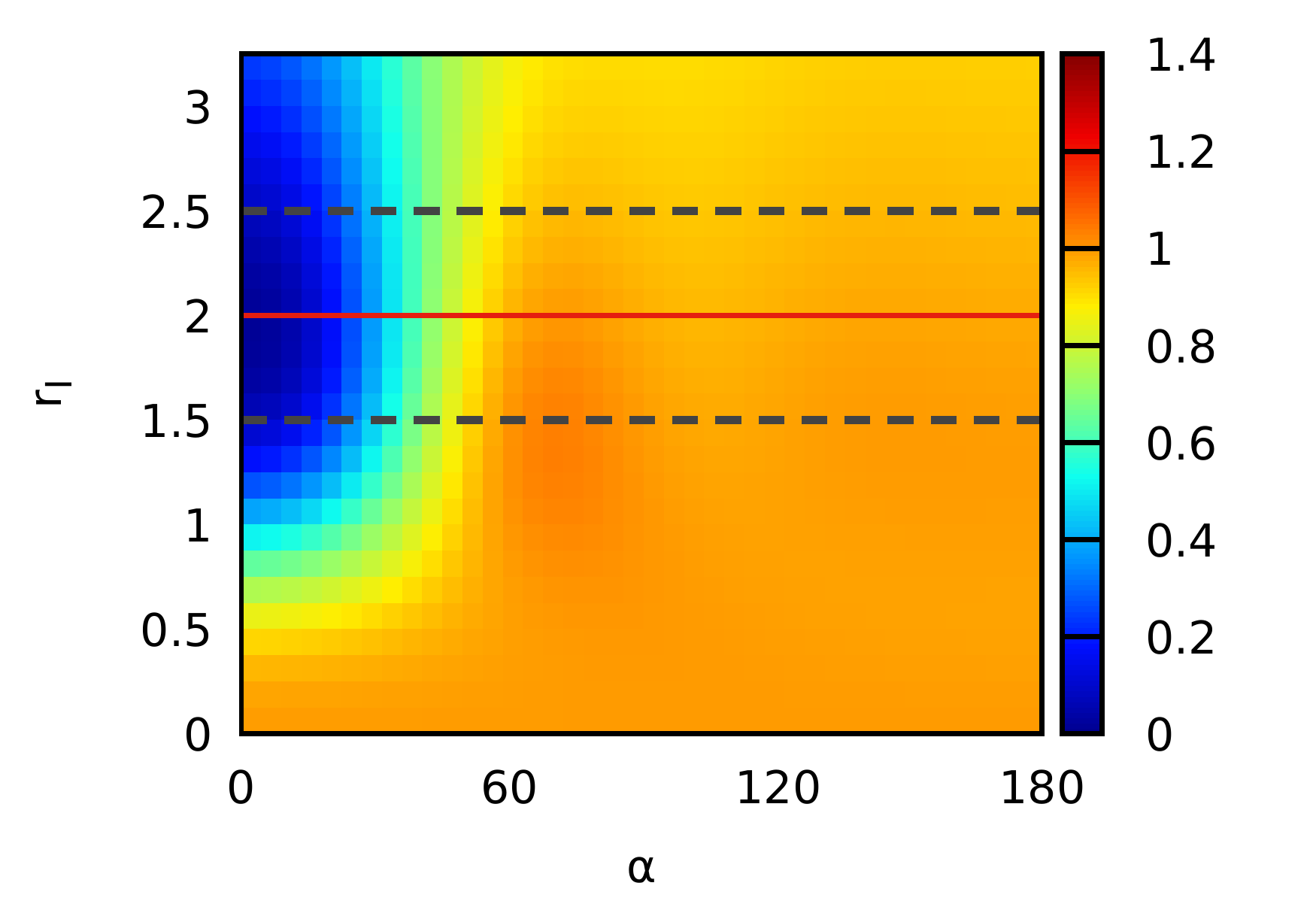}
\includegraphics[width=0.40447\textwidth]{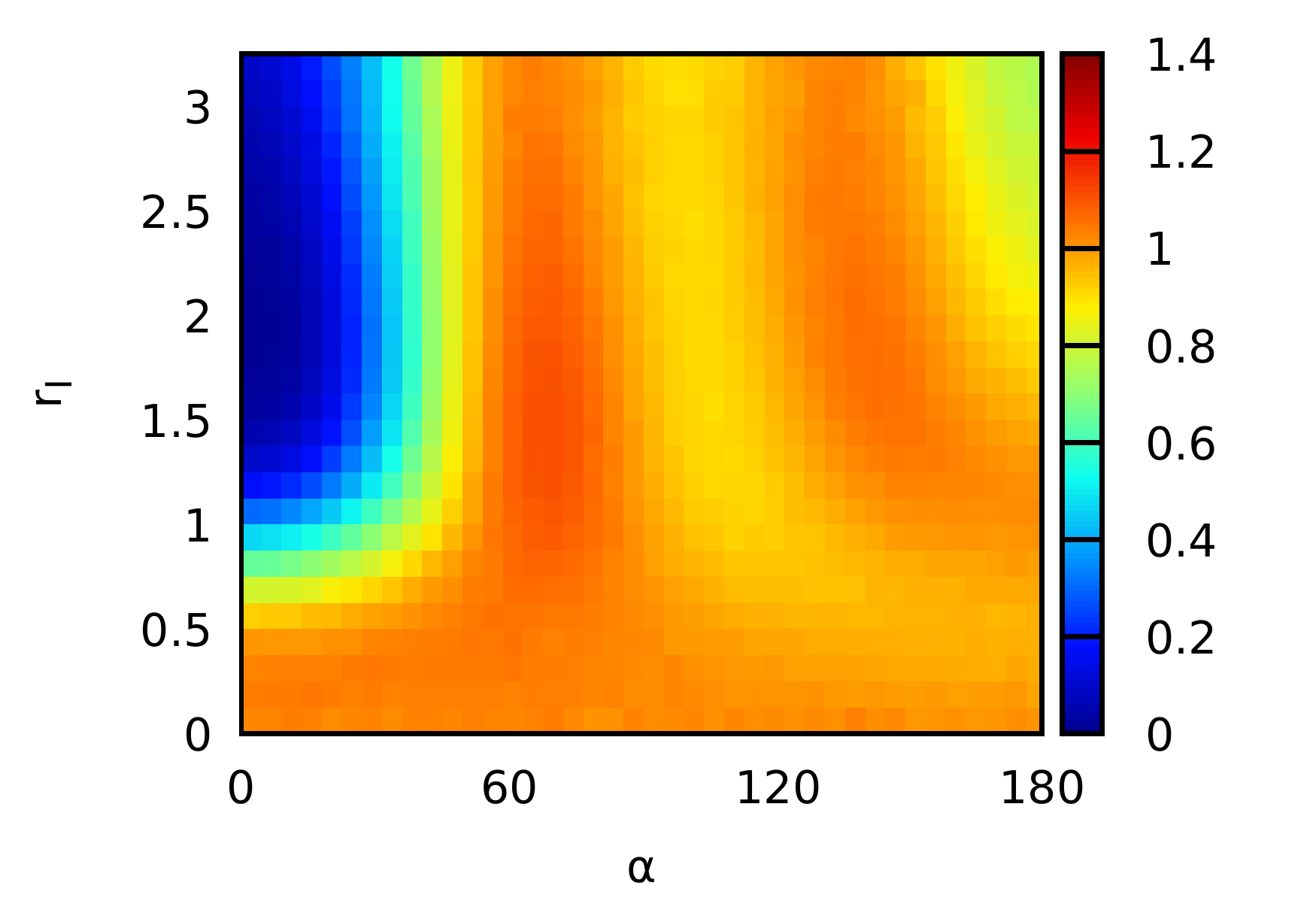}\\
\includegraphics[width=0.40447\textwidth]{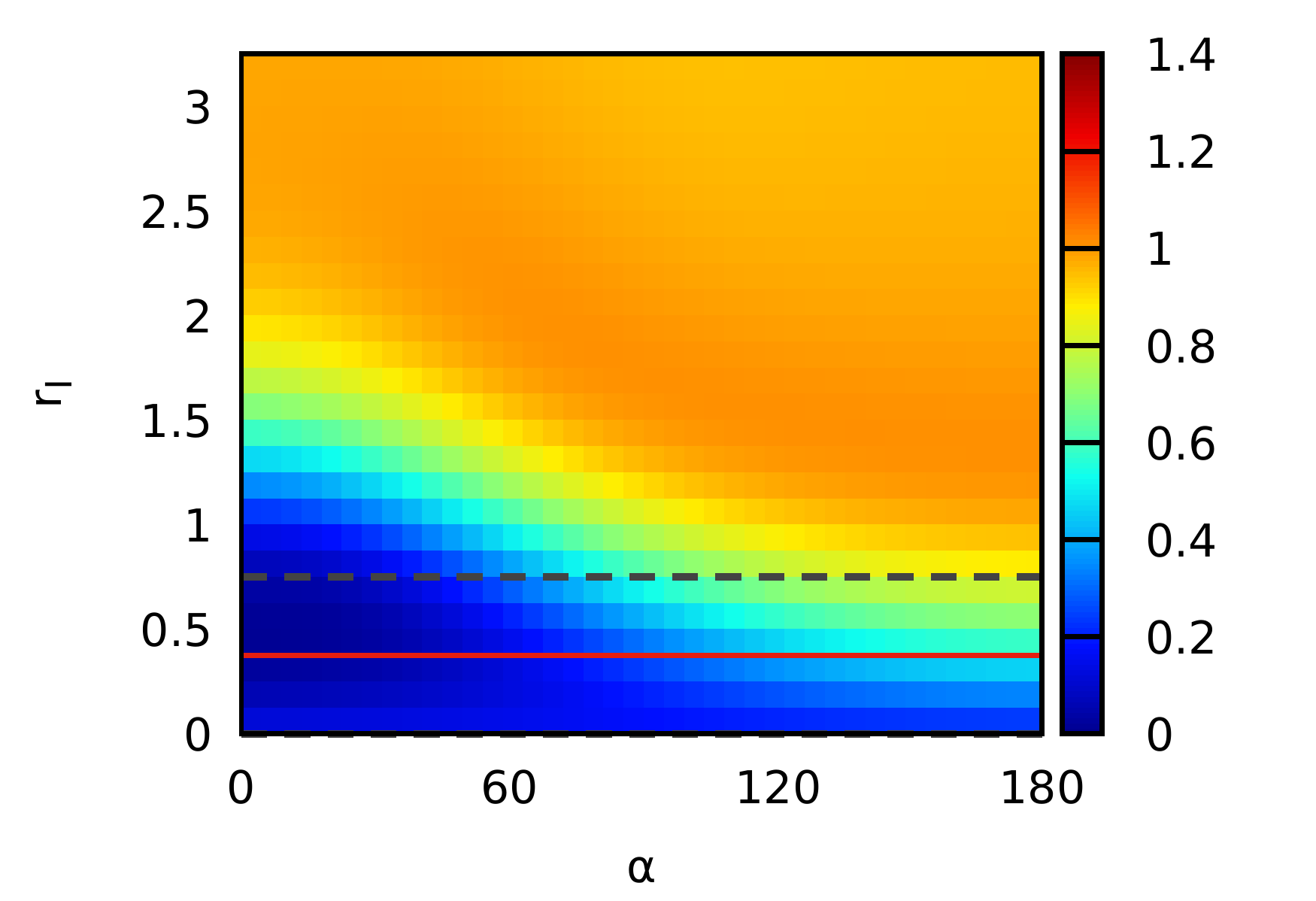}
\includegraphics[width=0.40447\textwidth]{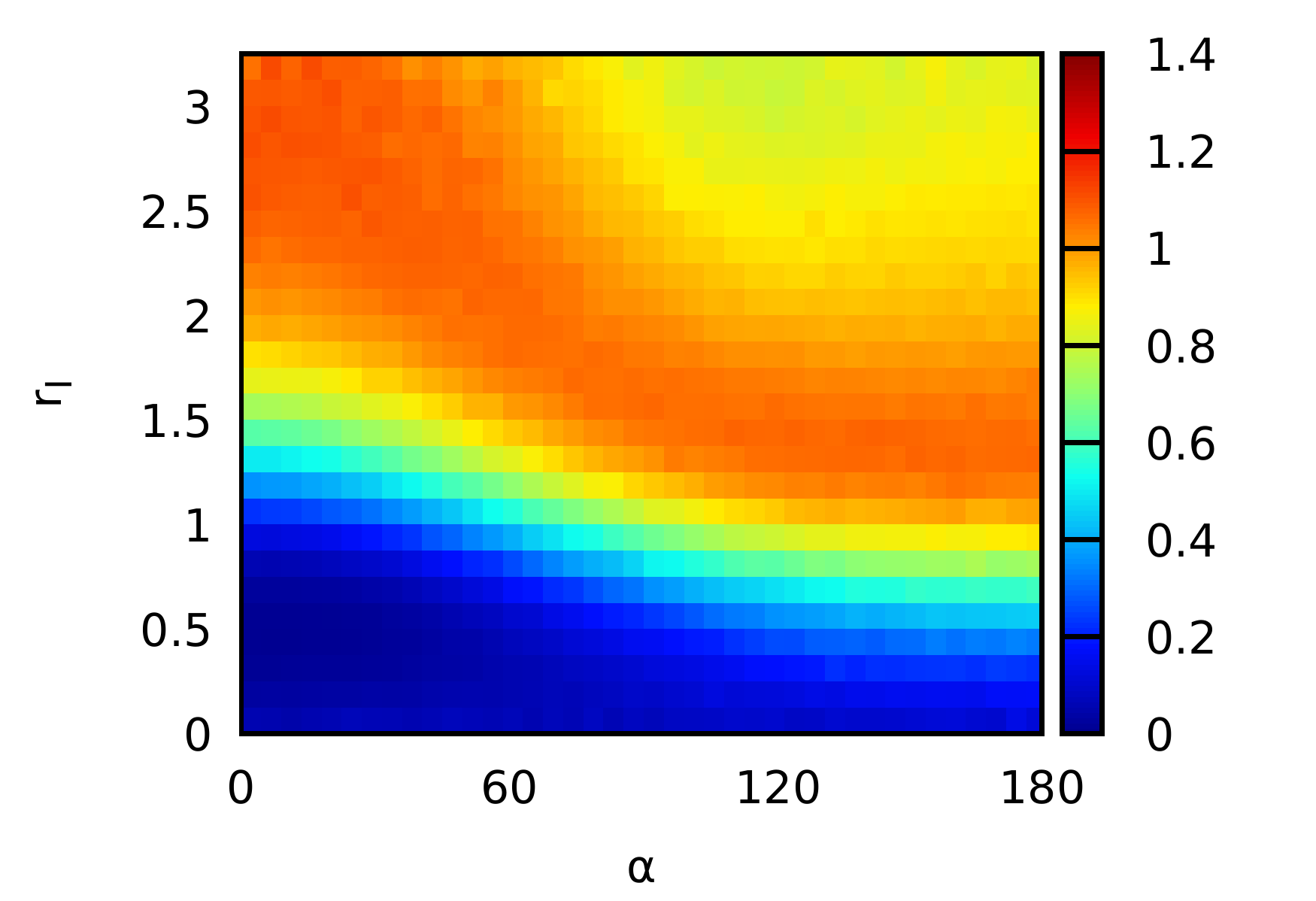}
\caption{\label{fig:c2p_lambda1_N6}
Integrated center-two particle correlation function $g^\textnormal{int}_\textnormal{c2p}(r_1,\alpha,r_{2,\textnormal{min}},r_{2,\textnormal{max}})$ [cf.~Eq.~(\ref{eq:c2p_int})] for $N=6$, at $\beta=5$ and $\lambda=1$. The left and right columns correspond to Bose- and Fermi-statistics. Top row: outer shell, $1.5\leq r_2 \leq 2.5$ (see the dark grey dashed lines); bottom row: inner shell, $0\leq r_2 \leq0.75$. The solid red lines correspond to scan lines shown in Fig.~\ref{fig:Scanline_N6_lambda1_beta5}.
}
\end{figure*}

Let us for now ignore the crosses and focus on the red (black) circles corresponding to the fermionic (bosonic) data at $\beta=7$. 
In particular, the red curve exhibits a pronounced minimum around $\alpha\approx100$, followed by a second maximum at $\alpha\approx140$, and even a second minimum at $\alpha=180$ (i.e., at the opposite end of the trap), which are almost absent in the black data set. We thus conclude that, at moderate coupling and low temperature, Fermi statistics effectively enhance the impact of the dipole repulsion on the structural properties of the system. This is in contrast to the weak-coupling regime (see Fig.~\ref{fig:c2p_N4_lambda} and the corresponding discussion), where we observed the opposite effect, as the dipole interaction was essentially masked by the exchange hole, and the fermionic system closely resembled the ideal case for comparatively larger values of $\lambda$.

In the bottom panel of Fig.~\ref{fig:c2p_lambda3_N6}, we show the same data for the integrated C2P, but at a relatively high temperature, $\beta=1$. In this case, the effect of quantum statistics does not only vanish in the radial density (cf.~the right panel of Fig.~\ref{fig:densty_beta_N6_lambda3}), but also cannot be resolved in $g^\textnormal{int}_\textnormal{c2p}(r_1,\alpha,1.5,2.5)$. Again, this becomes especially clear in the scan line over $r_1=2$ shown as the crosses in Fig.~\ref{fig:Scanline_N6_lambda3}.

\begin{figure}
\includegraphics[width=0.4447\textwidth]{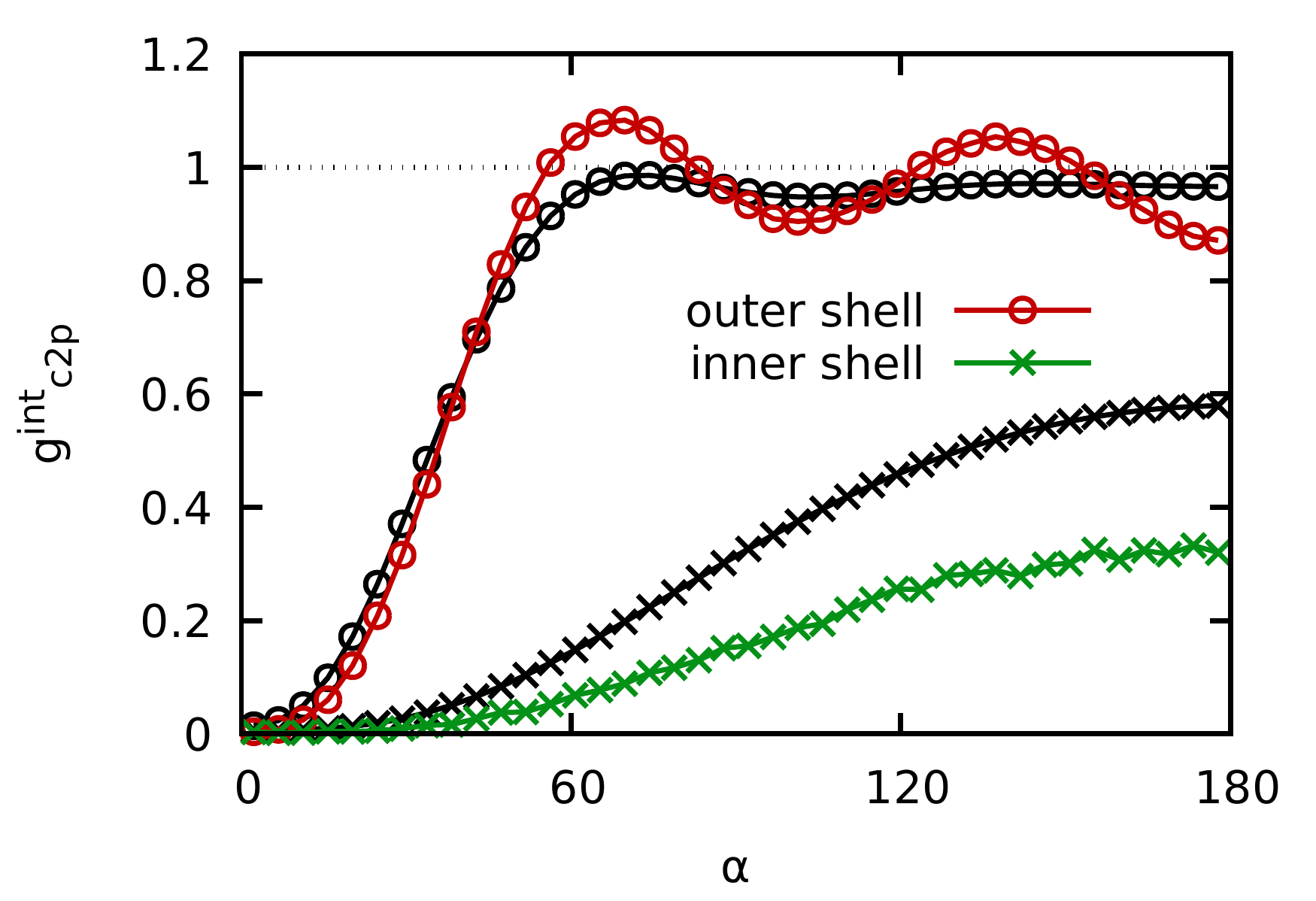}
\caption{\label{fig:Scanline_N6_lambda1_beta5}
Scanline of the integrated center-two particle correlation function shown in Fig.~\ref{fig:c2p_lambda1_N6}, evaluated at $r_1=1$ (circles) and $r_1=3/8$ (crosses). The colored and black symbols refer to Fermi- and Bose-statistics, respectively.
}
\end{figure}

As a final example for the utility of the integrated C2P regarding the investigation of the structural properties of trapped quantum systems, we show $g^\textnormal{int}_\textnormal{c2p}(r_1,\alpha,r_{2,\textnormal{min}},r_{2,\textnormal{max}})$ for $N=6$ and $\lambda=1$ in Fig.~\ref{fig:c2p_lambda1_N6}. In particular, the top row has been obtained by integrating over $1.5 \leq r_2 \leq 2.5$ (see the dashed dark grey lines in the top left panel), which corresponds to particles in the outer shell for fermions, and the outer region for bosons, cf.~Fig.~\ref{fig:densty_beta_N6_lambda3}. As usual, the left panel shows the results for Bose statistics, and we find a pronounced correlation hole around $\alpha=0$, but no pronounced features beyond. In stark contrast, the fermionic data exhibit in addition to the exchange-correlation hole a distinct structure for all angles $\alpha$, i.e., beyond next-neighbour effects and throughout the entire system. 

Again, this can be seen best from a scan line over $r_2=2$ (solid red line in the top left panel), which is shown in Fig.~\ref{fig:Scanline_N6_lambda1_beta5}. The bosonic curve (black circles) remains nearly flat for $\alpha\gtrsim 60$, whereas there are pronounced oscillations in the fermionic curve (red circles) even for $\alpha=180$.

\begin{figure}
\includegraphics[width=0.4447\textwidth]{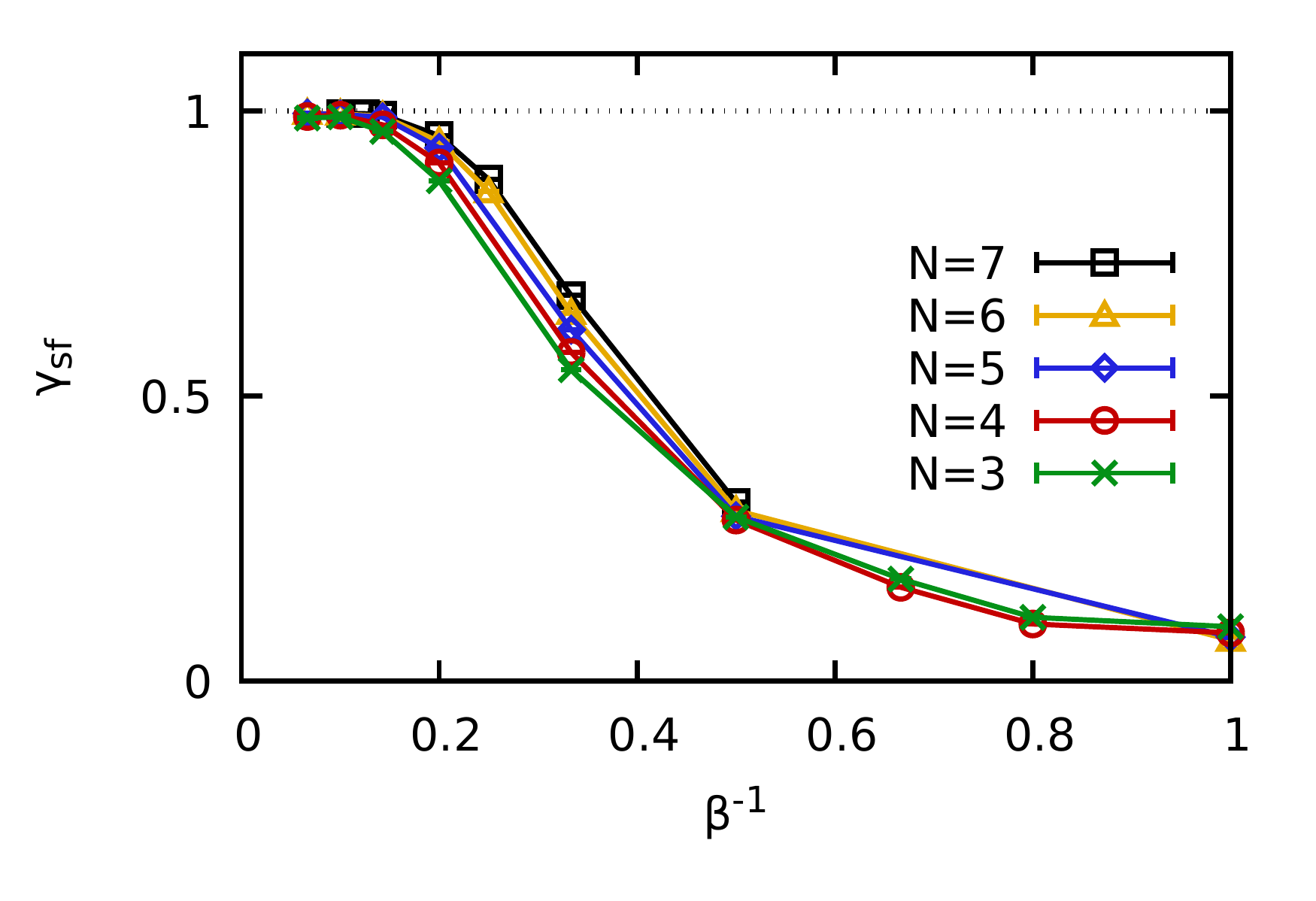}\\
\includegraphics[width=0.4447\textwidth]{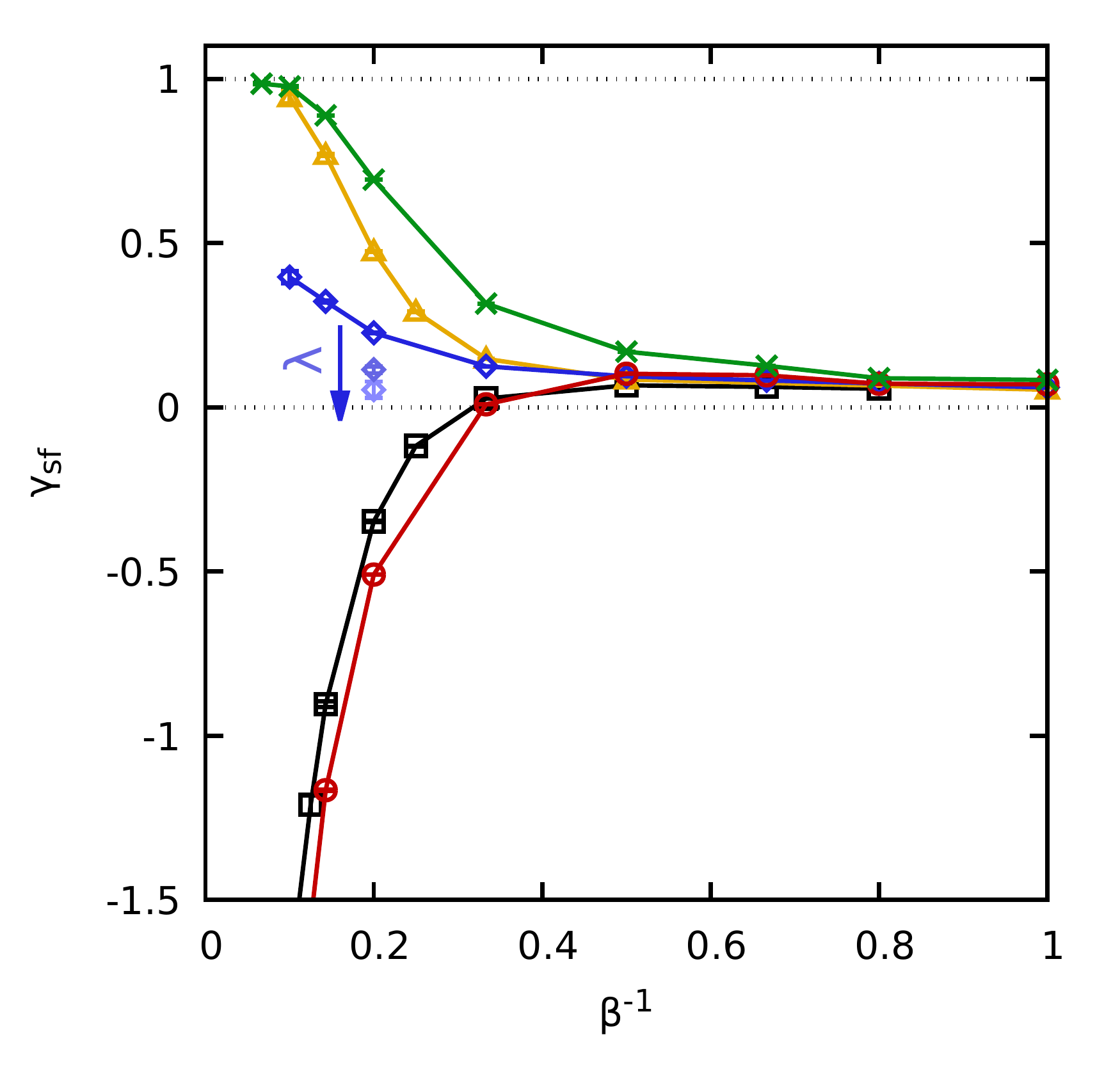}
\caption{\label{fig:SF_lambda3}
PIMC results for the temperature dependence of the superfluid fraction for Bose- (top) and Fermi-statistics (bottom) at $\lambda=3$. The black squares, yellow triangles, blue diamonds, red circles, and green crosses distinguish PIMC data for $N=7,6,5,4$ and $3$. The blue arrow in the bottom panel indicates the decreasing coupling strength, and the two light blue points at $\beta=5$ correspond to $\lambda=1$ and $\lambda=0.7$ (in descending order) for $N=5$.
}
\end{figure}

Finally, the bottom row of Fig.~\ref{fig:c2p_lambda1_N6} shows the integrated C2P for $0\leq r_2 \leq 0.75$, which measures the correlation towards a particle in the inner region. Overall, we observe similar trends as in the outer region, with the fermionic data exhibiting correlations throughout the entire system. Moreover, the scanline depicted in Fig.~\ref{fig:Scanline_N6_lambda1_beta5} shows that it is much less likely to find a second particle in the inner shell for fermions as compared to bosons in the first place.

In a nutshell, we have used the recent integrated C2P to analyze the structural properties of harmonically confined quantum dipole systems. Our three key findings are i) the comparably much later impact (i.e., for larger $\lambda$) of the dipole interaction for fermions as compared to bosons, ii) the emergence of a shell structure in the density profile and system-wide correlations in the C2P for fermions at intermediate coupling strength, and iii) the value of the integrated C2P as a tool for the investigation of correlated quantum systems, which is interesting in its own right.

\subsection{Superfluidity and non-classical rotational inertia\label{sec:superfluid_results}}

Let us conclude our investigation of ultracold atoms in a harmonic trap with a systematic study of the impact of the dipole interaction on the moment of inertia. To this end, we plot PIMC results for the superfluid fraction versus the temperature $T=\beta^{-1}$ in Fig.~\ref{fig:SF_lambda3} for $\lambda=3$ (i.e., intermediate coupling strength) and five different particle numbers $N$ (different symbols and colors). The top panel has been obtained for Bose-statistics and all five curves exhibit the expected crossover from a classical system with $\gamma_\textnormal{sf}=0$ to a quantum system where $I$ vanishes ($\gamma_\textnormal{sf}=1$). Moreover, we find that the curves are ordered with increasing the system size $N$ just as in the case of ideal bosons, cf.~Fig.~\ref{fig:ideal_sf}.

\begin{figure}
\includegraphics[width=0.4447\textwidth]{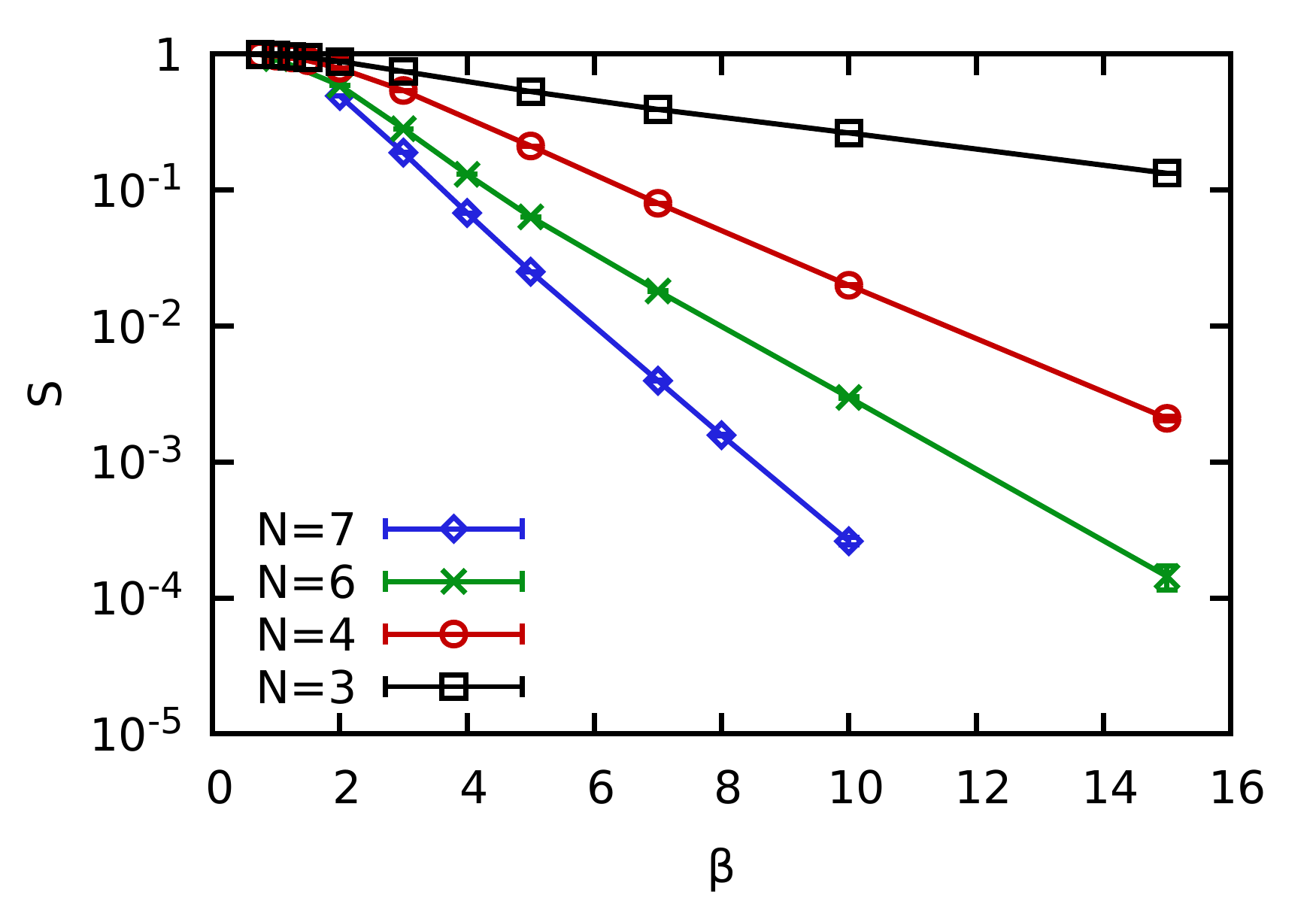}
\caption{\label{fig:Sign_lambda3}
Temperature dependence of the average sign at $\lambda=3$. Shown are PIMC results for the sign $S$ plotted versus the inverse temperature $\beta$ for $N=7$ (blue diamonds), $N=6$ (green crosses), $N=4$ (red circles), and $N=3$ (black squares).
}
\end{figure}

\begin{figure*}
\includegraphics[width=0.4447\textwidth]{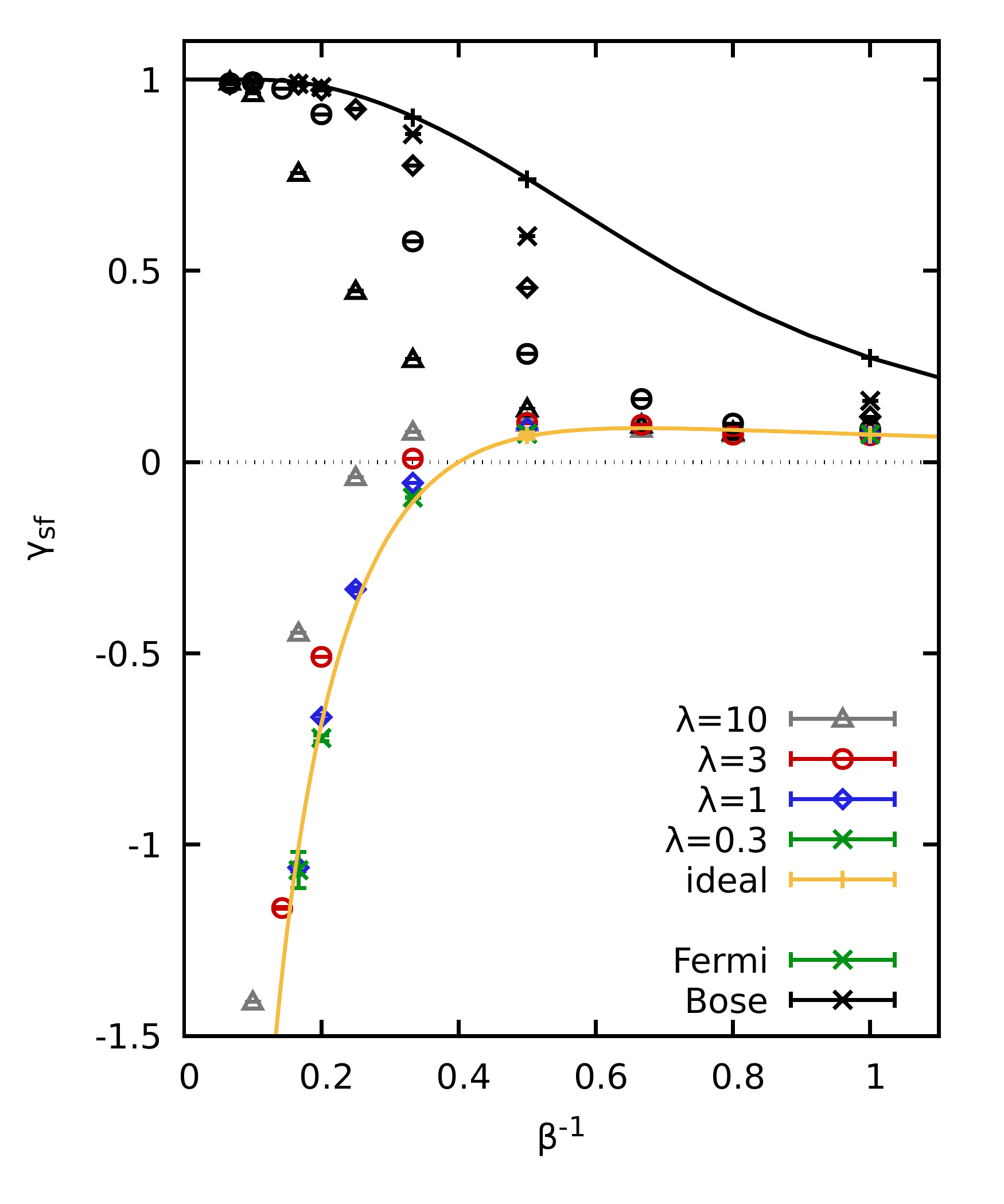}
\includegraphics[width=0.4447\textwidth]{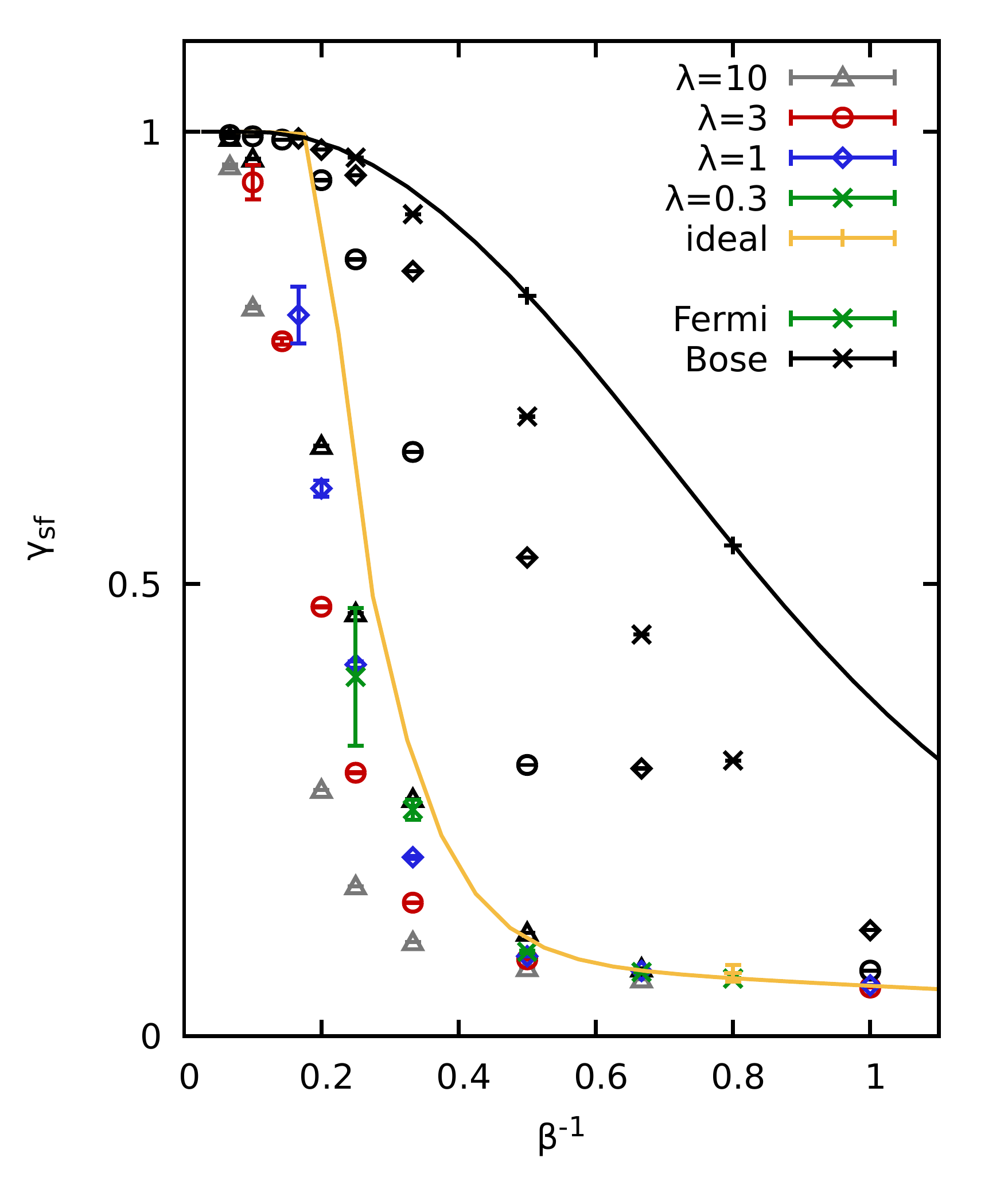}
\caption{\label{fig:SF_N}
Temperature dependence of the superfluid fraction $\gamma_\textnormal{sf}$ for $N=4$ (left) and $N=6$ (right) ultracold atoms. The colored symbols depict PIMC data for different values of the coupling parameter $\lambda$, and the yellow and black solid lines to the exact ideal result for Fermi- and Bose-statistics, respectively.
}
\end{figure*}

The bottom panel shows the same information for fermions. For $N=3$ (green crosses) and $N=6$ (yellow triangles) we observe a qualitatively similar crossover as in the case of Bose statistics, whereas for $N=4$ (red circles) and $N=7$ (black squares), $\gamma_\textnormal{sf}$ diverges towards negative infinity. This is precisely the behaviour exhibited by ideal fermions (see Sec.~\ref{sec:ideal_results}), which strongly indicates that the symmetry of the ground state wave function is not changed by the interaction~\cite{blume}. 
A seeming exception to this pattern is given by the blue diamonds corresponding to $N=5$: in the noninteracting case, the superfluid fraction diverges towards negative infinity, although it does so for lower temperature as compared to $N=6$; the $\lambda=3$ curve shown in Fig.~\ref{fig:SF_lambda3}, on the other hand, exhibits a monotonically increasing $\gamma_\textnormal{sf}$ for the depicted temperature range. Still, the system is not fully superfluid even for the lowest temperature point, and the low temperature limit cannot presently be resolved due to the fermion sign problem.

To further illuminate this issue, we have also performed PIMC simulations of the $N=5$ system at $\lambda=1$ and $\lambda=0.7$ for $\beta=5$, and the results are shown as the light blue points. In contrast to bosons, where $\gamma_\textnormal{sf}$ monotonically increases when the system becomes more ideal, the superfluid fraction drops in the present example and, eventually, becomes negative.

Let us now briefly revisit the sign problem. 
In Fig.~\ref{fig:Sign_lambda3}, we show our PIMC results for the average sign $S$ for the same conditions as in Fig.~\ref{fig:SF_lambda3}. First and foremost, we note that all curves exhibit the same expected qualitative behavior: at large temperature, the system is nearly ideal, quantum degeneracy and exchange effects are negligible and the average sign approaches one. In the path integral picture, this means that the probability to find a particle in a single-particle cycle ($(P(1)$, cf.~Fig.~\ref{fig:ideal_permutations}) becomes $100\%$. With increasing $\beta$, the single-particle wave functions become more extended and paths begin to overlap. Consequently, permutations of length $l>1$ start to appear with increasing frequency, and the average sign drops due to the cancellation of positive and negative weights. Since the Monte Carlo error bar is (in first approximation) inversely proportional to $S$ [cf.~Eq.~(\ref{eq:fsp_error})], the simulations become computationally more involved and eventually unfeasible for $S < 10^{-3}$. However, a more extensive discussion of the sign problem is beyond the scope of the present work, and the interested reader is referred to Ref.~\cite{dornheim_sign_problem} for a topical review.

In order to more systematically study the impact of the dipole interaction on the non-classical rotational inertia, we show the superfluid fraction of $N=4$ (left panel) and $N=6$ (right panel) interacting quantum dipole particles in Fig.~\ref{fig:SF_N} for different values of the coupling parameter $\lambda$. Let us start with a discussion of the bosonic results (black symbols and lines), which exhibit the same behavior for both particle numbers. The solid lines correspond to the exact ideal result known from theory (see Sec.~\ref{sec:ideal_theory}), and the pluses to PIMC data at the same conditions. Here, too, we find perfect agreement between theory and simulations, as it is expected. While we do observe the by now familiar crossover from the classical to the superfluid regime for all values of $\lambda$, $\gamma_\textnormal{sf}$ significantly decreases with increasing coupling strength for intermediate temperatures. Equivalently, it can be said that the crossover is shifted to substantially lower temperatures, which can be intuitively understood in the following way: at weak coupling, the paths corresponding to different particles in our PIMC simulations can overlap and form permutation cycles even when the temperature is relatively high and, consequently, $\lambda_\beta$ is small. With increasing $\lambda$, the particles are further pushed away from each other and the average inter-particle distance $\overline{r}$ becomes larger. Thus, it requires lower temperatures for $\overline{r}$ and $\lambda_\beta$ to be of comparable size, and the superfluid crossover happens at larger values of $\beta$.

The colored symbols show the same information as the black ones, but for fermions. For completeness, we mention that here, too, we find excellent agreement between our PIMC simulations and the theoretical curve for $\lambda=0$, although simulations are restricted to much smaller values of $\beta$ (in particular for $N=6$) due to the fermion sign problem. 
Furthermore, the $N=4$ curve approaches negative infinity for all considered coupling strengths, whereas the $N=6$ curve approaches one, cf.~the discussion of Fig.~\ref{fig:SF_lambda3} above. While the respective curves are shifted towards lower temperature with increasing $\lambda$, the effect of the coupling strength on $\gamma_\textnormal{sf}$ is much less pronounced than in the case of bosons. This is in good agreement to the trends reported in Sec.~\ref{sec:structural_results} that fermions react less strongly to the dipole interaction, which is effectively masked by the Pauli repulsion.

\begin{figure}
\includegraphics[width=0.4147\textwidth]{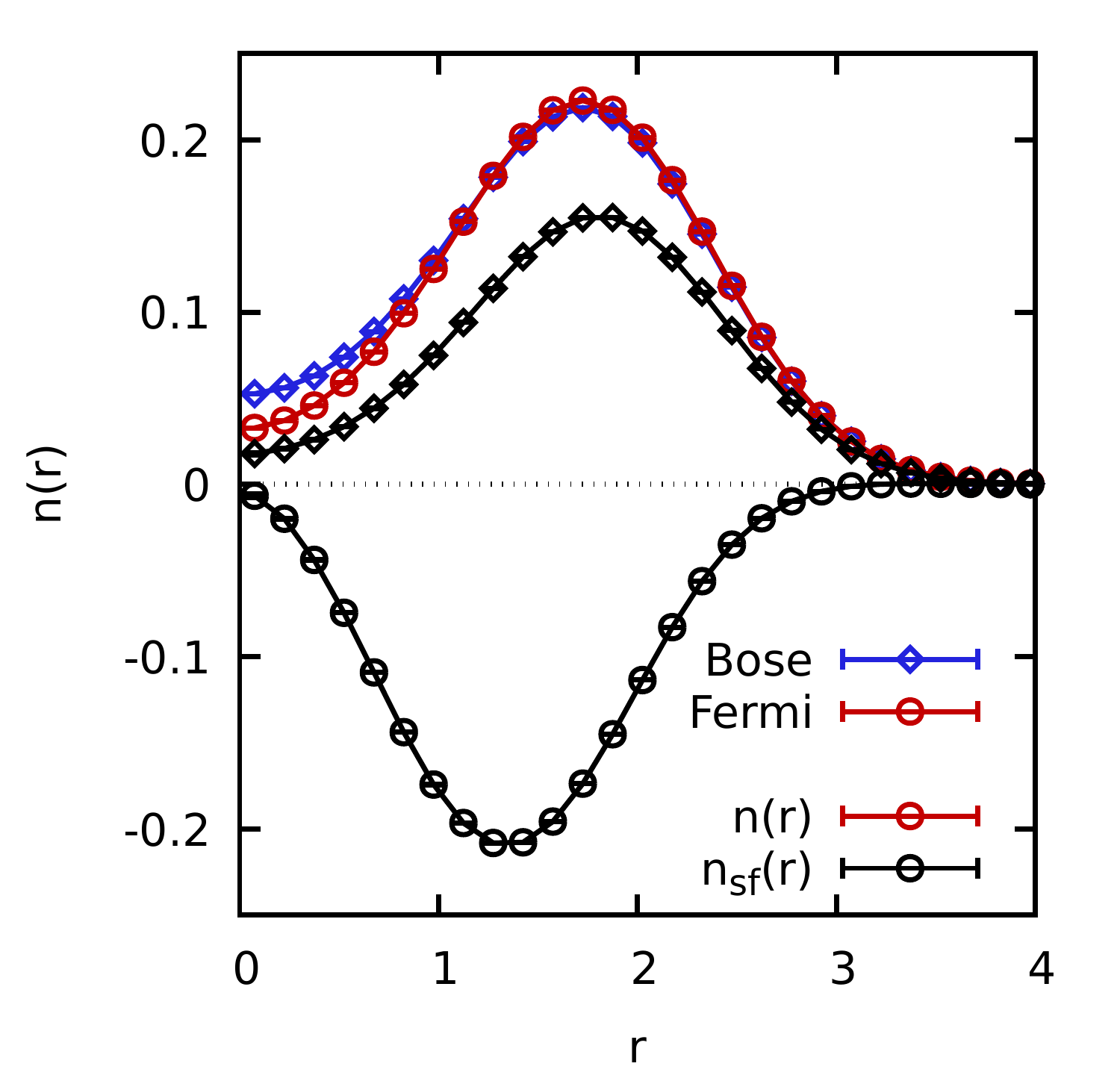}
\caption{\label{fig:LSF_SC}
PIMC results for the radial density $n(r)$ (colored) and superfluid density [black, see Eq.~(\ref{eq:nsf})] of $N=4$ ultracold atoms with $\lambda=10$ and $\beta=6$. The diamonds and circles correspond to bosons and fermions, and the respective total superfluid fractions are given by $\gamma_\textnormal{sf}\approx0.76$ and $\gamma_\textnormal{sf}\approx-0.45$. 
}
\end{figure}

A further interesting question is whether the drastic difference in the quantum mechanical moment of inertia in the case of $N=4$ are somehow reflected by the structural properties of the system. To address this issue, we show both the radial density distribution $n(r)$ (colored symbols) and the superfluid density [black symbols, see Eq.~(\ref{eq:nsf})] in Fig.~\ref{fig:LSF_SC} for $\lambda=10$ and $\beta=6$. First and foremost, we note that there appears hardly any difference in $n(r)$ between bosons (diamonds) and fermions (circles) apart from a somewhat more pronounced minimum around the center of the trap in the latter case. For completeness, we mention that we find an average sign of $S\approx0.57$.

In contrast, the superfluid density behaves entirely differently in both cases: for bosons, $n_\textnormal{sf}$ approximately follows the full density $n(r)$, and the maximum occurs at roughly the same position; for fermions, on the other hand, $n_\textnormal{sf}$ is negative over the entire $r$-range and the minimum is shifted significantly towards smaller $r$ as compared to the maximum of $n(r)$. This is in qualitative agreement to the results reported in Ref.~\cite{blume} for a noninteracting system.

\begin{figure*}
\includegraphics[width=0.4447\textwidth]{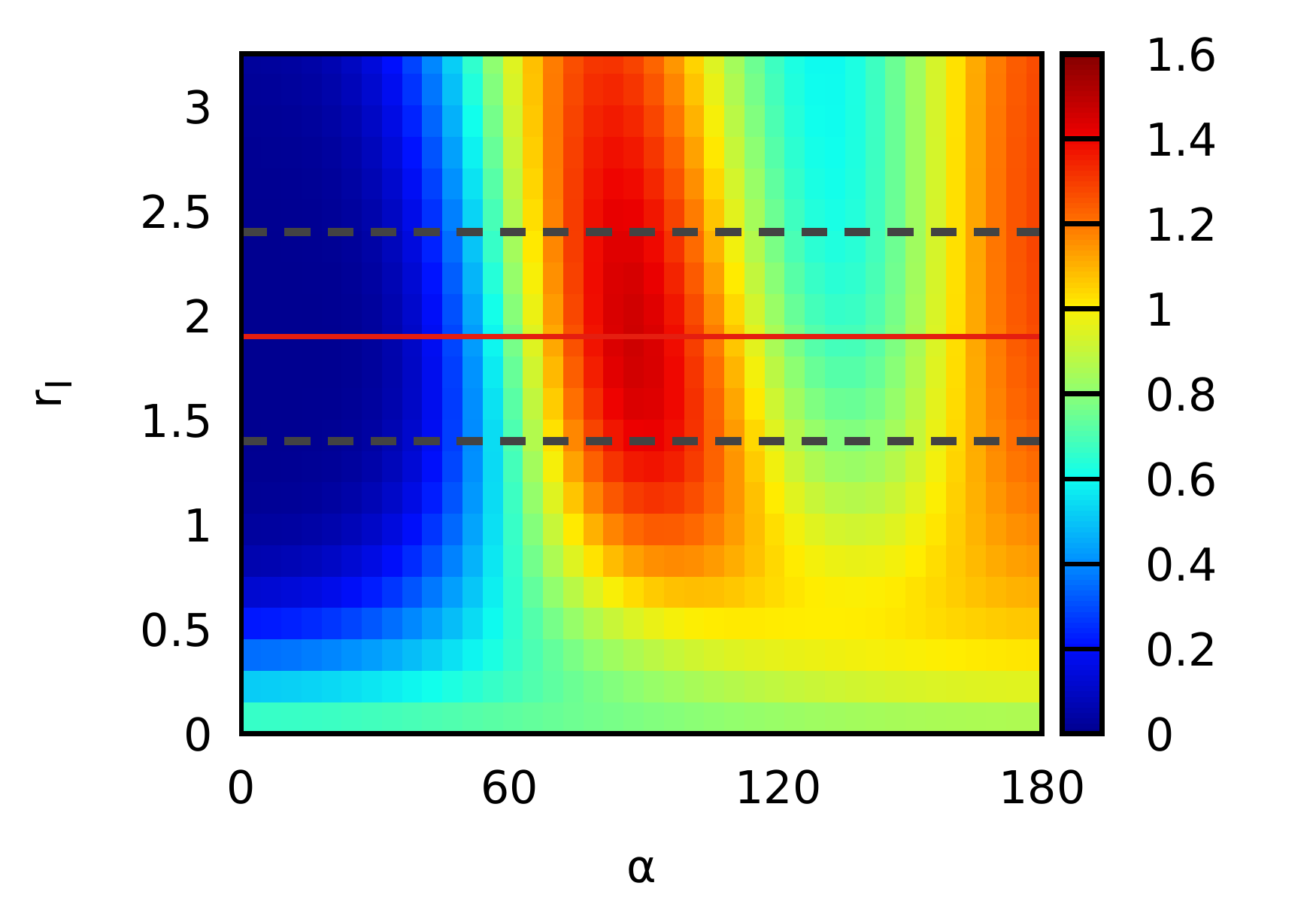}
\includegraphics[width=0.4447\textwidth]{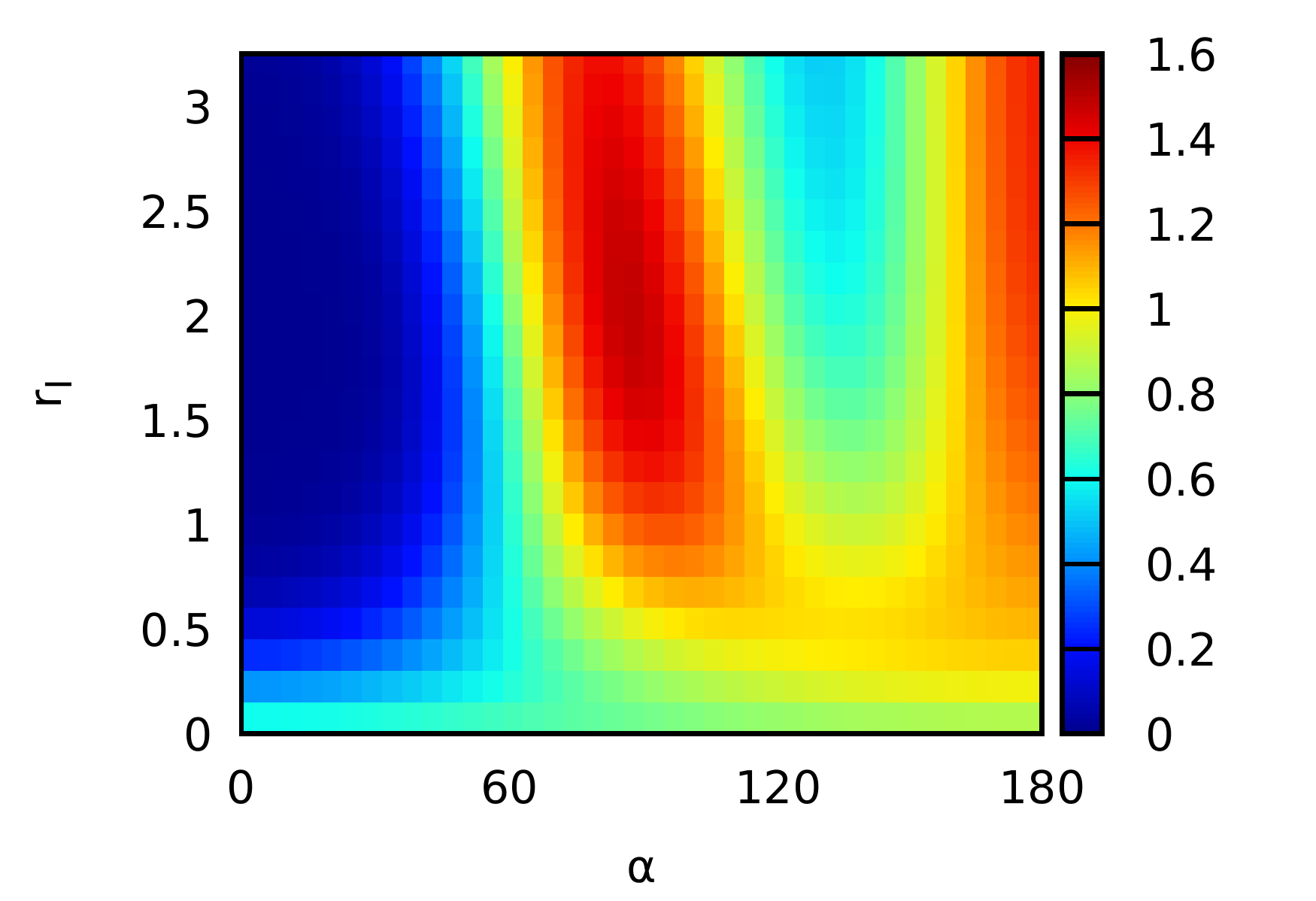}
\caption{\label{fig:C2P_SC}
Integrated center-two particle correlation function $g^\textnormal{int}_\textnormal{c2p}(r_1,\alpha,1.4,2.4)$ [cf.~Eq.~(\ref{eq:c2p_int})] for $N=4$, $\lambda=10$, and $\beta=6$. The dashed dark grey lines in the left panel indicate the integration boundaries for the reference particle, and the solid red line the scan line depicted in Fig.~\ref{fig:Scan_SC}.
}
\end{figure*}

\begin{figure}
\includegraphics[width=0.4447\textwidth]{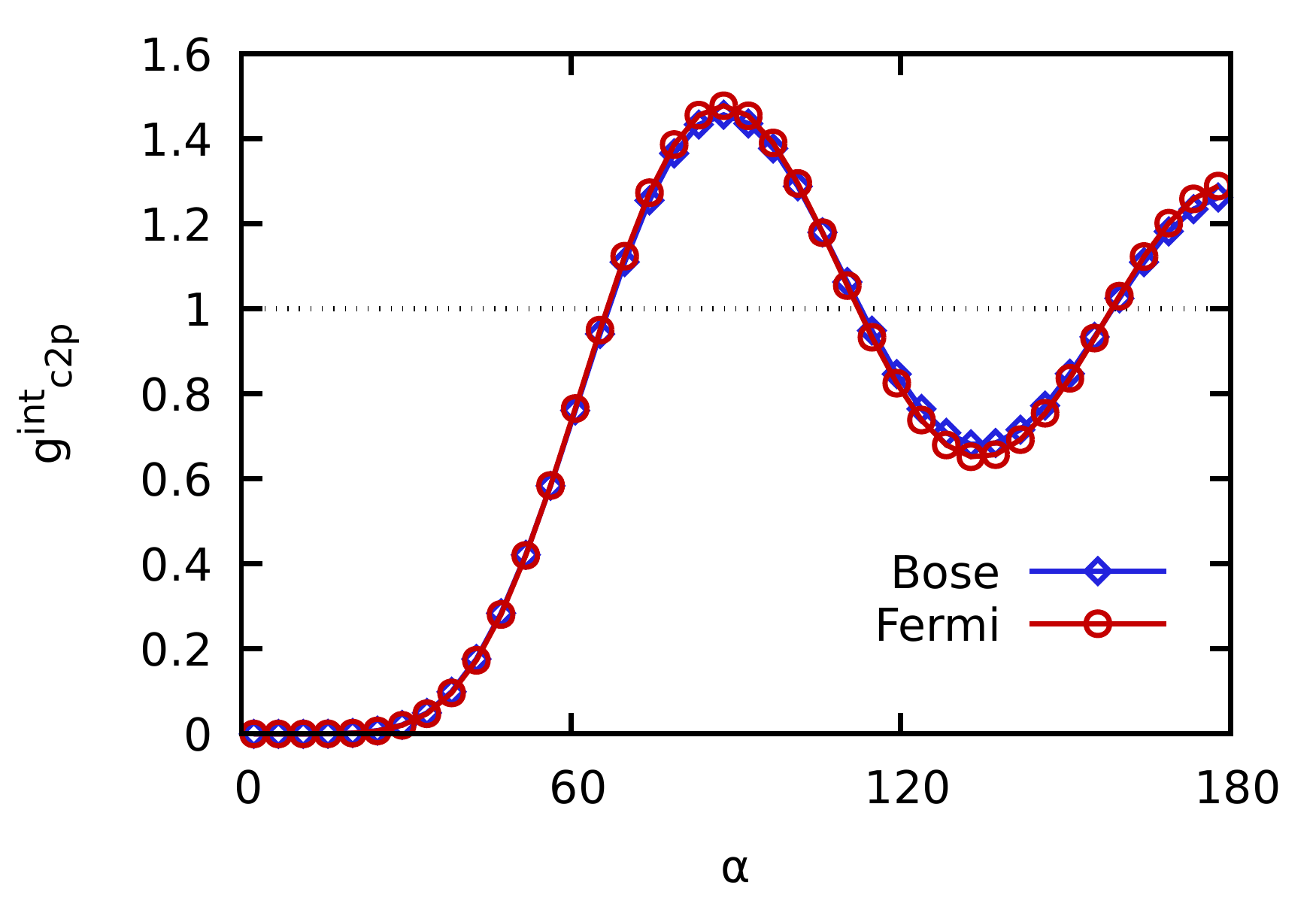}
\caption{\label{fig:Scan_SC}
Scanline of the integrated center-two particle correlation function shown in Fig.~\ref{fig:C2P_SC}, evaluated at $r_1=1.9$. The blue diamons and red circles correspond to bosons and fermions, respectively.
}
\end{figure}

While the onset of negative superfluidity is evidently not connected to a divergence from the bosonic results in the radial density $n(r)$, more subtle pair correlation effects might be resolved using the integrated C2P that was introduced and applied in the previous sections. In Fig.~\ref{fig:C2P_SC}, we show PIMC results for $g^\textnormal{int}_\textnormal{c2p}(r_1,\alpha,1.4,2.4)$ for the same conditions as in Fig.~\ref{fig:LSF_SC}. Remarkably, here, too, we do not find any substantial impact of the type of quantum statistics. This is further confirmed by the scanline shown in Fig.~\ref{fig:Scan_SC}: both the bosonic (blue diamonds) and fermionic (red circles) curve can hardly be distinguished with the naked eye and exhibit the same structure with a pronounced exchange-correlated hole around $\alpha=0$, followed by a first maximum, a minimum, and a subsequent second maximum around $\alpha=180$, i.e., at the opposite end of the system.

We thus conclude that the onset of negative superfluidity does not leave a distinct signature on the structural properties of the system.


\section{Summary and Outlook\label{sec:summary}}

In summary, we have presented extensive \textit{ab initio} PIMC results for quantum dipole systems in a harmonic confinement, taking into account both Bose- and Fermi-statistics. More specifically, we have briefly revisited the noninteracting case, which was used to benchmark the implementation of our simulation scheme and to demonstrate the utility of the integrated C2P as a diagnostic for the impact of quantum statistical effects on the structural properties of the system. 
Subsequently, we have investigated correlated quantum dipole systems, starting with an analysis of the emergence of the exchange--correlation hole upon increasing the coupling parameter $\lambda$. Here we have found that bosons sensitively react even for a small degree of nonideality, whereas coupling effects are effectively masked by the Pauli exclusion principle for fermions. This indicates that mean-field theories and other perturbative methods might perform better for Fermi- as compared to Bose-systems. Moreover, we have investigated radial density profiles, where we were able to clearly resolve the impact of quantum statistics on the respective shell structure. In addition, it was shown that the integrated C2P can be used to measure quantum exchange effects even for parameters where they cannot be detected in averaged quantities like the radial density.

A further important question studied in this work is given by the nonclassical rotational inertia, and how it is affected by quantum statistics and the structural properties of the system. More specifically, we have found that the superfluid fraction of harmonically confined fermions with dipole--dipole interaction can be negative for certain particle numbers $N$, which is in good agreement to a previous study for a different type of pair-interaction~\cite{blume}, and can be explained by the topology of the density matrix. Remarkably, this effect does not seem to be influenced by the structural characteristics of the system that are nearly equal for both bosons and fermions, whereas the moments of inertia diverge from each other.

Let us conclude this work by outlining a few topics for future investigations. First and foremost, we mention that, while the previous study was restricted to finite systems in a harmonic trap, it is possible to extend these efforts to bulk systems in periodic boundary conditions~\cite{dynamic_alex1,dynamic_alex2}.
Here, possible research topics include the bosonization for fermionic bilayer or multilayer systems or the investigation of collective exciations that can be obtained from PIMC data for imaginary-time correlation functions~\cite{berne1} via a subsequent analytical continuation~\cite{jarrell,supersolid_spectrum,dynamic_alex1,dynamic_alex2}. Despite being seemingly ambitious in the light of the FSP, the latter project was recently achieved for correlated electrons in the warm dense matter regime~\cite{dornheim_dynamic,dynamic_folgepaper,dynamic_FSC}, where the sign problem is expected to be even more severe~\cite{dornheim_sign_problem}. Similarly, fermionic PIMC simulations can directly be used to study the static density response~\cite{dornheim_ML,dornheim_electron_liquid,dornheim_HEDP} of bulk quantum dipole systems.

In addition, the investigation of trapped quantum systems is very interesting in its own right, and yields many additional topics for future research. For example, PIMC simulations can be used to estimate the quantum breathing mode~\cite{JWA1,JWA2}, which is possible for different types of pair potentials, including dipole--dipole interaction. Moreover, we mention the study of both crystallization~\cite{hanno,boninsegni_dipole_crystal,crystal_review} and also \emph{quantum melting}~\cite{mezza_melting1,mezza_melting2}, which is not trivial for Monte-Carlo simulations of finite systems. More specifically, the widely used Lindemann-type criteria for melting~\cite{jens_melting} are problematic as they depend on the unphysical dynamics within the respective Markov chain. A better alternative is given by the C2P studied above (or a triple-correlation function in $3D$), as it allows for the estimation of a reduced entropy that signals the onset of different crossovers while also being a proper observable that can potentially be measured in experiments~\cite{thomsen_c2p,ott,Hauke_PHD}.

Lastly, we mention that it is straightforward to extend the present investigation to other geometries~\cite{dyuti1,dyuti2} or pair-potentials, with electrons in $2D$ quantum dots~\cite{alex_wigner,reimann_wigner,reimann_review,dornheim,egger,ghosal,ilkka} constituting a particularly interesting application.

 \section*{Acknowledgments}

This work was partially  funded by the Center of Advanced Systems Understanding (CASUS) which is financed by Germany’s Federal Ministry of Education and Research (BMBF) and by the Saxon Ministry for Science, Culture and Tourism (SMWK) with tax funds on the basis of the budget approved by the Saxon State Parliament.

All PIMC calculations were carried out on the clusters \emph{hypnos} and \emph{hemera} at Helmholtz-Zentrum Dresden-Rossendorf (HZDR) and on a Bull Cluster at the Center for Information Services and High Performance Computing (ZIH) at Technische Universit\"at Dresden.


\section*{References}
\begin{thebibliography}{10}











\bibitem{snoke} D.~Snoke, Spontaneous Bose Coherence of Excitons and Polaritons, \href{https://science.sciencemag.org/content/298/5597/1368?casa_token=77MinSWdRGwAAAAA:yBfw_DlzES6B9hmuQAWs0pT6KDBKcblLRTXVjHrX47tLitVkzK1Yd5lcOiwIMo7TNTRtaDrDLfN1-w}{\textit{Nature} \textbf{298}, 1368-1372} (2002)

\bibitem{cohen} K.~Cohen, R.~Rapaport, and P.V.~Santos, Remote Dipolar Interactions for Objective Density Calibration and Flow Control of Excitonic Fluids, \href{https://journals.aps.org/prl/abstract/10.1103/PhysRevLett.106.126402}{\textit{Phys.~Rev.~Lett.}~\textbf{106}, 126401} (2011)


 
\bibitem{exciton1} J.~B\"oning, A.~Filinov, and M.~Bonitz, Crystallization of an exciton superfluid, \href{https://journals.aps.org/prb/abstract/10.1103/PhysRevB.84.075130}{\textit{Phys.~Rev.~B} \textbf{84}, 075130} (2011)

\bibitem{alex_PRL} A.~Filinov, N.V.~Prokof'ev, and M.~Bonitz, Berezinskii-Kosterlitz-Thouless Transition in Two-Dimensional Dipole Systems, \href{https://journals.aps.org/prl/abstract/10.1103/PhysRevLett.105.070401}{\textit{Phys.~Rev.~Lett.}~\textbf{105}, 070401} 
\bibitem{exciton2} Yu.E.~Lozovik, S.A.~Verzakov, and M.~Willander, Superfluidity of indirect excitons in a quantum dot, \href{https://www.sciencedirect.com/science/article/pii/S0375960199005435}{\textit{Phys.~Lett.~A} \textbf{260}, 400-405} (1999)

\bibitem{exciton3} A.~Filinov, M.~Bonitz, P.~Ludwig, and Yu.E.~Lozovik, Path integral Monte Carlo results for Bose condensation of mesoscopic indirect excitons, \href{https://onlinelibrary.wiley.com/doi/abs/10.1002/pssc.200668038}{\textit{phys.~stat.~sol.~c}~\textbf{3}, 2457-2460} (2006)

\bibitem{exciton4} G.E.~Astrakharchik, J.~Boronat, I.L.~Kurbakov, and Yu.E.~Lozovik, Quantum Phase Transition in a Two-Dimensional System of Dipoles, \href{https://journals.aps.org/prl/abstract/10.1103/PhysRevLett.98.060405}{\textit{Phys.~Rev.~Lett.}~\textbf{98}, 060405} (2007)





\bibitem{Ryd1} G.~Pupillo, A.~Micheli, M.~Boninsegni, I.~Lesanovsky, and P.~Zoller, Strongly Correlated Gases of Rydberg-Dressed Atoms: Quantum and Classical Dynamics, \href{https://journals.aps.org/prl/abstract/10.1103/PhysRevLett.104.223002}{\textit{Phys.~Rev.~Lett.}~\textbf{104}, 223002} (2010)

\bibitem{Ryd2} M.~Saffman, T.G.~Walker, and K.~M{\o}lmer, Quantum information with Rydberg atoms, \href{https://journals.aps.org/rmp/abstract/10.1103/RevModPhys.82.2313}{\textit{Rev.~Mod.~Phys.}~\textbf{82}, 2313} (2010)













\bibitem{stuhler} J.~Stuhler, A.~Griesmaier, T.~Koch, M.~Fattori, T.~Pfau, S.~Giovanazzi, P.~Pedri, and L.~Santos, Observation of Dipole-Dipole Interaction in a Degenerate Quantum Gas, \href{https://journals.aps.org/prl/abstract/10.1103/PhysRevLett.95.150406}{\textit{Phys.~Rev.~Lett.}~\textbf{95}, 150406} (2005)



\bibitem{griesmaier} A.~Griesmaier, J.~Werner, S.~Hensler, J.~Stuhler, and T.~Pfau, Bose-Einstein Condensation of Chromium, \href{https://journals.aps.org/prl/abstract/10.1103/PhysRevLett.94.160401}{\textit{Phys.~Rev.~Lett.}~\textbf{94}, 160401} (2005)





\bibitem{dynamic_alex1} A.~Filinov and M.~Bonitz, Collective and single-particle excitations in two-dimensional dipolar Bose gases, \href{https://journals.aps.org/pra/abstract/10.1103/PhysRevA.86.043628}{\textit{Phys.~Rev.~A} \textbf{86}, 043628} (2012)


\bibitem{dynamic_alex2} A.~Filinov, Correlation effects and collective excitations in bosonic bilayers: Role of quantum statistics, superfluidity, and the dimerization transition, \href{https://journals.aps.org/pra/abstract/10.1103/PhysRevA.94.013603}{\textit{Phys.~Rev.~A} \textbf{94}, 013603} (2016)











\bibitem{cep} D.M.~Ceperley, Path integrals in the theory of condensed helium, \href{http://link.aps.org/doi/10.1103/RevModPhys.67.279}{\textit{Rev. Mod. Phys.} \textbf{67}, 279-355} (1995)



\bibitem{jain} P.~Jain, F.~Cinti, and M.~Boninsegni, Structure, Bose-Einstein condensation, and superfluidity of two-dimensional confined dipolar assemblies, \href{https://journals.aps.org/prb/abstract/10.1103/PhysRevB.84.014534}{\textit{Phys.~Rev.~B} \textbf{84}, 014534} (2011)



\bibitem{rotating} S.M.~Roccuzzo, A.~Gallemi, A.~Recati, and S.~Stringari, Rotating a supersolid dipolar gas, \href{https://arxiv.org/pdf/1910.08513.pdf}{\textit{arxiv:1910.08513}}




\bibitem{supersolid} A.E.~Golomedov, G.E.~Astrakharchik, and Yu.E.~Lozovik, Mesoscopic supersolid of dipoles in a trap, \href{https://journals.aps.org/pra/abstract/10.1103/PhysRevA.84.033615}{\textit{Phys.~Rev.~A} \textbf{84}, 033615} (2011)








\bibitem{boninsegni_dipole_crystal} M.~Boninsegni, Mesoscopic dipolar quantum crystals, \href{https://journals.aps.org/pra/abstract/10.1103/PhysRevA.87.063604}{\textit{Phys.~Rev.~A} \textbf{87}, 063604} (2013)

\bibitem{crystal_review} C.~Cazorla and J.~Boronat, Simulation and understanding of atomic and molecular quantum crystals, \href{https://journals.aps.org/rmp/abstract/10.1103/RevModPhys.89.035003}{\textit{Rev.~Mod.~Phys.}~\textbf{89}, 035003} (2017)









\bibitem{mean_field} M.~Mackie, K.-A.~Suominen, and J.~Javanainen, Mean-Field Theory of Feshbach-Resonant Interactions in $^85$Rb Condensates, \href{https://journals.aps.org/prl/abstract/10.1103/PhysRevLett.89.180403}{\textit{Phys.~Rev.~Lett.}~\textbf{89}, 180403} (2002)

\bibitem{mean_field2} V.D.~Snyder, S.J.J.M.F.~Kokkelmans, and L.D.~Carr, Hartree-Fock-Bogoliubov model and simulation of attractive and repulsive Bose-Einstein condensates, \href{https://journals.aps.org/pra/abstract/10.1103/PhysRevA.85.033616}{\textit{Phys.~Rev.~A} \textbf{85}, 033616} (2012)







\bibitem{bec_review} M.~Combescot, R.~Combescot, and F.~Dubin, Bose-Einstein condensation and indirect excitons: a review, \href{https://iopscience.iop.org/article/10.1088/1361-6633/aa50e3/pdf}{\textit{Rep.~Prog.~Phys.}~\textbf{80}, 066501} (2017)






\bibitem{new_wdm_paper} M.~Bonitz, T.~Dornheim, Zh.A.~Moldabekov, S.~Zhang, P.~Hamann, H.~K\"ahlert, A.~Filinov, K.~Ramakrishna, and J.~Vorberger, \textit{Ab initio} simulation of warm dense matter, \href{https://aip.scitation.org/doi/full/10.1063/1.5143225\%40php.2020.DPP61.issue-1}{\textit{Phys.~Plasmas} \textbf{DPP61}, 042710} (2020)



\bibitem{wdm_book} F.~Graziani, M.P.~Desjarlais, R.~Redmer, and S.B.~Trickey (eds.), Frontiers and Challenges in Warm Dense Matter, Springer International Publishing (2014)











\bibitem{fortov_review} V.E.~Fortov, Extreme states of matter on Earth and in space, \href{https://www.turpion.org/php/paper.phtml?journal_id=pu&paper_id=6821}{\textit{Phys.-Usp.}~\textbf{52}, 615--647} (2009)



\bibitem{falk_wdm} K.~Falk, Experimental methods for warm dense matter research, \href{https://www.cambridge.org/core/journals/high-power-laser-science-and-engineering/article/experimental-methods-for-warm-dense-matter-research/7205AE1029BEA0061044F84875F1CEDB}{\textit{High Power Laser Sci. Eng.}~\textbf{6}, e59} (2018)











\bibitem{berne3} M.F.~Herman, E.J.~Bruskin, and B.J.~Berne, On path integral Monte Carlo simulations, \href{https://aip.scitation.org/doi/abs/10.1063/1.442815}{\textit{J.~Chem.~Phys.}~\textbf{76}, 5150} (1982)








\bibitem{boninsegni1} M.~Boninsegni, N.V.~Prokofev, and B.V.~Svistunov, Worm algorithm and diagrammatic Monte Carlo: A new approach to continuous-space path integral Monte Carlo simulations, \href{https://journals.aps.org/pre/abstract/10.1103/PhysRevE.74.036701}{\textit{Phys.~Rev.~E}~\textbf{74}, 036701} (2006)

\bibitem{boninsegni2} M.~Boninsegni, N.V.~Prokofev, and B.V.~Svistunov, Worm Algorithm for Continuous-Space Path Integral Monte Carlo Simulations, \href{https://journals.aps.org/prl/abstract/10.1103/PhysRevLett.96.070601}{\textit{Phys.~Rev.~Lett.}~\textbf{96}, 070601} (2006)








\bibitem{sf2} E.L.~Pollock and D.M.~Ceperley, Path-integral computation of superfluid densities, \href{https://journals.aps.org/prb/abstract/10.1103/PhysRevB.36.8343}{\textit{Phys.~Rev.~B} \textbf{36}, 8343} (1987)

\bibitem{sindzingre} P.~Sindzingre, M.L.~Klein, and D.M.~Ceperley, Path Integral Monte Carlo study of low-temperature $^4$He Clusters, \href{https://journals.aps.org/prl/abstract/10.1103/PhysRevLett.63.1601}{\textit{Phys.~Rev.~Lett.}~\textbf{63}, 1601} (1989)






\bibitem{supersolid_spectrum} S.~Saccani, S.~Moroni, and M.~Boninsegni, Excitation Spectrum of a Supersolid, \href{https://journals.aps.org/prl/abstract/10.1103/PhysRevLett.108.175301}{\textit{Phys.~Rev.~Lett.}~\textbf{108}, 175301} (2012)




\bibitem{dornheim_dynamic} T.~Dornheim, S.~Groth, J.~Vorberger, and M.~Bonitz, \textit{Ab initio} Path Integral Monte Carlo Results for the Dynamic Structure Factor of Correlated Electrons: From the Electron Liquid to Warm Dense Matter, \href{https://journals.aps.org/prl/abstract/10.1103/PhysRevLett.121.255001}{\textit{Phys.~Rev.~Lett.}~\textbf{121}, 255001} (2018)


\bibitem{dynamic_folgepaper} S.~Groth, T.~Dornheim, and J.~Vorberger, \textit{Ab Initio} Path Integral Monte Carlo Approach to the Static and Dynamic Density Response of the Uniform Electron Gas, \href{https://link.aps.org/doi/10.1103/PhysRevB.99.235122}{\textit{Phys.~Rev.~B} \textbf{99}, 235122} (2019)

\bibitem{dynamic_FSC} T.~Dornheim and J.~Vorberger, Finite-size effects in the reconstruction of dynamic properties from ab initio path integral Monte-Carlo simulations, \href{https://arxiv.org/abs/2004.13429}{arXiv:2004.13429}









\bibitem{loh} E.Y.~Loh, J.E.~Gubernatis, R.T.~Scalettar, S.R.~White, D.J.~Scalapino and R.L.~Sugar, Sign problem in the numerical simulation of many-electron systems, \href{http://link.aps.org/doi/10.1103/PhysRevB.41.9301}{\textit{Phys. Rev. B} \textbf{41}, 9301-9307} (1990)


\bibitem{troyer} M.~Troyer and U.J.~Wiese, Computational Complexity and Fundamental Limitations to Fermionic Quantum Monte Carlo Simulations, \href{http://link.aps.org/doi/10.1103/PhysRevLett.94.170201}{\textit{Phys. Rev. Lett.} \textbf{94}, 170201} (2005)







\bibitem{dornheim_sign_problem} T.~Dornheim, Fermion sign problem in path integral Monte Carlo simulations: Quantum dots, ultracold atoms, and warm dense matter, \href{https://journals.aps.org/pre/abstract/10.1103/PhysRevE.100.023307}{\textit{Phys.~Rev.~E} \textbf{100}, 023307} (2019)


















\bibitem{cpimc_original} T.~Schoof, M.~Bonitz, A.~Filinov, D.~Hochstuhl, and J.W.~Dufty, Configuration Path Integral Monte Carlo, \href{https://onlinelibrary.wiley.com/doi/abs/10.1002/ctpp.201100012}{\textit{Contrib.~Plasma Phys.}~\textbf{51}, 687-697} (2011)


\bibitem{brown_ethan} E.W.~Brown, B.K.~Clark, J.L.~DuBois, and D.M.~Ceperley, Path-Integral Monte Carlo Simulation of the Warm Dense Homogeneous Electron Gas, \href{https://journals.aps.org/prl/abstract/10.1103/PhysRevLett.110.146405}{\textit{Phys.~Rev.~Lett.}~\textbf{110}, 146405} (2013)

\bibitem{blunt1} N.S.~Blunt, T.W.~Rogers, J.S.~Spencer, and W.M.C.~Foulkes, Density-matrix quantum Monte Carlo method, \href{https://journals.aps.org/prb/abstract/10.1103/PhysRevB.89.245124}{\textit{Phys.~Rev.~B}~\textbf{89}, 245124} (2014)

\bibitem{schoof_prl} T.~Schoof, S.~Groth, J.~Vorberger, and M.~Bonitz, \textit{Ab Initio} Thermodynamic Results for the Degenerate Electron Gas at Finite Temperature, \href{https://journals.aps.org/prl/abstract/10.1103/PhysRevLett.115.130402}{\textit{Phys.~Rev.~Lett.}~\textbf{115}, 130402} (2015)

\bibitem{malone1} F.D.~Malone, N.S.~Blunt, J.J.~Shepherd, D.K.K.~Lee, J.S.~Spencer, and W.M.C.~Foulkes, Interaction picture density matrix quantum Monte Carlo, \href{https://aip.scitation.org/doi/abs/10.1063/1.4927434}{\textit{J.~Chem.~Phys.}~\textbf{143}, 044116} (2015)

\bibitem{blunt2} N.S.~Blunt, A.~Alavi, and G.H.~Booth, Krylov-Projected Quantum Monte Carlo Method, \href{https://journals.aps.org/prl/abstract/10.1103/PhysRevLett.115.050603}{\textit{Phys.~Rev.~Lett.}~\textbf{115}, 050603} (2015)

\bibitem{malone2} F.D.~Malone, N.S.~Blunt, E.W.~Brown, D.K.K.~Lee, J.S.~Spencer, W.M.C.~Foulkes, and J.J.~Shepherd, Accurate Exchange-Correlation Energies for the Warm Dense Electron Gas, \href{https://journals.aps.org/prl/abstract/10.1103/PhysRevLett.117.115701}{\textit{Phys.~Rev.~Lett.}~\textbf{117}, 115701} (2016)


\bibitem{dornheim} T.~Dornheim, S.~Groth, A.~Filinov and M.~Bonitz, Permutation blocking path integral Monte Carlo: a highly efficient approach to the simulation of strongly degenerate non-ideal fermions, \href{ http://iopscience.iop.org/1367-2630/17/7/073017 }{ \textit{New J. Phys.} \textbf{17}, 073017} (2015)

\bibitem{dornheim2} T.~Dornheim, T.~Schoof, S.~Groth, A.~Filinov, and M.~Bonitz, Permutation Blocking Path Integral Monte Carlo Approach to the Uniform Electron Gas at Finite Temperature, \href{ http://scitation.aip.org/content/aip/journal/jcp/143/20/10.1063/1.4936145 }{ \textit{ J. Chem. Phys.} \textbf{143}, 204101} (2015)


\bibitem{vladimir_UEG} V.S.~Filinov, V.E.~Fortov, M.~Bonitz, and Zh.A.~Moldabekov, Fermionic path-integral Monte Carlo results for the uniform electron gas at finite temperature, \href{https://journals.aps.org/pre/abstract/10.1103/PhysRevE.91.033108}{\textit{Phys.~Rev.~E} \textbf{91}, 033108} (2015)


\bibitem{groth} S.~Groth, T.~Schoof, T.~Dornheim, and M.~Bonitz, \textit{Ab Initio} Quantum Monte Carlo Simulations of the Uniform Electron Gas without Fixed Nodes, \href{ http://link.aps.org/doi/10.1103/PhysRevB.93.085102 }{ \textit{Phys.~Rev.~B} \textbf{93}, 085102} (2016)


\bibitem{dornheim3} T.~Dornheim, S.~Groth, T.~Schoof, C.~Hann, and M.~Bonitz, \textit{Ab initio} quantum Monte Carlo simulations of the Uniform electron gas without fixed nodes: The unpolarized case, \href{ http://link.aps.org/doi/10.1103/PhysRevB.93.205134 }{ \textit{Phys.~Rev.~B} \textbf{93}, 205134 } (2016)


\bibitem{dornheim_prl} T.~Dornheim, S.~Groth, T.~Sjostrom, F.D.~Malone, W.M.C.~Foulkes, and M.~Bonitz, \textit{Ab Initio}   Quantum Monte Carlo Simulation of the Warm Dense Electron Gas in the Thermodynamic Limit, \href{ http://link.aps.org/doi/10.1103/PhysRevLett.117.156403}{\textit{Phys.~Rev.~Lett.}~\textbf{117}, 156403} (2016)




\bibitem{groth_prl} S.~Groth, T.~Dornheim, T.~Sjostrom, F.D.~Malone, W.M.C.~Foulkes, and M.~Bonitz, \textit{Ab initio} Exchange--Correlation Free Energy of the Uniform Electron Gas at Warm Dense Matter Conditions, \href{https://journals.aps.org/prl/abstract/10.1103/PhysRevLett.119.135001}{\textit{Phys.~Rev.~Lett.}~\textbf{119}, 135001} (2017)


\bibitem{dubois} J.L.~DuBois,  E.W.~Brown,  and  B.J.~Alder,  Overcoming the
Fermion Sign Problem in Homogeneous Systems, in E.~Schwegler,
B.M.~Rubenstein, and S.B.~Libby (Eds.),
Advances in the Computational Sciences-Symposium in Honor of Dr Berni Alder's
90th Birthday, World Scientific, Singapore (2017)




\bibitem{claes} J.~Claes and B.K.~Clark, Finite-temperature properties of strongly correlated systems via variational Monte Carlo, \href{https://journals.aps.org/prb/abstract/10.1103/PhysRevB.95.205109}{\textit{Phys.~Rev.~B} \textbf{95}, 205109} (2017)




\bibitem{dornheim_cpp} T.~Dornheim, S.~Groth, and M.~Bonitz, Ab initio results for the Static Structure Factor of the Warm Dense Electron Gas, \href{https://onlinelibrary.wiley.com/doi/full/10.1002/ctpp.201700096}{\textit{Contrib.~Plasma Phys.}~\textbf{57}, 468-478} (2017)





\bibitem{brenda} Y.~Liu, M.~Cho, and B.~Rubenstein, \textit{Ab Initio} Finite Temperature Auxiliary Field Quantum Monte Carlo, \href{https://pubs.acs.org/doi/abs/10.1021/acs.jctc.8b00569}{\textit{J.~Chem.~Theory~Comput.}~\textbf{14}, 4722-4732} (2018)


\bibitem{universe} V.~Filinov and A.~Larkin, Quantum Dynamics of Charged Fermions in the Wigner Formulation of Quantum Mechanics, \href{https://www.mdpi.com/2218-1997/4/12/133/htm}{\textit{Universe} \textbf{4}, 133} (2018)

\bibitem{dornheim_neu} T.~Dornheim, S.~Groth, and M.~Bonitz, Permutation Blocking Path Integral Monte Carlo Simulations of Degenerate Electrons at Finite Temperature, \href{https://onlinelibrary.wiley.com/doi/full/10.1002/ctpp.201800157}{\textit{Contrib.~Plasma Phys.}~\textbf{59}, e201800157} (2019)





\bibitem{dornheim_pop} T.~Dornheim, S.~Groth, F.D.~Malone, T. Schoof, T.~Sjostrom, W.M.C.~Foulkes, and M.~Bonitz, \textit{Ab Initio}   Quantum Monte Carlo Simulation of the Warm Dense Electron Gas, \href{ http://aip.scitation.org/doi/full/10.1063/1.4977920}{ \textit{Phys.~Plasmas}  {\bf 24}, 056303} (2017)














\bibitem{blume_review} D.~Blume, Few-body physics with ultracold atomic and molecular systems in traps, \href{https://iopscience.iop.org/article/10.1088/0034-4885/75/4/046401/meta}{\textit{Rep.~Prog.~Phys.}~\textbf{75}, 046401} (2012)













\bibitem{JWA1} J.W.~Abraham, M.~Bonitz, C.~McDonald, G.~Orlando, and T.~Brabec, Quantum breathing mode of trapped systems in one and two dimensions, \href{https://iopscience.iop.org/article/10.1088/1367-2630/16/1/013001}{\textit{New J.~Phys.}~\textbf{16}, 013001} (2014)


\bibitem{JWA2} J.W.~Abraham and M.~Bonitz, Quantum Breathing Mode of Trapped Particles: From Nanoplasmas to Ultracold Gases, \href{https://onlinelibrary.wiley.com/doi/abs/10.1002/ctpp.201300066}{\textit{Contrib.~Plasma Phys.}~\textbf{54}, 27-99} (2014)



\bibitem{dornheim_c2p} T.~Dornheim, H.~Thomsen, P.~Ludwig, A.~Filinov, and M.~Bonitz, Analyzing Quantum Correlations Made Simple, \href{https://onlinelibrary.wiley.com/doi/abs/10.1002/ctpp.201500120}{\textit{Contrib.~Plasma Phys.}~\textbf{56}, 371-379} (2016)




\bibitem{alex_wigner} A.V.~Filinov, M.~Bonitz, Yu.E.~Lozovik, Wigner Crystallization in Mesoscopic 2D Electron Systems, \href{https://journals.aps.org/prl/abstract/10.1103/PhysRevLett.86.3851}{\textit{Phys.~Rev.~Lett.}~\textbf{86}, 3851} (2001)

\bibitem{reimann_wigner} S.M.~Reimann, M.~Koskinen, and M.~Manninen, Formation of Wigner molecules in small quantum dots, \href{https://journals.aps.org/prb/abstract/10.1103/PhysRevB.62.8108}{\textit{Phys.~Rev.~B} \textbf{62}, 8108} (2000)

\bibitem{reimann_review} S.M.~Reimann and M.~Manninen, Electronic structure of quantum dots, \href{https://journals.aps.org/rmp/abstract/10.1103/RevModPhys.74.1283}{\textit{Rev.~Mod.~Phys.}~\textbf{74}, 1283} (2002)

\bibitem{egger} R.~Egger, W.~H\"ausler, C.H.~Mak, and H.~Grabert, Crossover from Fermi Liquid to Wigner Molecule Behavior in Quantum Dots, \href{https://journals.aps.org/prl/abstract/10.1103/PhysRevLett.82.3320}{\textit{Phys.~Rev.~Lett.}~\textbf{82}, 3320} (1999)

\bibitem{ghosal} A.~Ghosal, A.D.~G\"ucl\"u, C.J.~Umrigar, D.~Ullmo, and H.U.~Baranger, Correlation-induced inhomogeneity in circular quantum dots, \href{https://www.nature.com/articles/nphys293}{\textit{Nature Phys.}~\textbf{2}, 336-340} (2006)

\bibitem{ilkka} I.~Kyl\"anp\"a\"a and E.~R\"as\"anen, Path integral Monte Carlo benchmarks for two-dimensional quantum dots, \href{https://journals.aps.org/prb/abstract/10.1103/PhysRevB.96.205445}{\textit{Phys.~Rev.~B} \textbf{96}, 205445} (2017)




\bibitem{dyuti1} D.~Bhattacharya, A.V.~Filinov, A.~Ghosal, and M.~Bonitz, Role of confinements on the melting of Wigner molecules in quantum dots, \href{https://link.springer.com/article/10.1140/epjb/e2016-60448-5}{\textit{Eur.~Phys.~J.~B} \textbf{89}, 60} (2016)

\bibitem{dyuti2} D.~Bhattacharya and A.~Ghosal, Melting of Coulomb-interacting classical particles in 2D irregular traps, \href{https://link.springer.com/article/10.1140/epjb/e2013-40568-2}{\textit{Eur.~Phys.~J.~B} \textbf{86}, 499} (2013)













\bibitem{yushi} Y.~Shi, Superfluidity or supersolidity as a consequence of off-diagonal long-range order, \href{https://journals.aps.org/prb/abstract/10.1103/PhysRevB.72.014533}{\textit{Phys.~Rev.~B} \textbf{72}, 014533} (2005)




\bibitem{kwon_lsf} Y.~Kwon, F.~Paesani, and K.B.~Whaley, Local superfluidity in inhomogeneous quantum fluids, \href{https://journals.aps.org/prb/abstract/10.1103/PhysRevB.74.174522}{\textit{Phys.~Rev.~B} \textbf{74}, 174522} (2006)

\bibitem{mezza} F.~Mezzacapo and M.~Boninsegni, Local Superfluidity of Parahydrogen Clusters, \href{https://journals.aps.org/prl/abstract/10.1103/PhysRevLett.100.145301}{\textit{Phys.~Rev.~Lett.}~\textbf{100}, 145301} (2008)

\bibitem{filinov_lsf} A.~Filinov, J.~B\"oning, M.~Bonitz, and Yu.~Lozovik, Controlling the spatial distribution of superfluidity in radially ordered Coulomb clusters \href{https://journals.aps.org/prb/abstract/10.1103/PhysRevB.77.214527}{\textit{Phys.~Rev.~B} \textbf{77}, 214527} (2008)





\bibitem{lsf} B.~Kulchytskyy, G.~Gervais, and A.~Del Maestro, Local superfluidity at the nanoscale, \href{https://journals.aps.org/prb/abstract/10.1103/PhysRevB.88.064512}{\textit{Phys.~Rev.~B} \textbf{88}, 064512} (2013)



\bibitem{dornheim_superfluid} T.~Dornheim, A.~Filinov, and M.~Bonitz, Superfluidity of strongly correlated bosons in two- and three-dimensional traps, \href{https://journals.aps.org/prb/abstract/10.1103/PhysRevB.91.054503}{\textit{Phys.~Rev.~B} \textbf{91}, 054503} (2015)












\bibitem{blume} Y.~Yan and D.~Blume, Abnormal Superfluid Fraction of Harmonically Trapped Few-Fermion Systems, \href{https://journals.aps.org/prl/abstract/10.1103/PhysRevLett.112.235301}{\textit{Phys.~Rev.~Lett.}~\textbf{112}, 235301} (2014)
















\bibitem{thomsen_c2p} H.~Thomsen and M.~Bonitz, Resolving structural transitions in spherical dust clusters, \href{https://journals.aps.org/pre/abstract/10.1103/PhysRevE.91.043104}{\textit{Phys.~Rev.~E} \textbf{91}, 043104} (2015)

\bibitem{ott} T.~Ott, H.~Thomsen, J.W.~Abraham, T.~Dornheim, and M.~Bonitz, Recent progress in the theory and simulation of strongly correlated plasmas: phase transitions, transport, quantum, and magnetic field effects, \href{https://link.springer.com/article/10.1140/epjd/e2018-80385-7}{\textit{Eur.~Phys.~J.~D} \textbf{72}, 84} (2018)



\bibitem{Hauke_PHD} H.~Thomsen, Melting Processes and Laser Manipulation of Strongly Coupled Yukawa Systems, \href{http://www.theo-physik.uni-kiel.de/bonitz/theses/thomsen_15.pdf}{\textit{PhD thesis}}, Christian-Albrechts-Universit\"a zu Kiel (2015)












\bibitem{hirshberg} B.~Hirshberg, M.~Invernizzi, and M.~Parrinello, Path Integral Molecular Dynamics for Fermions: Alleviating the Sign Problem with the Bogoliubov Inequality, \href{https://arxiv.org/abs/2003.10317}{arXiv:2003.10317}













\bibitem{tune1} S.~Giovanazzi, A.~G\"orlitz, and T.~Pfau, Tuning the Dipolar Interaction in Quantum Gases, \href{https://journals.aps.org/prl/abstract/10.1103/PhysRevLett.89.130401}{\textit{Phys.~Rev.~Lett.}~\textbf{89}, 130401} (2002)


\bibitem{tune2} Ph.~Courteille, R.S.~Freeland, D.J.~Heinzen, F.A.~van Abeelen, and B.J.~Verhaar, Observation of a Feshbach Resonance in Cold Atom Scattering, \href{https://journals.aps.org/prl/abstract/10.1103/PhysRevLett.81.69}{\textit{Phys.~Rev.~Lett.}~\textbf{81}, 69} (1998)













\bibitem{dornheim_permutation_cycles} T.~Dornheim, S.~Groth, A.~Filinov, and M.~Bonitz, Path Integral Monte Carlo Simulation of Degenerate Electrons: Permutation-Cycle Properties, \href{https://aip.scitation.org/doi/10.1063/1.5093171}{\textit{J.~Chem.~Phys.}~\textbf{151}, 014108} (2019)



\bibitem{trotter} H.~De Raedt and B.~De Raedt, Applications of the generalized Trotter formula, \href{https://journals.aps.org/pra/abstract/10.1103/PhysRevA.28.3575}{\textit{Phys.~Rev.~A} \textbf{28}, 3575} (1983)



\bibitem{metropolis} N.~Metropolis, A.W.~Rosenbluth, M.N.~Rosenbluth, A.H.~Teller, and E.~Teller, Equation of State Calculations by Fast Computing Machines, \href{https://aip.scitation.org/doi/abs/10.1063/1.1699114}{\textit{J.~Chem.~Phys.}~\textbf{21}, 1087} (1953)




\bibitem{ceperley_fermions} D.M.~Ceperley, Path Integral Monte Carlo Methods for Fermions, \textit{Monte Carlo and Molecular Dynamics of Condensed Matter Systems}, K.~Binder and G.~Ciccotti (Eds.), Bologna (Italy) (1996)







\bibitem{draeger} E.W.~Draeger and D.M.~Ceperley, Superfluidity in a Doped Helium Droplet, \href{https://journals.aps.org/prl/abstract/10.1103/PhysRevLett.90.065301}{\textit{Phys.~Rev.~Lett.}~\textbf{90}, 065301} (2003)






\bibitem{krauth_book} W.~Krauth,
Statistical  Mechanics: Algorithms  and  Computations, \textit{Oxford Master Series in Physics}, Oxford University Press, 2006







\bibitem{greiner_book} W.~Greiner, D.~Rischke, L.~Neise, and H.~St\"ocker, Thermodynamics and Statistical Mechanics, Springer, New York (2012)







\bibitem{janke} W.~Janke and T.~Sauer, Optimal energy estimation in path-integral Monte Carlo simulations, \href{https://aip.scitation.org/doi/abs/10.1063/1.474309}{\textit{J.~Chem.~Phys.}~\textbf{107}, 5821} (1997)











\bibitem{berne1} D.~Thirumalai and B.J.~Berne, On the calculation of time correlation functions in quantum systems: Path integral techniques, \href{https://aip.scitation.org/doi/abs/10.1063/1.445597}{\textit{J.~Chem.~Phys.}~\textbf{79}, 5029} (1983)


\bibitem{jarrell} M.~Jarrell and J.E.~Gubernatis, Bayesian inference and the analytic continuation of imaginary-time quantum Monte Carlo data, \href{https://www.sciencedirect.com/science/article/abs/pii/0370157395000747}{\textit{Phys.~Reports}~\textbf{269}, 133-195} (1996)











\bibitem{dornheim_ML} T.~Dornheim, J.~Vorberger, S.~Groth, N.~Hoffmann, Zh.A.~Moldabekov, and M.~Bonitz, The Static Local Field Correction of the Warm Dense Electron Gas: An ab Initio Path Integral Monte Carlo Study and Machine Learning Representation, \href{https://aip.scitation.org/doi/full/10.1063/1.5123013}{\textit{J.~Chem.~Phys.}~\textbf{151}, 194104} (2019)




\bibitem{dornheim_electron_liquid} T.~Dornheim, T.~Sjostrom, S.~Tanaka, and J.~Vorberger, Strongly Coupled Electron Liquid: ab initio Path Integral Monte Carlo Simulations and Dielectric Theories, \href{https://journals.aps.org/prb/abstract/10.1103/PhysRevB.101.045129}{\textit{Phys.~Rev.~B} \textbf{101}, 045129} (2020)



\bibitem{dornheim_HEDP} T.~Dornheim, Zh.A.~Moldabekov, J.~Vorgerber, and S.~Groth, Ab initio path integral Monte Carlo simulation of the uniform electron gas in the high energy density regime \href{https://iopscience.iop.org/article/10.1088/1361-6587/ab8bb4/meta}{\textit{Plasma Phys.~Control.~Fusion}} (in press)










\bibitem{hanno} H.~K\"ahlert and M.~Bonitz, How Spherical Plasma Crystals Form, \href{https://journals.aps.org/prl/abstract/10.1103/PhysRevLett.104.015001}{\textit{Phys.~Rev.~Lett.}~\textbf{104}, 015001} (2010)






\bibitem{mezza_melting1} F.~Mezzacapo and M.~Boninsegni, Superfluidity and Quantum Melting of p-$\textnormal{H}_2$ Clusters, \href{https://journals.aps.org/prl/abstract/10.1103/PhysRevLett.97.045301}{\textit{Phys.~Rev.~Lett.}~\textbf{97}, 045301} (2006)

\bibitem{mezza_melting2} F.~Mezzacapo and M.~Boninsegni, Structure, superfluidity, and quantum melting of hydrogen clusters, \href{https://journals.aps.org/pra/abstract/10.1103/PhysRevA.75.033201}{\textit{Phys.~Rev.~A} \textbf{75}, 033201} (2007)



\bibitem{jens_melting} J.~B\"oning, A.~Filinov, P.~Ludwig, H.~Baumgartner, M.~Bonitz, and Yu.E.~Lozovik, Melting of trapped few-particle systems, \href{https://journals.aps.org/prl/abstract/10.1103/PhysRevLett.100.113401}{\textit{Phys.~Rev.~Lett.}~\textbf{100}, 113401} (2008)



















































































































\end{thebibliography}
\end{document}